\documentclass[a4paper,12pt]{article}
\usepackage{amsmath}
\usepackage{amssymb}
\usepackage{amsfonts}
\usepackage{amssymb,amsmath}
\usepackage{cancel}
\usepackage{graphicx}
\usepackage{epsfig}
\usepackage{verbatim}
\usepackage{fleqn}
\usepackage[footnotesize]{caption}
\usepackage{mathrsfs}

\def\beq{\begin{equation}}
\def\eeq{\end{equation}}
\def\bea{\begin{eqnarray}}
\def\eea{\end{eqnarray}}

\def\<{\left\langle}
\def\>{\right\rangle}

\newcommand{\bc}{\begin{center}}
\newcommand{\ec}{\end{center}}
\newcommand{\bd}{\begin{displaymath}}
\newcommand{\ed}{\end{displaymath}}
\newcommand{\be}{\begin{equation}}
\newcommand{\ee}{\end{equation}}
\newcommand{\ba}{\begin{array}}
\newcommand{\ea}{\end{array}}
\newcommand{\bt}{\begin{tabular}}
\newcommand{\et}{\end{tabular}}

\newcommand{\ds}{\displaystyle}

\begin{document}

\bibliographystyle{OurBibTeX}

\begin{titlepage}

%\vspace*{-15mm}
%\begin{flushright}
%SHEP-\\
%\end{flushright}
%\vspace*{5mm}

\begin{flushright}
DFTT 38/2009\\
\end{flushright}

\begin{center}
{ \sffamily \Large The Constrained Exceptional
Supersymmetric Standard Model }
\\[8mm]
P.~Athron$^{a}$,
%\footnote{E-mail: \texttt{p.athron@physics.gla.ac.uk}}
S.F.~King$^{b}$,
%\footnote{E-mail: \texttt{sfk@hep.phys.soton.ac.uk}}
D.J.~Miller$^{c}$,
%\footnote{E-mail: \texttt{d.miller@physics.gla.ac.uk}}
S.~Moretti$^{b,\,d}$
%\footnote{E-mail: \texttt{stefano@phys.soton.ac.uk}}
and
R.~Nevzorov$^{c}$\footnote{On leave of absence from the Theory Department,
ITEP, Moscow, Russia
}\\[3mm]
{\small\it
$^a$ Institut f\"ur Kern und Teilchenphysik, TU Dresden, Dresden, D-01062, Germany.\\[2mm]
$^b$ School of Physics and Astronomy, University of Southampton,\\
Southampton, SO17 1BJ, U.K.\\[2mm]
$^c$ Department of Physics and Astronomy, University of Glasgow,\\
Glasgow G12 8QQ, U.K.\\[2mm]
$^d$Dipartimento di Fisica Teorica,
Universita' degli Studi di Torino,\\
Via Pietro Giuria 1,
10125 Torino,
Italy}
\\[1mm]
\end{center}
\vspace*{0.75cm}

\begin{abstract}
\noindent
We propose and study a constrained version of the Exceptional
Supersymmetric Standard Model (E$_6$SSM), which we call the cE$_6$SSM,
based on a universal high energy scalar mass $m_0$, trilinear scalar
coupling $A_0$ and gaugino mass $M_{1/2}$.  We derive the Renormalisation
Group (RG) Equations for the cE$_6$SSM, including the extra $U(1)_{N}$
gauge factor and the low energy matter content involving three $27$
representations of $E_6$. We perform a numerical RG analysis for the
cE$_6$SSM, imposing the usual low energy experimental constraints
and successful Electro-Weak Symmetry Breaking (EWSB). Our analysis
reveals that the sparticle spectrum of the cE$_6$SSM
involves a light gluino, two light neutralinos and a light chargino.
Furthermore, although the squarks, sleptons and $Z'$
boson are typically heavy, the exotic quarks and squarks can also 
be relatively light. We finally specify a set of benchmark points which 
correspond to particle spectra, production modes and decay patterns
peculiar to the cE$_6$SSM, altogether
 leading to spectacular new physics
signals at the Large Hadron Collider (LHC).
\end{abstract}

\end{titlepage}
\newpage
\setcounter{footnote}{0}

\section{Introduction}

Supersymmetry (SUSY) provides an attractive framework that allows one
to link gravity with the other fundamental forces of nature. Indeed,
it is well known that local SUSY (Supergravity) leads to a partial
unification of the Electro-Weak (EW), strong and gravitational
interactions \cite{sugra}. At some high energy scale local SUSY in
Supergravity (SUGRA) models can be spontaneously broken in a hidden
sector. Then the low--energy limit of such a theory is described by a
global SUSY Lagrangian plus a set of soft SUSY--breaking terms
\cite{soft-terms-1} which do not induce quadratic divergences, thus
preserving the Supersymmetric solution to the hierarchy problem
\cite{1} (for a recent review see \cite{Chung:2003fi}). A set of soft
SUSY--breaking terms involves gaugino masses $M_a$, soft scalar masses
$m^2_i$, plus bilinear ($B_i$) and trilinear ($A_i$) scalar couplings
\cite{soft-terms-2}. If the SUSY--breaking scale is within a few TeV
then the $SU(3)_C$, $SU(2)_W$ and $U(1)_Y$ gauge couplings converge to
a common value near the scale $M_X\simeq 2-3\cdot 10^{16}\,{\rm GeV}$
\cite{gauge-unif}, which allows one to embed SUSY extensions of the
Standard Model (SM) into Grand Unified Theories (GUTs)
\cite{GUTs}. The rational $U(1)_Y$ charges, which are postulated {\it
ad hoc} in the SM, then appear in a natural way in the context of SUSY
GUT models after the breakdown of the extended symmetry -- such as
$SU(5)$, $SO(10)$ or $E_6$ -- at the scale $M_X$.

However, the incorporation of the simplest SUSY extension of the SM
--- the Minimal Supersymmetric Standard Model (MSSM) --- into SUGRA or
SUSY GUT models leads to the $\mu$--problem
\cite{Chung:2003fi}. The Superpotential of the MSSM contains
one bilinear term $\mu \hat{H}_d\hat{H}_u$ that can be present before
SUSY is broken.  One would naturally expect the parameter
$\mu$ to be either zero or of the order of the Planck scale. On the
one hand, if $\mu\simeq M_{\rm{Pl}}$ then the Higgs scalars acquire a huge
positive contribution $\sim\mu^2$ to their squared masses and EW
Symmetry Breaking (EWSB) does not occur.  On the other hand, if
$\mu=0$ at some scale $Q$ the mixing between Higgs doublets is not
generated at any scale below $Q$ due to non--renormalisation
theorems \cite{10} so that $\langle H_d \rangle =0$ and down--type
quarks and charged leptons remain massless.  The correct pattern of
EWSB requires $\mu$ to be of the order of the SUSY--breaking (or EW)
scale.

An elegant solution to the $\mu$--problem naturally arises in the
framework of Superstring inspired $E_6$ models. Ten--dimensional
heterotic Superstring theory based on $E_8\times E'_8$ \cite{5} can
play a role in the ultraviolet completion of the non--renormalisable
SUGRA models.  In the strong coupling regime of an $E_8\times E'_8$
heterotic string theory, which is described by eleven dimensional
Supergravity (M--theory) \cite{6}, the string scale can be compatible
with the unification scale $M_X$ \cite{7}. Compactification of the
extra dimensions results in the breakdown of $E_8$ down to $E_6$ or
one of its subgroups in the observable sector \cite{8}. The remaining
$E'_8$ couples to the usual matter representations of the $E_6$ group
only by virtue of gravitational interactions and comprises a hidden
sector that gives rise to spontaneous breakdown of local SUSY. At low
energies the hidden sector decouples from the observable one. The only
signal it produces is a set of soft SUSY--breaking terms characterised
by the gravitino mass ($m_{3/2}$) scale\footnote{In the most general
case a complete set of expressions for the soft SUSY--breaking
parameters can be found in \cite{9}.}  which spoil the degeneracy
between bosons and fermions within one Supermultiplet.

At the string scale, $E_6$ can be broken via the Hosotani mechanism
\cite{11}.  The breakdown of the $E_6$ symmetry results in several
models based on rank--5 or rank--6 gauge groups. Therefore Superstring
inspired $E_6$ models may lead to low--energy gauge groups with one or
two additional $U(1)'$ factors in comparison to the SM. In particular,
$E_6$ can be broken directly to the rank--6 subgroup $SU(3)_C\times
SU(2)_W\times U(1)_Y\times U(1)_{\psi}\times U(1)_{\chi}$. Two
anomaly-free $U(1)_{\psi}$ and $U(1)_{\chi}$ symmetries of the rank-6
model are defined by \cite{12}: $E_6\to SO(10)\times
U(1)_{\psi},~SO(10)\to SU(5)\times U(1)_{\chi}$. This rank--6 model
can be reduced further to an effective rank--5 model with only one
extra gauge symmetry $U(1)'$ which is a linear combination of
$U(1)_{\chi}$ and $U(1)_{\psi}$:
\be
U(1)'=U(1)_{\chi}\cos\theta+U(1)_{\psi}\sin\theta\,.
\label{cessm1}
\ee
If $\theta\ne 0\,\mbox{or}\,\,\pi$ the extra $U(1)'$ gauge symmetry
forbids an elementary $\mu$ term but allows an interaction of the
extra SM singlet Superfield $\hat{S}$ with the Higgs Supermultiplets
$\hat{H}_d$ and $\hat{H}_u$ in the Superpotential:
$\lambda\hat{S}\hat{H}_d\hat{H}_u$.  After EWSB the scalar component
of the SM singlet Superfield $\hat{S}$ acquires a non-zero VEV
breaking $U(1)'$ and an effective $\mu$--term of the required size is
automatically generated \cite{Athron:2007en}.  Thus in Superstring
inspired $E_6$ models the $\mu$--problem is solved in a similar way to
the Next--to--Minimal Supersymmetric Standard Model (NMSSM)
\cite{nmssm}, but without the accompanying problems of singlet
tadpoles or domain walls \cite{14}.

$E_6$ inspired SUSY models with an extra $U(1)'$ have been extensively
studied \cite{12}, \cite{141}. In general the models predict extra
exotic matter beyond the MSSM and NMSSM. The large couplings of exotic
quarks ($D, \bar{D}$) to the SM singlet $S$ of the form $\kappa
S(D\overline{D})$ may induce radiative breakdown of the extra $U(1)'$
symmetry \cite{15}, \cite{Langacker:1998tc}--\cite{Daikoku:2000ep}.
An important feature of $E_6$ inspired SUSY models is that the mass of
the lightest Higgs particle can be substantially larger in these
scenarios than in the MSSM and NMSSM \cite{Daikoku:2000ep}.
Previously, the implications of $E_6$ inspired SUSY models with an
additional $U(1)'$ gauge symmetry have been studied for EWSB
\cite{Langacker:1998tc}--\cite{Daikoku:2000ep}, neutrino physics
\cite{Kang:2004ix}--\cite{Ma:1995xk}, leptogenesis
\cite{Hambye:2000bn}--\cite{King:2008qb}, EW baryogenesis
\cite{baryogen}, muon anomalous magnetic moment \cite{g-2}, electric
dipole moment of electron \cite{Suematsu:1997tv} and tau lepton \cite{GutierrezRodriguez:2006hb}, lepton flavour violating processes like $\mu\to e\gamma$ \cite{Suematsu:1997qt} and CP-violation in the Higgs sector \cite{Ham:2008fx}.  Such models have also been proposed as the solution to the tachyon problems of anomaly mediated SUSY breaking, via $U(1)^\prime$ D-term contributions \cite{Asano:2008ju}, and used in combination with a generation symmetry to construct a model explaining fermion mass hierarchy and mixing \cite{Stech:2008wd}.

Recent publications have focused on a particular $E_6$ inspired SUSY
model with an extra $U(1)_{N}$ gauge symmetry in which right handed
neutrinos do not participate in the gauge interactions. This
corresponds to $\theta=\arctan\sqrt{15}$. Only in this Exceptional
Supersymmetric Standard Model (E$_6$SSM)
\cite{King:2005jy}--\cite{King:2005my} right--handed neutrinos may be
superheavy, shedding light on the origin of the mass hierarchy in the
lepton sector and providing a mechanism for the generation of lepton
and baryon asymmetry of the universe
\cite{Hambye:2000bn}--\cite{King:2008qb}. Supersymmetric models with
an additional $U(1)_{N}$ gauge symmetry in which right--handed
neutrinos have zero charge have been studied in \cite{Ma:1995xk} in
the context of non--standard neutrino models with extra singlets, in
\cite{Suematsu:1997au} from the point of view of $Z-Z'$ mixing, in
\cite{Keith:1997zb} and \cite{Suematsu:1997au}--\cite{Keith:1996fv}
where the neutralino sector was explored, in \cite{Keith:1997zb} where
the RG flow of couplings was examined and in
\cite{Suematsu:1994qm}--\cite{Daikoku:2000ep} where EWSB was studied.

In a recent letter \cite{Athron:2009ue} we presented
predictions from a constrained version of the above E$_6$SSM,
referred to as the cE$_6$SSM\footnote{See also Ref.~\cite{Athron:2008np}
for a preliminary account.}, in which the soft SUSY--breaking
scalar masses, gaugino masses and the trilinear scalar couplings are
each assumed to be universal at the scale $M_X$,
i.e.\ $m^2_i(M_X)=m_0^2$, $M_i(M_X)=M_{1/2}$ and $A_i(M_X)=A_0$.  We
discussed scenarios of the cE$_6$SSM with the lowest values of $m_0$
and $M_{1/2}$ consistent with both EWSB and experimental
constraints, leading to very light exotic quarks, inert
Higgs/Higgsinos and $Z'$ masses.  As such these represented
scenarios which could be discovered early at the LHC using ``first
data''.  Since the emphasis was on early discovery we did not
explore the cE$_6$SSM parameter space thoroughly and did not present
a set of benchmarks which represent all the qualitatively
different spectra of TeV scale cE$_6$SSM scenarios.  For brevity we
also omitted the renormalisation group equations (RGEs) used in our
analysis and did not provide full details of our mass spectra
calculations.

In this paper we provide a comprehensive study of the parameter space of
the cE$_6$SSM and the TeV scale predictions of the model.  We present
two--loop RGEs for the gauge and Yukawa couplings together with
two--loop RGEs for the gaugino masses and trilinear scalar couplings
as well as one--loop RGEs for the soft scalar masses, in order to
calculate the values of all masses and couplings at the EW scale for
each set of fundamental parameters at the GUT scale $M_X$.  Two--loop
corrections to the $\beta$--functions are important for the analysis
of the particle spectrum because in $E_6$ inspired SUSY models the
$\beta$--function of the $SU(3)$ gauge coupling and the gluino mass
vanish in the one--loop approximation.  We perform a numerical RG
analysis for the cE$_6$SSM, imposing the usual low energy experimental
constraints and enforcing successful EWSB.  Our analysis reveals that
there is a substantial part of the cE$_6$SSM parameter space where the
correct breakdown of the gauge symmetry can be achieved and all
experimental constraints can be satisfied.  We then perform a scan of
the parameter space of the cE$_6$SSM and specify a set of benchmark
points that highlight particular characteristics of the particle
spectrum within the cE$_6$SSM parameter space.  A general feature of
the benchmark spectra is a light sector of SUSY particles consisting
of a light gluino, two light neutralinos and a light chargino,
resulting from the relative smallness of the low energy gaugino masses
$M_i$ due to the stronger gauge running. Although the squarks,
sleptons and $Z'$ boson are typically much heavier, the exotic 
quarks and squarks can be also relatively light leading to spectacular 
new physics signals at the LHC.

The paper is organised as follows. In the next section we introduce
the E$_6$SSM and define the cE$_6$SSM. In section 3 we discuss the
breakdown of gauge symmetry in the cE$_6$SSM. In section 4 we provide 
analytical expressions for the mass matrices and masses of all new particles
appearing in our model. In section 5 we study the RG flow of all
masses and couplings and summarise the results of our studies of the
particle spectrum.  Section 6 is reserved for our conclusions and
outlook. Appendix A contains explicit expressions for the one--loop 
corrections to the mass matrix of the CP--even Higgs bosons calculated
in the leading approximation. In Appendix B we specify the complete system 
of RGEs that we use in our analysis.

\section{From the E$_6$SSM to the cE$_6$SSM}

The E$_6$SSM is based on the $SU(3)_C\times SU(2)_W\times U(1)_Y
\times U(1)_N$ gauge group which is a subgroup of $E_6$. The extra
$U(1)_N$ gauge symmetry is defined such that right--handed neutrinos
carry zero charges. The E$_6$SSM can originate from an $E_6$ GUT gauge
group which is broken at the GUT scale $M_X$.  In $E_6$ theories the
anomalies are cancelled automatically; all models that are based on
the $E_6$ subgroups and contain complete representations of $E_6$
should be anomaly--free. Consequently, in order to make a
Supersymmetric model with an extra $U(1)_{N}$ anomaly--free, one is
forced to augment the minimal particle spectrum by a number of exotics
which, together with ordinary quarks and leptons, form complete
fundamental $27$ representations of $E_6$. Thus the particle content
of the E$_6$SSM involves at least three fundamental representations of
$E_6$ at low energies. These multiplets decompose under the
$SU(5)\times U(1)_{N}$ subgroup of $E_6$ as follows:
\begin{eqnarray}
27_i\to \ds\left(10,\,\ds{1}\right)_i+\left(5^{*},\,\ds{2}\right)_i
+\left(5^{*},\,-\ds{3}\right)_i +\ds\left(5,-\ds{2}\right)_i
+\left(1,\ds{5}\right)_i+\left(1,0\right)_i\,.
\label{cessm2}
\end{eqnarray}
The first and second quantities in brackets are the $SU(5)$
representation and extra $U(1)_{N}$ charge respectively, while $i$ is
a family index that runs from 1 to 3. An ordinary SM family, which
contains the doublets of left--handed quarks $Q_i$ and leptons $L_i$,
right-handed up-- and down--quarks ($u^c_i$ and $d^c_i$) as well as
right--handed charged leptons, is assigned to
$\left(10,\,1\right)_i$ +
$\left(5^{*},\,2\right)_i$. Right-handed neutrinos
$N^c_i$ should be associated with the last term in Eq.~(\ref{cessm2}),
$\left(1,\, 0\right)_i$.  The next-to-last term,
$\left(1,\, 5\right)_i$, represents SM-singlet fields
$S_i$, which carry non-zero $U(1)_{N}$ charges and therefore survive
down to the EW scale.  The pair of $SU(2)_W$--doublets ($H^d_{i}$ and
$H^u_{i}$) that are contained in
$\left(5^{*},\,-3\right)_i$ and
$\left(5,\,-2\right)_i$ have the quantum numbers of
Higgs doublets. They form either Higgs or Inert Higgs $SU(2)_W$
multiplets\footnote{We use the terminology ``Inert Higgs'' to denote
Higgs--like doublets that do not develop VEVs.}.  Other components of
these $SU(5)$ multiplets form colour triplets of exotic quarks
$\overline{D}_i$ and $D_i$ with electric charges $-1/3$ and $+1/3$,
respectively. 

In addition to the complete $27_i$ multiplets the low energy matter
content of the E$_6$SSM is supplemented by an $SU(2)_W$ doublet $\hat{H}'$
and anti-doublet $\hat{\overline{H'}}$ from the extra $27'$ and
$\overline{27'}$, in order to preserve gauge coupling
unification. These components of the $E_6$ fundamental representation
originate from $\left(5^{*},\,2 \right)$ of $27'$
and $\left(5,\,-2 \right)$ of $\overline{27'}$ by
construction. The analysis performed in \cite{King:2007uj} shows that
the unification of gauge couplings in the E$_6$SSM can be achieved for
any phenomenologically acceptable value of $\alpha_3(M_Z)$ consistent
with the measured low energy central value, unlike in the MSSM which,
ignoring the effects of high energy threshold corrections, requires
significantly higher values of $\alpha_3(M_Z)$, well above the
experimentally measured central value.  The splitting of $27'$ and
$\overline{27'}$ multiplets can be naturally achieved, for example, in
the framework of orbifold GUTs \cite{orbifold-GUT}.

\begin{table}[ht]
\centering
\begin{tabular}{|c|c|c|c|c|c|c|c|c|c|c|c|c|c|}
\hline
 & $Q$ & $u^c$ & $d^c$ & $L$ & $e^c$ & $N^c$ & $S$ & $H_2$ & $H_1$ & $D$ &
 $\overline{D}$ & $H'$ & $\overline{H'}$ \\
 \hline
$\sqrt{\frac{5}{3}}Q^{Y}_i$
 & $\frac{1}{6}$ & $-\frac{2}{3}$ & $\frac{1}{3}$ & $-\frac{1}{2}$
& $1$ & $0$ & $0$ & $\frac{1}{2}$ & $-\frac{1}{2}$ & $-\frac{1}{3}$ &
 $\frac{1}{3}$ & $-\frac{1}{2}$ & $\frac{1}{2}$ \\
 \hline
$\sqrt{{40}}Q^{N}_i$
 & $1$ & $1$ & $2$ & $2$ & $1$ & $0$ & $5$ & $-2$ & $-3$ & $-2$ &
 $-3$ & $2$ & $-2$ \\
\hline
\end{tabular}
\caption{\it\small The $U(1)_Y$ and $U(1)_{N}$ charges of matter fields in the
    E$_6$SSM, where $Q^{N}_i$ and $Q^{Y}_i$ are here defined with the correct
$E_6$ normalisation factor required for the RG analysis.}
\label{charges}
\end{table}

The matter content of the E$_6$SSM with correctly normalized Abelian charges
of all matter fields is summarised in Table~\ref{charges}. 
Because right--handed neutrinos $\hat{N}^c$ do not participate in gauge 
interactions they are expected to gain masses at some intermediate scale
after the breakdown of $E_6$ \cite{King:2005jy},\cite{Howl:2008xz}.
The remaining matter survives down to the EW scale near which the gauge group 
$U(1)_N$ is broken.  Thus, in addition to a $Z'$ corresponding to the $U(1)_N$ 
symmetry, the E$_6$SSM involves extra matter beyond the MSSM with the quantum
numbers of three $5+5^{*}$ representations of $SU(5)$ plus three
$SU(5)$ singlets with $U(1)_N$ charges. The presence of a $Z'$ boson
and exotic quarks predicted by the E$_6$SSM provides spectacular new
physics signals at the LHC which were discussed in
\cite{King:2005jy}--\cite{King:2005my}, \cite{Accomando:2006ga}.

Since the right--handed neutrinos are heavy, the three known doublet
neutrinos $\nu_e$, $\nu_{\mu}$ and $\nu_{\tau}$, acquire small
Majorana masses via the see--saw mechanism. At the same time 
the heavy Majorana right-handed neutrinos may decay into final states 
with lepton number $L=\pm 1$, thereby creating a lepton asymmetry 
in the early universe. In the E$_6$SSM the Yukawa couplings of exotic 
particles are not constrained by neutrino oscillation data.
As a result substantial values of the CP--asymmetries can be
induced even for a relatively small mass of the lightest right--handed
neutrino ($M_1 \sim 10^6\,\mbox{GeV}$) so that successful thermal
leptogenesis may be achieved without encountering a gravitino problem
\cite{King:2008qb}. 

In general $E_6$ symmetry does not forbid lepton and baryon number violating 
operators that result in rapid proton decay. Moreover, exotic particles in 
$E_6$ inspired SUSY models give rise to new Yukawa interactions that induce
unacceptably large non--diagonal flavour transitions. To suppress these effects
in the E$_6$SSM an approximate $Z^{H}_2$ symmetry is imposed.
Under this symmetry all superfields except one pair of $H_{1i}$ and $H_{2i}$ 
(say $H_d\equiv H_{13}$ and $H_u\equiv H_{23}$) and one SM-type singlet 
field ($S\equiv S_3$) are odd. The $Z^{H}_2$ symmetry reduces the structure 
of the Yukawa interactions to (see \cite{King:2005jy})
\begin{eqnarray}
W_{\rm E_6SSM} &\longrightarrow &
\lambda_i \hat{S}(\hat{H}^d_{i}\hat{H}^u_{i})+\kappa_i \hat{S}(\hat{D}_i\hat{\overline{D}}_i)+
f_{\alpha\beta}\hat{S}_{\alpha}(\hat{H}_d \hat{H}^u_{\beta})+ 
\tilde{f}_{\alpha\beta}\hat{S}_{\alpha}(\hat{H}^d_{\beta}\hat{H}_u) \nonumber\\[2mm]
&&+\dfrac{1}{2}M_{ij}\hat{N}^c_i\hat{N}^c_j+\mu'(\hat{H}'\hat{\overline{H'}})+
h^{E}_{4j}(\hat{H}_d \hat{H}')\hat{e}^c_j+h_{4j}^N (\hat{H}_{u} \hat{H}')\hat{N}_j^c 
\nonumber \\[2mm]
&& + W_{\rm{MSSM}}(\mu=0),
\label{cessm7}
\end{eqnarray}
where $\alpha,\beta=1,2$ and $i,j=1,2,3$\,. In Eq.~(\ref{cessm7}) we
choose the basis $H^d_{\alpha}$, $H^u_{\alpha}$, $D_i$ and
$\overline{D}_i$ so that the Yukawa couplings of the singlet field $S$
have flavour diagonal structure. The $SU(2)_W$ doublets $\hat{H}_u$
and $\hat{H}_d$, that are even under the $Z^{H}_2$ symmetry, play the
role of Higgs fields generating the masses of quarks and leptons after
EWSB. The singlet field $S$ must also acquire a large VEV in order to
induce sufficiently large masses for the exotic charged fermions and
$Z'$ boson and avoid conflict with direct particle searches at present
and past accelerators. This requires the Yukawa couplings $\lambda_i$
and $\kappa_i$ to be reasonably large. If $\lambda_i$ or $\kappa_i$
are large at the GUT scale they affect the evolution of the soft
scalar mass $m_S^2$ of the singlet field $S$ rather strongly resulting
in negative values of $m_S^2$ at low energies that triggers the
breakdown of the $U(1)_{N}$ symmetry.

Because $H_u$, $H_d$ and $S$ generate masses of all quarks, leptons and
exotic fermions, it is natural to assume that only these fields
acquire non--zero VEVs. To guarantee this, a certain hierarchy between
the Yukawa couplings must exist. Defining $\lambda \equiv
\lambda_3$, we impose $\kappa_i\sim\lambda_3\gtrsim\lambda_{1,2}\gg
f_{\alpha\beta},\,\tilde{f}_{\alpha\beta},\,h^{E}_{4j},\,h_{4j}^N$.
Although $f_{\alpha\beta}$ and $\tilde{f}_{\alpha\beta}$ are expected
to be considerably smaller than $\lambda_i$ and $\kappa_i$, they
cannot be negligibly small since the fermion components of the
Superfields $\hat{S}_1$ and $\hat{S}_2$ would become extremely light. The induced
masses of singlinos $\tilde{S}_1$ and $\tilde{S}_2$ should be as large
as a few MeV, otherwise the extra states could contribute to the
universe expansion rate prior to nucleosynthesis, thereby changing
nuclear abundances.

Although $Z^{H}_2$ eliminates any problem related with baryon number
violation and non-diagonal flavour transitions it also forbids all 
Yukawa interactions that would allow the exotic quarks to decay. 
Since models with stable charged exotic particles are ruled out by 
different experiments \cite{42} the $Z^{H}_2$ symmetry must be broken. 
But the breakdown of $Z^{H}_2$ should not give rise to the operators 
leading to rapid proton decay. There are two ways to overcome this
problem: the Lagrangian must be invariant with respect to either a
$Z_2^L$ symmetry, under which all Superfields except lepton ones are
even (Model I), or a $Z_2^B$ discrete symmetry, which implies that
exotic quark and lepton Superfields are odd whereas the others remain
even (Model II). If the Lagrangian is invariant under the $Z_2^L$ 
symmetry transformations then the terms in the superpotential which 
permit exotic quarks to decay and are allowed by the $E_6$ symmetry
can be written in the following form
\begin{equation}
W_1=g^Q_{ijk}\hat{D}_{i} (\hat{Q}_j \hat{Q}_k)+
g^{q}_{ijk}\hat{\overline{D}}_i \hat{d}^c_j \hat{u}^c_k\,.
\label{cessm3}
\end{equation}
that implies that exotic quarks are diquarks. If $Z_2^B$ is imposed 
then the following couplings are allowed:
\begin{equation}
W_2=g^E_{ijk} \hat{e}^c_i \hat{D}_j \hat{u}^c_k+
g^D_{ijk} (\hat{Q}_i \hat{L}_j) \hat{\overline{D}}_k\,,
\label{cessm4}
\end{equation}
In this case the baryon number conservation requires exotic quarks
to be leptoquarks.

Since $Z^{H}_2$ violating operators lead to non--diagonal flavour
interactions, the corresponding Yukawa couplings are expected to be
small, and must preserve either the $Z_2^B$ or $Z_2^L$ symmetry to
ensure proton stability. In order to guarantee that the contribution
of new particles and interactions to $K^0-\overline{K}^0$ oscillations
and to the muon decay $\mu\to e^{-}e^{+}e^{-}$ are suppressed in
accordance with experimental limits, it is necessary to assume that
the Yukawa couplings of exotic particles to ordinary quarks and
leptons are less than $10^{-3}-10^{-4}$. In this case, they do not
affect the RG flow of other masses and couplings and can safely be
ignored in our analysis of the particle spectrum.

The hierarchical structure of the Yukawa interactions allows one to
simplify the Superpotential substantially. Integrating out heavy
Majorana right--handed neutrinos and keeping only Yukawa interactions
whose couplings are allowed to be of order unity we find
\begin{eqnarray}
W_{\rm E_6SSM}&\simeq &\lambda \hat{S} (\hat{H}_{d} \hat{H}_{u})+
\lambda_{\alpha} \hat{S}(\hat{H}^d_{\alpha} \hat{H}^u_{\alpha})+
\kappa_i \hat{S} (\hat{D}_i\hat{\overline{D}}_i) \nonumber \\[2mm]
&&+h_t(\hat{H}_{u}\hat{Q})\hat{t}^c+h_b(\hat{H}_{d}\hat{Q})\hat{b}^c+ 
h_{\tau}(\hat{H}_{d}\hat{L})\hat{\tau}^c+ \mu'(\hat{H}'\hat{\overline{H'}})\,,
\label{cessm8}
\end{eqnarray}
where the Superfields $\hat{L}=\hat{L}_3$, $\hat{Q}=\hat{Q}_3$, 
$\hat{t}^c=\hat{u}^c_3$, $\hat{b}^c=\hat{d}^c_3$ and $\tau^c=e^c_3$ 
belong to the third generation. The Superpotential (\ref{cessm8}) includes only 
one bilinear term which is solely responsible for the masses of the 
charged and neutral components of $\hat{H}'$ and $\hat{\overline{H'}}$.  
The corresponding mass term is not suppressed by the $E_6$ symmetry 
and is not involved in the process of the EWSB. Therefore, the parameter 
$\mu'$ remains arbitrary. Gauge coupling unification requires $\mu'$ 
to be within $100\,\mbox{TeV}$ \cite{King:2007uj}. The simplified 
Superpotential (\ref{cessm8}) that we use in our analysis of 
the cE$_6$SSM contains seven new couplings compared to the MSSM 
with $\mu=0$: the parameter $\mu'$ and six new Yukawa couplings 
$\lambda_i$ and $\kappa_i$.

The most general scalar potential of the E$_6$SSM that ensures soft
SUSY--breaking can be presented as a sum
\be
V=V_F+V_D+V_{soft}\,,
\label{cessm9}
\ee
where $V_F$ and $V_D$ are the contributions of $F$ and $D$ terms
respectively, while $V_{soft}$ contains a set of soft SUSY--breaking
couplings:
\begin{eqnarray}
V_{soft}
&=&
m^2_{S_i}|S_i|^2+m_{H^u_{i}}^2|H^u_{i}|^2+m_{H^d_{i}}^2|H^d_{i}|^2+m_{D_i}^2|D_i|^2+
m_{\overline{D}_i}^2|\overline{D}_i|^2 +m_{Q_i}^2|Q_i|^2 \nonumber \\
&+& m_{u^c_i}^2|u^c_i|^2+m_{d^c_i}^2|d^c_i|^2+m_{L_i}^2|L_i|^2+m_{e^c_i}^2|e^c_i|^2
+m^2_{H'}|H^{'2}|+m^2_{\overline{H^{'}}} |\overline{H^{'}}|^2 \nonumber \\
&+& \biggl[B'\,\mu' (H^{'}\overline{H^{'}})+h.c.\biggr]+
\biggl[\lambda_i A_{\lambda_i} S(H^d_{i} H^u_{i})+\kappa_i A_{\kappa_i} S(D_i\overline{D}_i) \nonumber \\
&+& h_t A_t(H_{u}Q)t^c+h_b A_b(H_{d}Q)b^c+h_{\tau} A_{\tau}(H_{d}L)\tau^c+h.c.\,\biggr]\,.
\label{cessm10}
\end{eqnarray}
The soft breakdown of SUSY gives rise to many new couplings.  The six
additional Yukawa couplings are accompanied by six extra trilinear
scalar couplings, $A_{\lambda_i}$ and $A_{\kappa_i}$
(\ref{cessm10}). Soft SUSY--breaking also induces the bilinear scalar
coupling $B'$ that corresponds to the mass term $\mu'\hat{H'}\hat{\overline{H'}}$
in the Superpotential (\ref{cessm8}). In addition, the scalar
potential of the E$_6$SSM includes 15 extra soft scalar masses: six
masses of exotic squarks $m_{\tilde{D}_i}$ and
$m_{\tilde{\overline{D}}_i}$, four masses of Inert Higgs fields
$m_{H^d_{\alpha}}$ and $m_{H^u_{\alpha}}$, two soft
scalar masses of $H'$ and $\overline{H}'$ and three masses of SM
singlet scalar fields $m^2_{S_i}$. Due to the extra Yukawa couplings,
the parameter $\mu'$ and the new trilinear scalar and bilinear scalar
couplings (that can be complex), even the simplified version of the
$Z^{H}_2$--symmetric E$_6$SSM considered here involves 43 new
parameters in comparison to the MSSM with $\mu=0$. Fourteen of them
are phases, some of which (but not all) can be eliminated by an
appropriate redefinition of the new fields.
However, the number of fundamental parameters reduces drastically in
the cE$_6$SSM, defined at the GUT scale $M_X$, where all gauge
couplings coincide, i.e.\ $g_1(M_X) \simeq g_2(M_X)$ $ \simeq
g_3(M_X) \simeq g'_1(M_X)$, while the off--diagonal gauge coupling
$g_{11}(M_X)$ vanishes.  Constrained SUSY models impose extra
unification constraints on the soft SUSY--breaking parameters.  In
particular, all soft scalar masses are set to be equal to $m_0^2$ at
the scale $M_X$. Gaugino masses $M_i(M_X)$ are equal to an overall
gaugino mass $M_{1/2}$ at the GUT scale and all trilinear and bilinear
scalar couplings coincide at this scale, i.e.\ $A_i(M_X)=A_0$ and
$B_i(M_X)=B$. Thus the cE$_6$SSM is uniquely characterised by the set
of Yukawa couplings $\lambda_i(M_X)$, $\kappa_i(M_X)$, $h_t(M_X)$,
$h_b(M_X)$ and $h_{\tau}(M_X)$, the universal soft scalar mass $m_0$,
the universal gaugino mass $M_{1/2}$ and the universal trilinear
scalar coupling $A_0$. The phases of the dimensionless couplings in the
Superpotential are selected by appropriate field redefinitions and are
chosen so that all the dimensionless couplings are real. In order to
guarantee correct EWSB, $m_0^2$ has to be positive.  To simplify our
analysis we also assume that $A_0$ is real and $M_{1/2}$ is positive ---
this then naturally leads to real VEVs of the Higgs fields.

The set of parameters mentioned above should be in principle
supplemented by $B'$ and $\mu'$. However, since $\mu'$ is not
constrained by EWSB and the term $\mu'\hat{H}'\hat{\overline{H'}}$ in the
Superpotential (\ref{cessm8}) is not suppressed by the $E_6$ symmetry,
the parameter $\mu'$ can be as large as $10\,\mbox{TeV}$. Therefore we
assume that the scalar and fermion components of the Superfields $\hat{H}'$
and $\hat{\overline{H'}}$ are very heavy so that they decouple from the rest
of the particle spectrum. As a consequence the
parameters $B'$ and $\mu'$, that determine the masses of the survival
components of $27'$ and $\overline{27'}$, are irrelevant for our
analysis.

\section{EWSB and $Z$--$Z'$ mixing}

As described in the previous section, the Higgs sector of the model
involves two Higgs doublets $H_u$ and $H_d$, as well as the
SM--singlet field $S$. The corresponding Higgs effective potential 
can be written as,
\begin{eqnarray}
V&=&\lambda^2|S|^2(|H_d|^2+|H_u|^2)+\lambda^2|(H_d H_u)|^2+
\ds\frac{g_2^2}{8}\left(H_d^\dagger \sigma_a H_d+H_u^\dagger \sigma_a H_u\right)^2 \nonumber \\
&+&\ds\frac{{g'}^2}{8}\left(|H_d|^2-|H_u|^2\right)^2+
\ds\frac{g^{'2}_1}{2}\left(\tilde{Q}_1|H_d|^2+\tilde{Q}_2|H_u|^2+\tilde{Q}_S|S|^2\right)^2 \nonumber \\
&+& m_{S}^2|S|^2+m_1^2|H_d|^2+m_2^2|H_u|^2+\biggl[\lambda A_{\lambda}S(H_u H_d)+h.c.\biggr]+\Delta V\,,
\label{cessm11}
\end{eqnarray}
where $g'=\sqrt{3/5} g_1$ is the low energy (non-GUT normalised) gauge
coupling and $\tilde{Q}_1$, $\tilde{Q}_2$ and $\tilde{Q}_S$ are the
effective $U(1)_{N}$ charges of $H_d$, $H_u$ and $S$ defined below.
The first two terms in Eq.~(\ref{cessm11}) correspond to F--term
contributions while the subsequent three represent $D$--term
contributions associated with $SU(2)_W$, $U(1)_Y$ and $U(1)_N$ gauge
interactions. 
The term in Eq.~(\ref{cessm11}) proportional to ${g'_1}^2$ corresponds to 
the $D$--term contribution due to the extra $U(1)_{N}$ interaction, which 
is not present in the MSSM or NMSSM. The value of $g'_1$ at the EW scale 
can be determined by assuming gauge coupling unification.

The last term in Eq.~(\ref{cessm11}) $\Delta V$ represents the
contribution of loop corrections to the Higgs effective potential. 
Here we take into account only the dominant
contribution to $\Delta V$ that comes from loop diagrams involving
the top--quark and its Superpartners. In the leading one--loop
approximation we find
\be
\ba{c}
\Delta
V=\ds\frac{3}{32\pi^2}\left[m_{\tilde{t}_1}^4\left(
\ln\frac{m_{\tilde{t}_1}^2}{Q^2}-\frac{3}{2}\right)+
m_{\tilde{t}_2}^4\left(\ln\frac{m_{\tilde{t}_2}^2}{Q^2}-\frac{3}{2}\right)
-2m_t^4\left(\ln\frac{m_t^2}{Q^2}-\frac{3}{2}\right)\right]
\label{cessm12}
\ea
\ee
where $m_t$, $m_{\tilde{t}_1}$, $m_{\tilde{t}_2}$ are the masses of
the top quark and its Superpartners.  The analytical expressions for
$m_{\tilde{t}_1}$ and $m_{\tilde{t}_2}$ are specified in the next
section.

At the physical minimum of the scalar potential (\ref{cessm11}) the Higgs fields develop VEVs
\be
\langle H_d \rangle =\ds\frac{1}{\sqrt{2}}\left(
\begin{array}{c}
v_1\\ 0
\end{array}
\right) , \qquad
\langle H_u \rangle=\ds\frac{1}{\sqrt{2}}\left(
\begin{array}{c}
0\\ v_2
\end{array}
\right) ,\qquad
\langle S \rangle =\ds\frac{s}{\sqrt{2}}.
\label{cessm13}
\ee
The equations for the extrema of the Higgs boson potential are:
\begin{eqnarray}
\ds\frac{\partial V}{\partial s}&=&\ds m_{S}^2 s-\frac{\lambda
A_{\lambda}}{\sqrt{2}}v_1v_2+\frac{\lambda^2}{2}(v_1^2+v_2^2)s \nonumber \\
&& \qquad \qquad \qquad +\ds\frac{g^{'2}_1}{2}\biggl(\tilde{Q}_1v_1^2+\tilde{Q}_2v_2^2+\tilde{Q}_S
s^2\biggr)\tilde{Q}_S s+\ds\frac{\partial\Delta V}{\partial
s}\:\:=\:\:0\,,
\label{cessm14a} \\
\ds\frac{\partial V}{\partial v_1}&=&\ds
m_1^2v_1-\frac{\lambda A_{\lambda}}{\sqrt{2}}s v_2
+\frac{\lambda^2}{2}(v_2^2+s^2)v_1+\frac{\bar{g}^2}{8}\biggl(v_1^2-v_2^2)\biggr)v_1 \nonumber \\
&&\qquad \qquad \qquad +\ds\frac{g^{'2}_1}{2}\biggl(\tilde{Q}_1v_1^2+\tilde{Q}_2v_2^2+\tilde{Q}_Ss^2\biggr)\tilde{Q}_1
v_1+\ds\frac{\partial\Delta V}{\partial v_1}\:\:=\:\:0\,,
\label{cessm14b} \\
\ds\frac{\partial V}{\partial v_2}&=&\ds m_2^2v_2-\frac{\lambda A_{\lambda}}{\sqrt{2}}s v_1+
\frac{\lambda^2}{2}(v_1^2+s^2)v_2+\frac{\bar{g}^2}{8}\biggl(v_2^2-v_1^2\biggr)v_2 \nonumber \\
&&\qquad \qquad \qquad +\ds\frac{g^{'2}_1}{2}\biggl(\tilde{Q}_1v_1^2+\tilde{Q}_2v_2^2+\tilde{Q}_Ss^2\biggr)\tilde{Q}_2 v_2+
\ds\frac{\partial\Delta V}{\partial v_2}\:\:=\:\:0\,,
\label{cessm14c}
\end{eqnarray}
where $\bar{g}=\sqrt{g_2^2+g'^2}$. Instead of $v_1$ and $v_2$, it is more convenient
to use $\tan\beta=v_2/v_1$ and $v=\sqrt{v_1^2+v_2^2}=246\,\mbox{GeV}$.

The VEVs of the Higgs fields (\ref{cessm13}) induce masses for the
gauge bosons and lead to $Z$--$Z'$ mixing. In this context, note that
the $U(1)_Y$ and $U(1)_N$ mix at low energies even before EWSB
because gauge symmetries do not forbid a mixing term in the E$_6$SSM Lagrangian,
\be
\mathcal{L}_{mix}^{kin}=-\frac{\sin\chi}{2}F^{Y}_{\mu\nu}F^{N}_{\mu\nu}\,,
\label{cessm15}
\ee
where $F_{\mu\nu}^Y$ and $F_{\mu\nu}^{N}$ are field strengths for the
$U(1)_Y$ and $U(1)_{N}$ gauge interactions. The parameter $\sin\chi$ 
is expected to be equal to zero at the GUT scale. Nevertheless a small 
value of $\sin\chi$ is generated at low energies due to loop effects. 
The mixing in the gauge kinetic part of the Lagrangian (\ref{cessm15})
can be eliminated by means of a non--unitary transformation of the two $U(1)$
gauge fields \cite{Langacker:1998tc}, \cite{Suematsu:1998wm}--\cite{48}.
In this case all physical phenomena related to the gauge kinetic term mixing
can be described by using effective $U(1)_{N}$ charges
\be
\tilde{Q}_i\equiv Q^{N}_i+Q^{Y}_i\delta\,,
\label{cessm19}
\ee
where $\delta=g_{11}/g'_1$ 
\be 
g_1=g_Y\,, \qquad
g'_1=g_{N}/\cos\chi\,,\qquad g_{11}=-g_Y\tan\chi\,,
\label{cessm18}
\ee 
while all $U(1)_Y$ charges remain the same.

Initially the EWSB sector involves ten degrees of freedom. However, four of
them are massless Goldstone modes which are eaten by the $W^{\pm}$, $Z$
and $Z'$ gauge bosons. The charged $W^{\pm}$ bosons gain masses via the
interaction with the neutral components of the Higgs doublets in the same
way as in the MSSM so that $M_W=\ds\frac{g_2}{2}v$. In contrast, 
neutral gauge bosons get mixed leading to the formation of two 
mass eigenstates $Z_1$ and $Z_2$. Letting $Z'$ be the gauge boson associated 
with $U(1)_{N}$ we get
\be
\ba{c}
Z_1=Z\cos\alpha_{ZZ'}+Z'\sin\alpha_{ZZ'}\,,\qquad\qquad Z_2=-Z\sin\alpha_{ZZ'}+Z'\cos\alpha_{ZZ'}\,,\\[1mm]
M^2_{Z_1,\,Z_2}=\ds\frac{1}{2}\left[M_Z^2+M^2_{Z'}\mp\sqrt{(M_Z^2-M_{Z'}^2)^2+4\Delta^4}\right]\,,
\ea
\label{cessm21}
\ee
where
\be
\ba{c}
M_Z^2=\ds\frac{\bar{g}^2}{4}v^2\,,\qquad\qquad \Delta^2=\ds\frac{\bar{g}g'_1}{2}v^2\biggl(
\tilde{Q}_1\cos^2\beta-\tilde{Q}_2\sin^2\beta\biggr)\,,\\[1mm]
M^2_{Z'}=g^{'2}_1 v^2\biggl(\tilde{Q}_1^2\cos^2\beta+\tilde{Q}_2^2\sin^2\beta\biggr)+g^{'2}_1\tilde{Q}^2_S s^2\,,\\[1mm]
\alpha_{ZZ'}=\ds\frac{1}{2}\arctan\left(\frac{2\Delta^2}{M^2_{Z'}-M_Z^2}\right)\,.
\ea
\label{cessm22}
\ee
Phenomenological constraints typically require the mixing angle $\alpha_{ZZ'}$ to be less than
$1-2\times 10^{-3}$ \cite{49} and the mass of the extra neutral gauge boson to be heavier than
$860\,\mbox{GeV}$ \cite{Grivaz:2008tf}. A suitable mass hierarchy and mixing between $Z$ and $Z'$ are
maintained if the field $S$ acquires a large VEV $s \gtrsim 1.5-2\,\mbox{TeV}$. Then the mass of
the lightest neutral gauge boson $Z_1$ is very close to $M_Z$ whereas the mass of $Z_2$ is set by
the VEV of the singlet field $M_{Z_2}\simeq M_{Z'}\approx g'_1\tilde{Q}_S\, s$.

\section{Particle spectrum}
\label{Sec:ParticleSpec}
\subsection{The squarks and sleptons}

In Supersymmetric theories, each quark and lepton state with a
specific chirality has a scalar Superpartner. In principle, all
scalars with the same electric charge, R--parity and colour quantum
numbers can mix with one another. This means that the mass eigenstates
of the squarks and sleptons should be obtained by diagonalising three
$6\times 6$ squared--mass matrices for up--type squarks
$(\tilde{u}_L,\,\tilde{c}_L,\,\tilde{t}_L,\,
\tilde{u}_R,\,\tilde{c}_R,\,\tilde{t}_R)$, down--type squarks
$(\tilde{d}_L,\,\tilde{s}_L,\,
\tilde{b}_L,\,\tilde{d}_R,\,\tilde{s}_R,\,\tilde{b}_R)$ and charged
leptons $(\tilde{e}_L,\,
\tilde{\mu}_L,\,\tilde{\tau}_L,\,\tilde{e}_R,\,\tilde{\mu}_R,\,\tilde{\tau}_R)$
and one $3\times 3$ matrix for sneutrinos
$(\tilde{\nu}_e,\,\tilde{\nu}_{\mu},\,\tilde{\nu}_{\tau})$.  However,
since the first and second family quarks and leptons have negligible
Yukawa couplings the mixing angles of the corresponding squark and
slepton states are very small so that their masses are set by the
appropriate diagonal entries. Thus one finds,
\begin{eqnarray}
m_{\tilde{d}_{L\,i}}^2&\simeq &m_{Q_i}^2+\left(-\dfrac{1}{2}+\dfrac{1}{3}\sin^2\theta_W\right)M_Z^2\cos 2\beta+\Delta_{Q}\,,
\label{cessm25a}\\
m_{\tilde{u}_{L\,i}}^2&\simeq &m_{Q_i}^2+\left(\dfrac{1}{2}-\dfrac{2}{3}\sin^2\theta_W\right)M_Z^2\cos 2\beta+\Delta_{Q}\,,\label{cessm25b}\\
m_{\tilde{u}_{R\,i}}^2&\simeq &m_{u^c_i}^2+\dfrac{2}{3} M_Z^2 \sin^2\theta_W \cos 2\beta+\Delta_{u^c}\,,\label{cessm25c}\\
m_{\tilde{d}_{R\,i}}^2&\simeq &m_{d^c_i}^2-\dfrac{1}{3} M_Z^2 \sin^2\theta_W \cos 2\beta+\Delta_{d^c}\,,\label{cessm25d}\\[2mm]
m_{\tilde{e}_{L\,i}}^2&\simeq &m_{L_i}^2+\left(-\dfrac{1}{2}+\sin^2\theta_W\right)M_Z^2\cos 2\beta+\Delta_{L}\,,\label{cessm25e}\\
m_{\tilde{\nu}_{i}}^2&\simeq &m_{L_i}^2+\dfrac{1}{2} M_Z^2\cos 2\beta+\Delta_{L}\,,\label{cessm25f}\\
m_{\tilde{e}_{R\,i}}^2&\simeq &m_{e^c_i}^2- M_Z^2 \sin^2\theta_W \cos 2\beta+\Delta_{e^c}\,.
\label{cessm25g}
\end{eqnarray}
The first terms on the right--hand side of
Eqs.~(\ref{cessm25a})-(\ref{cessm25g}) are soft scalar masses while
all other terms come from the $SU(2)_W$, $U(1)_Y$ and $U(1)_N$ D--term
quartic interactions in the scalar potential (\ref{cessm9}) when the
Higgs fields get VEVs. In particular,
\be
\Delta_{\phi}=\dfrac{g^{'2}_1}{2}$ $\biggl(\tilde{Q}_1 v_1^2$ +
$\tilde{Q}_2 v_2^2$ + $\tilde{Q}_S s^2\biggr) \tilde{Q}_{\phi},
\ee
are contributions of $U(1)_N$ D--term to the masses of squarks and
sleptons. 
In general the terms in Eqs.~(\ref{cessm25a})-(\ref{cessm25g}) which
are proportional to $M_Z^2$ or $g^{'2}_1 v^2$ are typically much
smaller than the soft scalar masses squared and $g^{'2}_1 s^2$. As a
consequence the D--term contributions to the squark and slepton masses
are governed by $\Delta_{\phi}$ which in the leading approximation are
given by
\be
\Delta_{Q}\simeq \Delta_{u^c}\simeq \Delta_{e^c}\simeq \dfrac{1}{10}M_{Z'}^2\,,\qquad\qquad
\Delta_{d^c}\simeq \Delta_{L}\simeq \dfrac{1}{5}M_{Z'}^2\,.
\label{cessm26}
\ee
We emphasise that the extra $U(1)_N$ D--term gives positive
contributions to the masses of squarks and sleptons because the
$U(1)_N$ charges of the SM-singlet Superfield $S$ and the charges of
quark and lepton Supermultiplets have the same sign.

Let us now consider the masses of squarks and sleptons of the third
generation. In contrast with the first two families the top quark
Yukawa coupling is always large at the EW scale resulting in
substantial mixing between left--handed and right--handed top
squarks. Diagonalising the $2\times 2$ top squark
mass matrix it is easy to see that
\begin{eqnarray}
\hspace{-20mm} m^2_{\tilde{t}_1,\tilde{t}_2}
&=&\dfrac{1}{2}\left\{ \lefteqn{\phantom{\sqrt{\left[\frac{1}{2}\right]^2}}}
m^2_{Q_3}+m^2_{u^c_3}+\dfrac{1}{2} M_Z^2 \cos2\beta+
\Delta_{Q}+\Delta_{u^c}+2m_t^2 \right. \nonumber \\
&& \mp \left. \sqrt{\left[m_{Q_3}^2-m_{u^c_3}^2+\left[\dfrac{1}{2}-\dfrac{4}{3}\sin^2 \theta_W\right]M_Z^2 \cos2\beta+
\Delta_{Q}-\Delta_{u^c}\right]^2+4 m_t^2 X_t^2}\, \right\}\,, \nonumber \\
\label{cessm27}
\end{eqnarray}
where $X_t=A_t-\dfrac{\lambda s}{\sqrt{2}\tan\beta}$ is a stop mixing
parameter. The large value of $X_t$ induces a significant mixing in
the stop sector which reduces the mass of the lightest top squark so
that it may become one of the lightest eigenstates in the sparticle
spectrum.

With increasing $\tan\beta$, the $b$--quark and $\tau$--lepton Yukawa
couplings grow. At large values of $\tan\beta\gg 10$ the couplings
$h_b$ and $h_{\tau}$ become comparable with the top quark Yukawa
coupling at the EW scale. This leads to substantial mixing between
left--handed and right--handed sbottoms as well as left--handed and
right--handed staus. The eigenvalues of the corresponding $2\times 2$
matrices are given by
\begin{eqnarray}
\hspace{-20mm}
m^2_{\tilde{b}_1,\tilde{b}_2}
&=&\dfrac{1}{2}\left\{ \lefteqn{\phantom{\sqrt{\left[\frac{1}{2}\right]^2}}}
m^2_{Q_3}+m^2_{d^c_3}-\dfrac{1}{2} M_Z^2 \cos2\beta+
\Delta_{Q}+\Delta_{d^c} \right. \nonumber\\
&& \mp \left. \sqrt{\left[m_{Q_3}^2-m_{d^c_3}^2+\left[-\dfrac{1}{2}+\dfrac{2}{3}\sin^2 \theta_W\right]M_Z^2 \cos2\beta+
\Delta_{Q}-\Delta_{d^c}\right]^2+4 m_b^2 X_b^2}\, \right\}\,, \nonumber \\
\label{cessm28a} \\[-5mm]
\hspace{-20mm} m^2_{\tilde{\tau}_1,\tilde{\tau}_2}
&=&\dfrac{1}{2}\left\{ \lefteqn{\phantom{\sqrt{\left[\frac{1}{2}\right]^2}}}
m^2_{L_3}+m^2_{e^c_3}-\dfrac{1}{2} M_Z^2 \cos2\beta+
\Delta_{L}+\Delta_{e^c} \right. \nonumber\\
&&\mp \left. \sqrt{\left[m_{L_3}^2-m_{e^c_3}^2+\left[-\dfrac{1}{2} + 2\sin^2\theta_W\right]M_Z^2 \cos2\beta+
\Delta_{L}-\Delta_{e^c}\right]^2+4 m_{\tau}^2 X_{\tau}^2}\, \right\}\,, \nonumber \\
\label{cessm28b}
\end{eqnarray}
where $X_b=A_b-\dfrac{\lambda s}{\sqrt{2}}\tan\beta$ and
$X_{\tau}=A_{\tau}-\dfrac{\lambda s}{\sqrt{2}}\tan\beta$.  From
Eqs.~(\ref{cessm28a})-(\ref{cessm28b}) one can see that the magnitude
and importance of mixing in the sbottom and stau sectors depend on
$\tan\beta$. If $\tan\beta$ is not too large ($\lesssim 10$) the
sbottoms and staus are not strongly effected by the mixing terms
because $m_b$ and $m_{\tau}$ are small. In this case the mass
eigenstates are very nearly the same as the gauge eigenstates
$\tilde{b}_L$, $\tilde{b}_{R}$, $\tilde{\tau}_L$ and
$\tilde{\tau}_{R}$.  while their masses can be calculated using
Eqs.~(\ref{cessm25a})-(\ref{cessm25g}). For larger values of
$\tan\beta$, the mixing effects are non--negligible, and the lightest
sbottom and stau mass eigenstates can be significantly lighter than
their first and second family counterparts.

\subsection{The gluino}

The gluino is a colour octet fermion. Therefore, it can not mix with
any other particle in SUSY models.  Since the gluino is strongly
interacting, its running mass $M_3$ changes rather quickly
with the renormalisation scale $Q$. Consequently, for an accurate estimate
of the gluino mass one should use the scale--independent mass
$M_{\tilde{g}}$ at which the renormalised gluino propagator has a
pole. Including one--loop corrections to the gluino propagator that
arise from gluon/gluino and quark/squark loops one finds that the
gluino's pole mass is given in terms of the running mass in the
$\overline{\mbox{DR}}$ scheme by
\be
M_{\tilde{g}}=M_3(Q)\biggl[1-\Delta_{\tilde{g}}(Q)\biggr]^{-1}\,,
\label{cessm29}
\ee
where,
\begin{eqnarray}
\Delta_{\tilde{g}}(Q)&=&\dfrac{g_3^2(Q)}{16\pi^2}\biggl\{
9\ln\left(\dfrac{Q^2}{M_3^2}\right)+15-\sum_{q'}\sum_{i=1}^2 B_1(M_3, m_{q'}, m_{\tilde{q}'_i})\nonumber\\
&&\qquad -\sum_{q}\dfrac{m_q}{M_3}\sin(2\theta_q)\biggl[B_0(M_3, m_q, m_{\tilde{q}_1})-
B_0(M_3, m_q, m_{\tilde{q}_2})\biggr]\biggr\}\,,
\label{cessm30}
\end{eqnarray}
and,
\begin{eqnarray}
B_1(p, m_1, m_2)&=&\dfrac{1}{2p^2}\biggl[A_0(m_2)-A_0(m_1)+(p^2+m_1^2-m_2^2)B_0(p, m_1, m_2)\biggr]\,,\\
A_0(m)&=& m^2\biggl[1-\ln\dfrac{m^2}{Q^2}\biggr]\,,\\
B_0(p, m_1, m_2)&=&-\ln\left(\dfrac{p^2}{Q^2}\right)-f_B(x_{+})-f_{B}(x_{-})\,,
\label{cessm31}
\end{eqnarray}
with,
$$
f_B(x)=\ln(1-x)-x\ln(1-x^{-1})-1\,,
\qquad\qquad
x_{\pm}=\dfrac{s\pm\sqrt{s^2-4p^2(m_1^2-i\varepsilon)}}{2p^2}\,,
$$
and $s=p^2-m_2^2+m_1^2$. This expression for the gluino's pole mass
(\ref{cessm29}) automatically incorporates the one--loop
renormalisation group resummation. The first two terms in the right
hand side of Eq.~(\ref{cessm30}) correspond to the gluon/gluino
one--loop contributions while other terms represent quark/squark
one--loop corrections to the gluino mass. Indices $q'$ and
$\tilde{q}'$ in Eq.~(\ref{cessm30}) denote light quarks and their
Superpartners. In the case of the light quarks we neglect the mixing
between left--handed and right--handed squark states. The sum over $q$
in the bottom line of Eq.~(\ref{cessm30}) includes only heavy quarks
for which mixing effects parametrised via the mixing angle
$\theta_q$ can not be ignored.  The corrections specified above
can be as large as $20\%-30\%$ because the gluino is strongly
interacting, with a large group theory factor due to its colour,
and because it couples to all of the squark--quark pairs.

\subsection{The charginos and neutralinos}

After EWSB, all Superpartners of the gauge and Higgs bosons acquire
non--zero masses. Since the Supermultiplets of the $Z'$ boson and
SM-singlet Higgs field $S$ are electromagnetically neutral they do not
contribute any extra particles to the chargino spectrum. Consequently
the chargino mass matrix and its eigenvalues remain the same as in the
MSSM, namely
\begin{eqnarray}
m^2_{\chi^{\pm}_{1,\,2}}
&=&\ds\frac{1}{2}\biggl[M_2^2+\mu_{\rm eff}^2+2
M^2_{W} \biggl. \nonumber \\
&&\biggr.\qquad\qquad
\pm \sqrt{(M_2^2+\mu^2_{\rm eff}+2M^2_{W})^2-4(M_2\mu_{\rm eff}-M^2_{W}\sin
2\beta)^2} \biggr]\,,
\label{cessm32}
\end{eqnarray}
where $M_2$ is the $SU(2)$ gaugino mass and
$\mu_{\rm eff}=\ds\frac{\lambda s}{\sqrt{2}}$. LEP searches for SUSY
particles including data collected at $\sqrt{s}$ between
$90\,\mbox{GeV}$ and $209\,\mbox{GeV}$ set a $95\%$ CL lower limit on
the chargino mass of about $100\,\mbox{GeV}$ \cite{51}. This lower
bound constrains the parameter space of the E$_6$SSM restricting the
absolute values of the effective $\mu$--term and $M_2$ from below,
i.e.\ $|M_2|$, $|\mu_{\rm eff}|\ge 90-100\,\mbox{GeV}$.

In the neutralino sector there are two extra neutralinos besides the
four MSSM ones. One is an extra gaugino coming from the $Z'$ vector
Supermultiplet. The other is an additional Higgsino $\tilde{S}$ (singlino). 
In the interaction basis $(\tilde{B},\,\tilde{W}_3,\,\tilde{H}^0_1,\,\tilde{H}^0_2,\,
\tilde{S},\,\tilde{B}')$ the neutralino mass matrix takes a form
\be M_{\tilde{\chi}^0}= \left(
\ba{cccccc} M_1 & 0 & -\ds\frac{1}{2}g'v_1 & \ds\frac{1}{2}g'v_2 & 0 &
0\\[2mm] 0 & M_2 & \ds\frac{1}{2}g v_1 & -\ds\frac{1}{2}g v_2 & 0 &
0\\[2mm] -\ds\frac{1}{2}g'v_1 & \ds\frac{1}{2}g v_1 & 0 & -\mu_{\rm eff} &
-\ds\frac{\lambda v_2}{\sqrt{2}} & \tilde{Q}_1 g'_1 v_1\\[2mm]
\ds\frac{1}{2}g'v_2 & -\ds\frac{1}{2}gv_2 & -\mu_{\rm eff} & 0 &
-\ds\frac{\lambda v_1}{\sqrt{2}} & \tilde{Q}_2 g'_1 v_2\\[2mm] 0 & 0 &
-\ds\frac{\lambda v_2}{\sqrt{2}} & -\ds\frac{\lambda v_1}{\sqrt{2}} &
0 & \tilde{Q}_S g'_1 s \\[2mm] 0 & 0 & \tilde{Q}_1 g'_1 v_1 &
\tilde{Q}_2 g'_1 v_2 & \tilde{Q}_S g'_1 s & M'_1 \ea \right)\,,
\label{cessm33}
\ee
where $M_1$, $M_2$ and $M'_1$ are the soft gaugino masses for
$\tilde{B}$, $\tilde{W}_3$ and $\tilde{B}'$ respectively.  In
Eq.~(\ref{cessm33}) we neglect the Abelian gaugino mass mixing
$M_{11}$ between $\tilde{B}$ and $\tilde{B}'$ that arises at low
energies as a result of the kinetic term mixing even if there is no
mixing in the initial values of the soft SUSY--breaking gaugino masses
near the GUT or Planck scale \cite{Suematsu:1998wm}. The top--left
$4\times 4$ block of the mass matrix (\ref{cessm33}) contains the
neutralino mass matrix of the MSSM where the parameter $\mu$ is
replaced by $\mu_{\rm eff}$. The lower right $2\times 2$ submatrix
represents the extra components of neutralinos. The neutralino sector in
$E_6$ inspired SUSY models was studied recently in
\cite{Keith:1997zb}, \cite{Suematsu:1997tv}--\cite{Suematsu:1997qt},
\cite{Suematsu:1997au}--\cite{Keith:1996fv},
\cite{Hesselbach:2001ri}--\cite{E6neutralino-higgs}.

As one can see from Eqs.~(\ref{cessm32})--(\ref{cessm33}) the masses
of charginos and neutralinos depend on $\lambda$, $s$, $\tan\beta$,
$M_1,\,M'_1$ and $M_2$. In SUGRA models with uniform gaugino masses at
the GUT scale, the RGE flow yields a relationship 
$ M'_1\simeq M_1\simeq 0.5 M_2$\,.
Due to stringent constraints on the mass of the $Z'$ boson, the VEV of
the SM singlet field $S$ has to be large ($s \gtrsim
2\,\mbox{TeV}$). This implies that $\tilde{Q}_S g'_1 s$ and $\mu_{\rm
eff}$ are much larger than other entries in the
neutralino mass matrix (\ref{cessm33}). As a result the mass matrix
(\ref{cessm33}) can be approximately diagonalised and the expressions
for the chargino masses (\ref{cessm32}) can be substantially
simplified.  In this case one chargino and two neutralinos
are almost degenerate with mass $|\mu_{\rm eff}|$, i.e.\
\be
|m_{\chi^{\pm}_2}|\simeq |m_{\chi^{0}_3}|\simeq |m_{\chi^{0}_4}|\simeq |\mu_{\rm eff}|\,.
\label{cessm35}
\ee
They are formed predominantly from the neutral and charged
Superpartners of the Higgs bosons. Two other neutralinos are mixtures
of the $U(1)_{N}$ gaugino $\tilde{B}'$ and singlino $\tilde{S}$. Their
masses are closely approximated by
\be
|m_{\chi^0_{5,6}}|\simeq\frac{1}{2}\biggl[\sqrt{M^{'2}_1+4M^2_{Z'}}\mp M'_1\biggr]\,.
\label{cessm36}
\ee
Since the masses of extra neutralino states are controlled by the $Z'$
boson mass they tend to be heavy $(\sim 1\,\mbox{TeV})$ so that their
direct observation is unlikely in the near future.  The Superpartners
of the $SU(2)$ gauge bosons compose another chargino and neutralino
whose masses are governed by $|M_2|$. Finally, the mass of the
neutralino state that is predominantly bino, $\tilde{B}$, is set by
$|M_1|$.

\subsection{The exotic particles}

In addition to the NMSSM-like particle content, the E$_6$SSM involves
exotic matter that forms three families of down--type quark
Superfields ($\overline{D}_i$ and $D_i$), two generations of Inert
Higgs Supermultiplets ($H^d_{\alpha}$ and $H^u_{\alpha}$), two
families of extra singlets $S_{\alpha}$ and a vector--like doublet
Superfield associated with the survival components of the extra $27'$
and $\overline{27}'$ ($H'$ and $\overline{H}'$) which manifest
themselves in the Yukawa interactions (\ref{cessm7}) as fields with
lepton number $L=\pm 1$. The masses of the fermion and scalar
components of this vector--like lepton Supermultiplet are set by $\mu'$
which is expected to be of the order of $10\,\mbox{TeV}$. Therefore
these exotic lepton fields are normally very heavy and decouple from
the rest of the particle spectrum. The masses of the fermion
components of the exotic quark and Inert Higgs Supermultiplets are
determined by the VEV of the SM-singlet field $S$ and by the Yukawa
couplings $\kappa_i$ and $\lambda_{\alpha}$. They are given by
\be
\mu_{D_i}=\dfrac{\kappa_i}{\sqrt{2}}\,s\,,\qquad\qquad 
\mu_{\tilde{H}_{\alpha}}=\dfrac{\lambda_{\alpha}}{\sqrt{2}}\,s\,,
\label{cessm37}
\ee
where $\mu_{D_i}$ are exotic quark masses, while $\mu_{\tilde{H}_{\alpha}}$ are
the masses of the Inert Higgsinos.  The experiments at LEP,
HERA and the Tevatron set stringent lower bounds on the masses of exotic
quarks and new charged particles, so the Yukawa couplings $\kappa_i$ and
$\lambda_{\alpha}$ cannot be negligibly small.

Relatively large masses of exotic quarks give rise to a substantial
mixing between the corresponding exotic squark states. Since we choose a
field basis such that the Yukawa couplings of $D_i$ and $\overline{D}_i$
to $S$ are flavour diagonal, the calculation of the exotic
squark masses reduces to the diagonalisation of three $2\times 2$
matrices whose eigenvalues can be written as
\begin{eqnarray}
\hspace{-10mm}  M^2_{D_{i\,1},\,D_{i\,2}}
&=&\dfrac{1}{2} \left\{ \lefteqn{\phantom{\sqrt{\left[\frac{1}{2}\right]^2}}}
m^2_{D_i}+m^2_{\overline{D}_i}+2\mu_{D_i}^2+
\Delta_{D}+\Delta_{\overline{D}} \right. \nonumber \\
&&\mp \left. \sqrt{\left[m^2_{D_i}-m^2_{\overline{D}_i}+\dfrac{2}{3} M_Z^2 \cos 2\beta \sin^2\theta_W
+\Delta_{D}-\Delta_{\overline{D}}\right]^2+4\mu_{D_i}^2 X_{D_i}^2}\, \right\}\,, \nonumber \\[-4mm]
\label{cessm38}
\end{eqnarray}
where $X_{D_i}=A_{\kappa_i}-\dfrac{\lambda}{2\sqrt{2}s}v^2\sin 2\beta$
and $\Delta_{\phi}=\dfrac{g^{'2}_1}{2} \biggl(\tilde{Q}_1
v_1^2+\tilde{Q}_2 v_2^2+\tilde{Q}_S s^2\biggr)
\tilde{Q}_{\phi}$. Relatively heavy Inert Higgsinos also lead to
significant mixing effects in the Inert Higgs boson sector. Once
again, the flavour diagonal structure of the Yukawa couplings of
$H^d_{\alpha}$ and $H^u_{\alpha}$ to the singlet field $S$, leads to
mixing only between the Inert Higgs bosons from the same family.
Diagonalising the appropriate $2\times 2$ mass matrices one finds,
\begin{eqnarray}
\hspace{-10mm}
m^2_{H^{0}_{\alpha\,1},\,H^{0}_{\alpha\,2}}
&=&\dfrac{1}{2} \left\{ \lefteqn{\phantom{\sqrt{\left[\frac{1}{2}\right]^2}}}
m^2_{H^d_{\alpha}}+m^2_{H^u_{\alpha}}+
2\mu_{\tilde{H}_{\alpha}}^2+\Delta_{H^d}+\Delta_{H^u} \right. \nonumber \\
&& \mp \left. \sqrt{\left[m^2_{H^d_{\alpha}}-m^2_{H^u_{\alpha}}+M_Z^2\cos 2\beta + \Delta_{H^d}-\Delta_{H^u}\right]^2+
4\mu_{\tilde{H}_{\alpha}}^2 X_{H_{\alpha}}^2}\, \right\}\,,
\label{cessm39} \\
\hspace{-10mm}
m^2_{H^{\pm}_{\alpha\,1},\,H^{\pm}_{\alpha\,2}}
&=&\dfrac{1}{2}\left\{ \lefteqn{\phantom{\sqrt{\left[\frac{1}{2}\right]^2}}}
m^2_{H^d_{\alpha}}+m^2_{H^u_{\alpha}}+
2\mu_{\tilde{H}_{\alpha}}^2+\Delta_{H^d}+\Delta_{H^u} \right. \nonumber \\
&&\mp \left. \sqrt{\left[m^2_{H^d_{\alpha}}-m^2_{H^u_{\alpha}}-M_Z^2\cos 2\beta \cos 2\theta_W + \Delta_{H^d}-\Delta_{H^u}\right]^2+
4\mu_{\tilde{H}_{\alpha}}^2 X_{H_{\alpha}}^2} \, \right\}\,, \nonumber \\
\label{cessm40}
\end{eqnarray}
where
$X_{H_{\alpha}}=A_{\lambda_{\alpha}}-\dfrac{\lambda}{2\sqrt{2}s}v^2\sin
2\beta$. The magnitude of the mixing in the exotic squark and Inert
Higgs sectors is governed by the mixing parameters $X_{D_i}$ and
$X_{H_{\alpha}}$ as well as by the Yukawa couplings $\kappa_i$ and
$\lambda_{\alpha}$. If the Yukawa couplings that determine the
mixing of the exotic scalar fields are large, the mixing effects can
be so substantial that the corresponding lightest exotic squarks
and/or Inert Higgs bosons may be among the lightest SUSY particles in
the spectrum of the E$_6$SSM.  Additionally, when $\kappa_i$ or
$\lambda_i$ are relatively small the appropriate exotic quarks or
Inert Higgsinos may be sufficiently light that they can be discovered
at the LHC.

Since we neglect the couplings $f_{\alpha\beta}$ and
$\tilde{f}_{\alpha\beta}$ in the Superpotential (\ref{cessm7}), the
scalar components of the SM-singlet Superfields $S_{\alpha}$ do not
mix with other scalar fields.  Their masses are given by
\be
M^2_{S_{\alpha}}=m^2_{S_{\alpha}}+\Delta_{S}\,,
\label{cessm41}
\ee
where $m^2_{S_{\alpha}}$ are soft scalar masses while $\Delta_{S}$ is
a $U(1)_N$ D--term contribution.  In the leading approximation, the
$U(1)_N$ D--term contributions to the masses of the exotic scalars
are set by $M_{Z'}^2$
\be
\Delta_{D}\simeq \Delta_{H^u}\simeq -\dfrac{1}{5}M_{Z'}^2\,,\qquad\quad
\Delta_{\overline{D}}\simeq \Delta_{H^d}\simeq -\dfrac{3}{10}M_{Z'}^2\,,\qquad\quad
\Delta_{S}\simeq \dfrac{1}{2} M_{Z'}^2\,.
\label{cessm42}
\ee
We emphasise that in contrast with the ordinary squarks and sleptons,
the $U(1)_N$ D--term gives negative contributions to the masses of
exotic squarks and Inert Higgs bosons because the $U(1)_N$ charge of
the SM-singlet Superfield $S$ and the $U(1)_N$ charges of the exotic
quarks and Inert Higgs Supermultiplets are opposite. The $U(1)_N$
D--term gives the largest contributions to the masses of the scalar
components of the SM-singlet Superfields $S_{\alpha}$, making these
fields rather heavy.

\subsection{The Higgs bosons}
\label{sec:Higgs}
Due to electric charge conservation the charged components of the
Higgs doublets do not mix with neutral Higgs fields. They form a
separate sector whose spectrum is described by a $2\times 2$ mass
matrix. Its determinant has zero value leading to the appearance of
two Goldstone states which are absorbed into the longitudinal degrees
of freedom of the $W^{\pm}$ gauge boson. Their orthogonal linear
combination gains mass
\be
m^2_{H^{\pm}}=\ds\frac{\sqrt{2}\lambda A_{\lambda}}{\sin 2\beta}s-\frac{\lambda^2}{2}v^2+\frac{g^2}{2}v^2+\Delta_{\pm}\,,
\label{cessm43}
\ee
where $\Delta_{\pm}$ represents the contribution of loop corrections
to the charged Higgs boson mass in the E$_6$SSM.

The imaginary parts of the neutral components of the Higgs doublets
and imaginary part of the SM-singlet field $S$ compose the CP--odd
Higgs sector of the model. This sector includes two
Goldstone modes $G_0,\, G'$ which are swallowed by the $Z$ and $Z'$
bosons after EWSB, leaving only one physical CP--odd Higgs state $A$
which acquires mass
\be
m^2_{A}=\ds\frac{\sqrt{2}\lambda A_{\lambda}}{\sin 2\varphi}v+\Delta_A\,,\qquad\qquad
\tan\varphi=\ds\frac{v}{2s}\sin2\beta\,,
\label{cessm44}
\ee
where $\Delta_A$ is the contribution of loop corrections. 

The CP--even Higgs sector involves $\mbox{Re}\,H_d^0$, $\mbox{Re}\,H_u^0$ and $\mbox{Re}\,S$. In the field space
basis $(h,\,H,\,N)$ rotated by an angle $\beta$ with respect to the initial one
\be
\ba{c}
Re\,H_d^0=(h \cos\beta- H
\sin\beta+v_1)/\sqrt{2}\,, \\[2mm] Re\,H_u^0=(h \sin\beta+ H
\cos\beta+v_2)/\sqrt{2}\,, \\[2mm] Re\,S=(s+N)/\sqrt{2}\,,
\ea
\label{cessm45}
\ee
the mass matrix of the Higgs scalars takes the form \cite{54}:
\be
M^2=
\left(
\ba{ccc}
\ds\frac{\partial^2 V}{\partial v^2}&
\ds\frac{1}{v}\frac{\partial^2 V}{\partial v \partial\beta}&
\ds\frac{\partial^2 V}{\partial v \partial s}\\[0.3cm]
\ds\frac{1}{v}\frac{\partial^2 V}{\partial v \partial\beta}&
\ds\frac{1}{v^2}\frac{\partial^2 V}{\partial^2\beta}&
\ds\frac{1}{v}\frac{\partial^2 V}{\partial s \partial\beta}\\[0.3cm]
\ds\frac{\partial^2 V}{\partial v \partial s}&
\ds\frac{1}{v}\frac{\partial^2 V}{\partial s \partial\beta}&
\ds\frac{\partial^2 V}{\partial^2 s}
\ea
\right)=\left(
\ba{ccc}
M_{11}^2 & M_{12}^2 & M_{13}^2\\
M_{21}^2 & M_{22}^2 & M_{23}^2\\
M_{31}^2 & M_{32}^2 & M_{33}^2
\ea
\right)\,.
\label{cessm46}
\ee
Taking second derivatives of the Higgs boson effective potential and substituting $m_1^2$, $m_2^2$, $m_{S}^2$
from the minimisation conditions (\ref{cessm14a})-(\ref{cessm14c}) one obtains,
\begin{eqnarray}
M_{11}^2&=&\ds\frac{\lambda^2}{2}v^2\sin^22\beta+\ds\frac{\bar{g}^2}{4}v^2\cos^22\beta+g^{'2}_1 v^2(\tilde{Q}_1\cos^2\beta+
\tilde{Q}_2\sin^2\beta)^2+\Delta_{11}\,,\nonumber \\
M_{12}^2&=&M_{21}^2=\ds\left(\frac{\lambda^2}{4}-\ds\frac{\bar{g}^2}{8}\right)v^2
\sin 4\beta \nonumber \\
&&\qquad \qquad +\ds\frac{g^{'2}_1}{2}v^2(\tilde{Q}_2-\tilde{Q}_1)
(\tilde{Q}_1\cos^2\beta+\tilde{Q}_2\sin^2\beta)\sin 2\beta+\Delta_{12}\, ,\nonumber \\
M_{22}^2&=&\ds\frac{\sqrt{2}\lambda A_{\lambda}}{\sin 2\beta}s+\left(\frac{\bar{g}^2}{4}-\ds\frac{\lambda^2}{2}\right)v^2
\sin^2 2\beta+\ds\frac{g^{'2}_1}{4}(\tilde{Q}_2-\tilde{Q}_1)^2 v^2 \sin^22\beta+\Delta_{22}\,,\nonumber \\
M_{23}^2&=&M_{32}^2=-\ds\frac{\lambda A_{\lambda}}{\sqrt{2}}v\cos 2\beta+\ds\frac{g^{'2}_1}{2}(\tilde{Q}_2-\tilde{Q}_1)\tilde{Q}_S
v s\sin 2\beta+\Delta_{23}\,,\nonumber \\
M_{13}^2&=&M_{31}^2=-\ds\frac{\lambda A_{\lambda}}{\sqrt{2}}v\sin 2\beta+\lambda^2 v s+g^{'2}_1(\tilde{Q}_1\cos^2\beta+
\tilde{Q}_2\sin^2\beta)\tilde{Q}_S v s+\Delta_{13}\,,\nonumber \\
M_{33}^2&=&\ds\frac{\lambda A_{\lambda}}{2\sqrt{2}s}v^2\sin 2\beta+g^{'2}_1\tilde{Q}_S^2s^2+\Delta_{33}\,.
\label{cessm47}
\end{eqnarray}
In Eqs.~(\ref{cessm47}) the $\Delta_{ij}$'s are loop corrections to
the mass matrix of the CP--even Higgs bosons in the E$_6$SSM.  The
explicit expressions for $\Delta_{ij}$, calculated in the leading
one--loop approximation, are given in Appendix A.

When the SUSY--breaking scale $M_S$ and VEV of the singlet field are
considerably larger than the EW scale, the mass matrix
(\ref{cessm46})--(\ref{cessm47}) has a hierarchical
structure. Therefore the masses of the heaviest Higgs bosons are
closely approximated by the diagonal entries $M_{22}^2$ and $M_{33}^2$
which are expected to be of the order of $M_S^2$ or even higher.  All
off--diagonal matrix elements are relatively small $\lesssim M_S
M_Z$. As a result the mass of one CP--even Higgs boson (approximately
given by $H$) is governed by $m_A$ while the mass of another one
(predominantly the $N$ singlet field) is set by $M_{Z'}$.  Since the
minimal eigenvalue of the mass matrix (\ref{cessm46})--(\ref{cessm47})
is always less than its smallest diagonal element at least one Higgs
scalar in the CP--even sector (approximately $h$) remains light even
when the SUSY--breaking scale tends to infinity, i.e.\
$m^2_{h_1}\lesssim M_{11}^2$. In contrast with the MSSM, the lightest 
Higgs boson in the E$_6$SSM can be heavier than $110-120\,\mbox{GeV}$ 
even at tree level. In the two--loop approximation the lightest Higgs 
boson mass does not exceed $150-155\,\mbox{GeV}$ 
\cite{King:2005jy}--\cite{King:2005my}. 
The Higgs sector in the $E_6$ inspired SUSY models was studied recently 
in \cite{King:2005jy}, \cite{E6neutralino-higgs}, \cite{E6-higgs}.

\section{Constructing realistic cE$_6$SSM scenarios}

\subsection{RG flow of couplings in the cE$_6$SSM}
%\subsection{Analysis of the RG flow}
Below the GUT scale, the RG flow causes the gauge couplings and the
soft SUSY--breaking parameters to split from the universal values
$g_0$, $m_0^2$, $M_{1/2}$ and $A_0$. This splitting is described
by the RGEs of the model, presented in Appendix \ref{Appendix:RGEs}.
For the gauge and Yukawa couplings two--loop RGEs are given as well as
two--loop RGEs for $M_a(\mu)$ and $A_i(\mu)$ and one--loop RGEs for $m_i^2(\mu)$.

This complete set of E$_6$SSM RGEs can be separated into two parts. The
first describes the evolution of gauge and Yukawa coupling constants
and is a nonlinear set of equations even in the one--loop
approximation. Therefore it is extremely difficult or even impossible
to find either exact or approximate solutions of these equations. The
remaining subset of RGEs describes the running of fundamental
parameters which break SUSY in a soft way. If the renormalisation
group flow of the gauge and Yukawa couplings is known, this part of the
RGEs can be considered as a set of linear differential equations for
the soft SUSY--breaking terms. To solve them, first
one integrates the equations for the gaugino masses $M_i$. In the
one--loop approximation we find,
\begin{equation}
M_i(t)=\ds\frac{g_i^2(t)}{g_0^2}\,M_{1/2},\qquad\qquad
M'_1(t)=\ds\frac{g^{'2}_1(t)}{g_0^2}\,M_{1/2},
\label{Eq:GaginoMasses_1loop}
\end{equation}
where the index $i$ runs from $1$ to $3$ and $t=\ln\ds\frac{Q}{M_X}$, with
$Q$ being the renormalisation scale at which
Eq.~(\ref{Eq:GaginoMasses_1loop}) holds true.

Next one integrates the one--loop RGEs for the trilinear scalar couplings
$A_i(t)$ which can be written as,
\begin{equation}
\frac{dA_i(t)}{dt}=S_{ij}(t)A_{j}(t)+F_i(t).
\label{cessm88}
\end{equation}
The dependence of $F_i$ on $t$ comes from the gaugino masses appearing
in the one--loop RGE of the trilinears.  One then finds the solution
of this system of linear differential equations,
\begin{equation}
\label{cessm99}
A_i(t)=\Phi_{ij}(t)A_j(0)+\Phi_{ik}(t)\int\limits_0^t\Phi_{kj}^{-1}(t')F_j(t')dt',
\end{equation}
where we have introduced $\Phi_{ij}(t)$, which is the solution of
the homogeneous equation \linebreak $d\Phi_{ij}(t)/dt=S_{ik}(t)\Phi_{kj}(t)$,
with the boundary conditions $\Phi_{ij}(0)=\delta_{ij}$. From the universality
constraint and exploiting Eq.~(\ref{Eq:GaginoMasses_1loop})
to write $F_i(t)\propto M_{1/2}$, the solution of the RGEs
for the trilinear scalar couplings takes the form
\begin{equation}
A_i(t)=e_i(t)A_0+f_i(t)M_{1/2}.
\label{Eq:NUHM_Trilinears}
\end{equation}
The obtained solution Eq.~(\ref{Eq:NUHM_Trilinears}) can be substituted into
the right--hand sides of the RGEs for the soft scalar masses which may be
presented in the following form,
\begin{equation}
\frac{dm_i^2(t)}{dt}=\tilde{S}_{ij}(t)m_{j}^2(t)+\tilde{F}_i(t).
\label{cessm111}
\end{equation}
Due to the scalar mass universality constraints and the fact that the functions
$\tilde{F}_i(t)$ contain terms which are proportional to $A_0^2$, $A_0 M_{1/2}$,
and $M_{1/2}^2$ the solution of the linear system of differential
Eq.~(\ref{cessm111}) reduces to,
\begin{equation}
m_i^2(t)=a_i(t)m_0^2+b_i(t)M_{1/2}^2+c_i(t)A_0 M_{1/2}+d_i(t)A_0^2.
\label{cessm:softscalars}
\end{equation}

Analytic expressions for $e_i(t)$, $f_i(t)$, $a_i(t)$, $b_i(t)$,
$c_i(t)$, and $d_i(t)$, which determine the evolution of $A_i(t)$ and
$m_i^2(t)$, are unknown, since an exact analytic solution of the
E$_6$SSM RGEs is not available.

The sensitivity of these functions to the Yukawa and gauge couplings at $M_X$
is again very strong. In particular it is important to reiterate that the
one--loop $\beta$--function for the gauge coupling of strong interactions is zero.
So the running of $g_3$ and $M_3$ is dictated solely by the two--loop contributions
and these two--loop $\beta$--functions can change the RG flow substantially.
In this study the two--loop $\beta$--functions for the gaugino masses and
trilinear couplings were included.  The solution of two--loop RGEs for the
$M_i(t)$ can be written as,
\begin{equation}
M_i(t)=p_i(t)A_0+q_i(t)M_{1/2}.
\label{cessm:2lpGauginos}
\end{equation}
One can see that in the two--loop approximation gaugino masses depend not only
on the universal gaugino mass, $M_{1/2}$, but also on the trilinear scalar coupling,
$A_0$. The numerical calculations show that the dependence of $M_i(t)$ on $A_0$ is
rather weak, i.e.\ $p_i(t_0)\ll 1$. However the change in the co-efficient $q_i(t)$
is substantial and at low--energies the gaugino masses change by 20--40\%.

The general form of the solutions of RGEs for $m_i^2(t)$ and $A_i(t)$ remains
intact after the inclusion of two--loop effects. At the same time some of the
coefficient functions $f_i(t)$, $b_i(t)$ and $c_i(t)$ change significantly.
The two--loop corrections to the $\beta$--functions have the strongest impact
on the RG flow of the soft SUSY--breaking terms which are sensitive to strong
interactions.

\begin{figure}\begin{minipage}[b]{0.5\linewidth}
\hspace*{5mm}{\large $\alpha_i(t)$}\\[1mm]
\includegraphics[height=45mm,keepaspectratio=true]{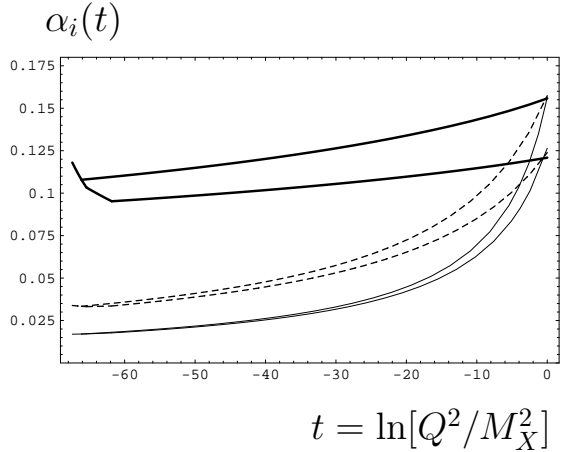}\\
{\hspace*{40mm}\large $t=\ln[Q^2/M_X^2]$}\end{minipage}
\begin{minipage}[b]{0.45\linewidth}\caption[Threshold dependence of
two--loop RG flow of gauge couplings within the E$_6$SSM]{Two--loop RG flow
of gauge couplings within the E$_6$SSM for $T_{MSSM}=T_{ESSM}=M_t=175\,\mbox{GeV}$
(upper lines) and $T_{MSSM}=250\,\mbox{GeV}$, $T_{ESSM}=1500\,\mbox{GeV}$
(lower lines). Here we fix $\tan\beta=10$ and $\alpha_3(M_Z)=0.118$.
\label{cessmgaugeRunning}} \vspace{1.5cm}
\end{minipage}
\end{figure}

The RG flow of the gauge couplings, $g_i(t)$, is also quite sensitive
to threshold effects. In Fig.~\ref{cessmgaugeRunning} the running of
$\alpha_i(t)$ is presented for two different sets of threshold scales,
$T_{MSSM}=T_{ESSM}=175\,\mbox{GeV}$ and $T_{MSSM}=250\,\mbox{GeV}$,
$T_{ESSM}=1500\,\mbox{GeV}$.  The threshold $T_{MSSM}$ is a common
scale for the sparticles of ordinary matter, while $T_{ESSM}$ is a
common mass scale for new exotic particles not present in the MSSM.
The unified gauge coupling at $M_X$ changes from 1.24 to 1.4 between
the two threshold choices. This result and also the value of $g_0^2$
for several other threshold choices, $T_{MSSM}$ and $T_{ESSM}$, are
summarised in Tab.~\ref{cessmThresholds}.
%
%%\begin{comment}
Since soft SUSY--breaking terms depend very strongly on the values of
the gauge couplings at the GUT scale, the uncertainty related to the
choice of the threshold scales limits the accuracy of our calculations of the
particle spectrum. The results of our numerical analysis presented in
Tab.~\ref{cessmThresholds} and Fig.~\ref{cessmgaugeRunning} indicate
that it is unrealistic to expect an accuracy, in the calculation of
the sparticle masses, better than 10\%.

\setlength{\tabcolsep}{2.5mm}
\begin{table}[t]
\centering
\begin{tabular}{|c|c|c|c|c|c|c|}
\hline
$T_{MSSM}$ (GeV)   & $250$          & $250$             & $250$             & $175$            & $175$          & $175$  \\
\hline
$T_{ESSM}$ (GeV)   & $1500$           & $800$            & $250$            & $1500$            & $250$          & $175$  \\
\hline
$g_0^2$ & $1.54$               & $1.60$               & $1.78$               & $1.61$               & $1.88$             & $1.96$   \\
\hline
$M_X$ (GeV)   & $3.5\cdot 10^{16}$  & $3.3\cdot 10^{16}$  & $3.5\cdot 10^{16}$  & $3.7\cdot 10^{16}$  & $4\cdot 10^{16}$  &
$4\cdot 10^{16}$  \\
\hline
\end{tabular}
\caption[Threshold dependence of $g_0$ and $M_X$ in the cE$_6$SSM]{The
dependence of $g_0^2$ and $M_X$ on the threshold effects in the
exceptional SUSY model. Here we fix $\tan\beta=10$ and
$\alpha_3(M_Z)=0.118$.}
\label{cessmThresholds}
\end{table}
%%\end{comment}

In our analysis thresholds are used only in the SUSY preserving sector
where full two--loop RGE are employed and are neglected in the soft
SUSY--breaking sector where only one--loop RGE are used for the
scalar masses. The thresholds are chosen before the spectrum is determined
and are therefore only an estimate.  A more accurate analysis is left for
a further study.  We chose \mbox{$T_{MSSM} = 600$ GeV}  and
$T_{ESSM} = 3$ TeV to be the mass scale of the unobserved particles of
the MSSM and the new exotic objects in the E$_6$SSM respectively,
based on preliminary studies where relatively heavy spectra were observed.

\subsection{Procedure of our analysis}
To calculate the particle spectrum within the cE$_6$SSM one must find
masses and couplings which are consistent with both the high scale
universality constraints and the low scale EWSB constraints.  To
evolve between these two scales we use two--loop renormalisation group
equations (RGEs), presented in Appendix \ref{Appendix:RGEs}, in a
modified version of SOFTSUSY 2.0.5 \cite{Allanach:2001kg}. The details
of the procedure we followed are summarized below.

1. The gauge and Yukawa couplings are determined independently of
the soft SUSY breaking mass parameters as follows:

(i) We select values for $s =  \sqrt{2}\langle S \rangle$ and
$\tan\beta=v_2/v_1$.

(ii) We set the gauge couplings $g_1$, $g_2$ and $g_3$ equal to
the experimentally measured values at $M_Z$.

(iii) We fix the low energy Yukawa couplings $h_t$, $h_b$, and
$h_{\tau}$ using the relations between the running masses of the
fermions of the third generation and VEVs of the Higgs fields,
i.e.  \be m_t(M_t)=\dfrac{h_t(M_t) v}{\sqrt{2}}\sin\beta,~~
m_b(M_t)=\dfrac{h_b(M_t) v}{\sqrt{2}}\cos\beta,~~
m_{\tau}(M_t)=\dfrac{h_{\tau}(M_t) v}{\sqrt{2}}\cos\beta .
\label{Fermion_Running_masses} \ee

(iv) The gauge and Yukawa couplings are then evolved up to the GUT
scale $M_X$. Using the beta functions for QED and QCD, the gauge
couplings are evolved up to $m_t$. Between $m_t$ and $T_{MSSM}$ we
evolve the gauge and Yukawa couplings with SM RGEs and between
$T_{MSSM}$ and $T_{ESSM}$ we employ the MSSM RGEs. At $T_{ESSM}$ the
values of E$_6$SSM gauge and Yukawa couplings, $g_1$, $g_2$, $g_3$,
$h_t$, $h_b$ and $h_\tau$, form a low energy boundary condition for
what follows. Initial low energy estimates of the new E$_6$SSM Yukawa
couplings, $\lambda_i$ and $\kappa_i$ are also input here, and all
SUSY preserving couplings are evolved up to the unification scale using the
two--loop E$_6$SSM RGEs.

(v) At the unification scale $M_X$ we set $g_1' = g_0$ and select values
for $\kappa_i(M_X)$ and $\lambda_i(M_X)$, which are input
parameters in our procedure. An iteration is then performed
between $M_X$ and the low energy scale to obtain the values of
all the gauge and Yukawa couplings which are consistent with our
input values for $\kappa_i(M_X)$, $\lambda_i(M_X)$, gauge coupling unification 
and our low scale boundary conditions, derived from experimental data.

2. Now that the values of the gauge and Yukawa couplings have been
obtained, the coefficients $e_i(t)$, $f_i(t)$, $a_i(t)$, $b_i(t)$,
$c_i(t)$, $d_i(t)$, $p_i(t)$ and $q_i(t)$, appearing in
Eq.~(\ref{Eq:NUHM_Trilinears}), Eq.~(\ref{cessm:softscalars}) and
Eq.~(\ref{cessm:2lpGauginos}), can be obtained for $t
=\ln[T_{ESSM}/M_X^2]$.  Low energy soft mass parameters are then
functions of the GUT scale values of $A_0$, $M_{1/2}$ and $m_0$.
These coefficients are determined numerically as follows:

(i) Set $A_0$ and $M_{1/2}$ to zero at $M_X$ while giving $m_0$ a
non-zero value and run the full set of E$_6$SSM parameters down to the
low scale to yield the coefficients proportional to $a_i(t)$ in the
expressions for each low energy scalar (mass)$^2$, $m_i^2$.

(ii) Repeat for $A_0$ and $M_{1/2}$ to obtain coefficients $b_i(t)$ and
$d_i(t)$ for each $m_i^2$; coefficients $e_i(t)$ and $f_i(t)$ for each
low energy trilinear soft mass $A_i$ and coefficients $p_i(t)$ and
$q_i(t)$ for each low energy gaugino soft mass $M_i$.

(iii) The coefficients, $c_i(t)$, of the $A_0 M_{1/2}$ terms appearing
in the semi-analytic expressions for each $m_i^2$ are then determined
using non-zero values of both $A_0$ and $M_{1/2}$ at $M_X$, using the
results in part (ii) to isolate this term.

3. The semi-analytic expressions for the soft masses from step 2 above
provide the set of low energy constraints on the soft masses coming
from our cE$_6$SSM universality conditions. These are then combined with
the conditions for correct EWSB, appearing in Eqs.~(\ref{cessm14a})-(\ref{cessm14c}),
at low energy and determine sets of $m_0$, $M_{1/2}$ and $A_0$ which
are consistent with EWSB, as follows:

(i) Working with the tree--level potential $V_0$ (to start with) we
impose the minimisation conditions $\dfrac{\partial V_0}{\partial s}=
\dfrac{\partial V_0}{\partial v_1}=\dfrac{\partial V_0}{\partial
v_2}=0$.  In the tree--level approximation each of the EWSB conditions
are quadratic functions of $\lambda_3(\mu)$, where $\mu$ is the energy
scale at which the EWSB conditions are imposed. Using the
semi-analytic approach described above to replace the third generation
soft Higgs and Singlet masses and $A_{\lambda_3}$ reveals that each
EWSB condition also has quadratic dependence on the soft unification
scale parameters $m_0$, $M_{1/2}$ and $A_0$.  With three constraints and
three soft mass parameters, the equations can be reduced to two second
order equations with respect to $A_0$ and $M_{1/2}$, or equivalently
one quartic equation with respect to $A_0$.  This equation is solved
numerically, and the resulting value for $A_0$ is used to obtain
$M_{1/2}$ and $m_0$.  For fixed values of gauge couplings, Yukawas and
VEVs (determined from choices of $\tan \beta$ and $s$ with $v$ known
from experiment) there are four sets of soft masses $A_0$, $M_{1/2}$ and
$m_0$, though some or all can in principle be complex.  Here we restrict
our consideration to the scenarios with real values of fundamental
parameters which do not induce any CP--violating effects. Therefore
our routine deals with between $0$ and $4$ sets of real solutions to
the soft masses.

(ii) For each solution $m_0$, $M_{1/2}$ and $A_0$ the low energy
stop soft mass parameters are determined and the one--loop
Coleman-Weinberg Higgs effective potential $V_1$ is calculated.
The new minimisation conditions for $V_1$ are then imposed, and
new solutions for $m_0$, $M_{1/2}$ and $A_0$ are obtained.

(iii) The procedure in (ii) is then iterated until we find stable
solutions.  For some values of $\tan\beta$, $s$ and
Yukawa couplings the solutions with real $A_0$, $M_{1/2}$ and
$m_0$ do not exist. There is a substantial part of the parameter
space where there are only two solutions with real values of
fundamental parameters. However, there are also some regions of the
parameters where all four solutions of the non--linear algebraic
equations are real.

Although correct EWSB is not guaranteed in the cE$_6$SSM,
remarkably, there are always solutions with real $A_0$, $M_{1/2}$
and $m_0$ for sufficiently large values of $\kappa_i$, which drive
$m_S^2$ negative. This is easy to understand since the $\kappa_i$
couple the singlet to a large multiplicity of coloured fields,
thereby efficiently driving its squared mass negative to trigger
the breakdown of the gauge symmetry.

4. Using the obtained solutions we calculate the masses of all
exotic and SUSY particles, using expressions given in section
\ref{Sec:ParticleSpec}, for each set of fundamental parameters.

Finally, at the last stage of our analysis we vary Yukawa
couplings, $\tan\beta$ and $s$ to establish the qualitative
pattern of the particle spectrum within the cE$_6$SSM.
To avoid any conflict with present and former collider experiments as
well as with recent cosmological observations we impose the set of
constraints specified in the next section. We then demonstrate
how these bounds restrict the allowed range of the parameter space
in the cE$_6$SSM by performing scans over our input parameters.

\subsection{Experimental and Theoretical Constraints}
%\subsection{Low Energy Constraints}
The experimental constraints applied in our analysis are: $m_h \geq
114$ GeV, all sleptons and charginos are heavier than
$100\,\mbox{GeV}$, all squarks and gluinos have masses above
$300\,\mbox{GeV}$ and the $Z'$ boson has a mass which is larger than
$860\,\mbox{GeV}$ \cite{Grivaz:2008tf}. We also impose the most conservative
bound on the masses of exotic quarks and squarks that comes from the
HERA experiments \cite{Aktas:2005pr}, by requiring that they are
heavier than $300\,\mbox{GeV}$.  Finally, we require that the Inert
Higgs and Inert Higgsinos are heavier than 100 GeV to evade limits from LEP.

In addition to setting bounds from the non--observation of new
particles in experiment, we impose some theoretical constraints. We
require that the Lightest Supersymmetric Particle (LSP) should be a
neutralino. We also restrict our consideration to the values of the
Yukawa couplings $\lambda_i(M_X)$, $\kappa_i(M_X)$, $h_t(M_X)$,
$h_b(M_X)$ and $h_{\tau}(M_X)$ less than 3 to ensure the
applicability of perturbation theory up to the GUT scale.

In our exploration of the cE$_6$SSM parameter space we looked at
scenarios with a universal coupling between exotic coloured
Superfields and the third generation singlet field $\hat{S}$,
$\kappa_{1,2,3}(M_X) = \kappa(M_X)$ and fixed the Inert Higgs
couplings $\lambda_{1,2}(M_X) = 0.1$. In fixing $\lambda_{1,2}$ like
this we are deliberately pre-selecting for relatively light Inert
Higgsinos. The third generation Yukawa $\lambda = \lambda_3$ was
allowed to vary along with $\kappa$.  Splitting $\lambda_3$ from
$\lambda_{1,2}$ seems reasonable since $\lambda_3$ plays a very
special role in E$_6$SSM models in forming the effective $\mu$--term
when $S$ develops a VEV.

\begin{figure}[h]
 \begin{center}
\resizebox{!}{6cm}{%
\includegraphics{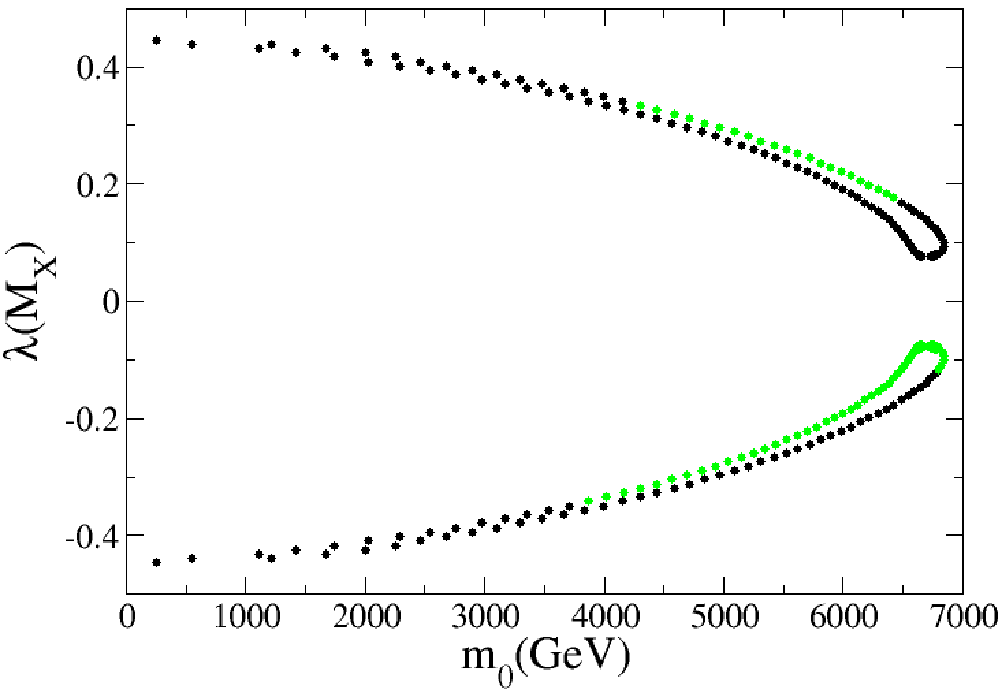}}  \\
\resizebox{!}{6cm}{%
\includegraphics{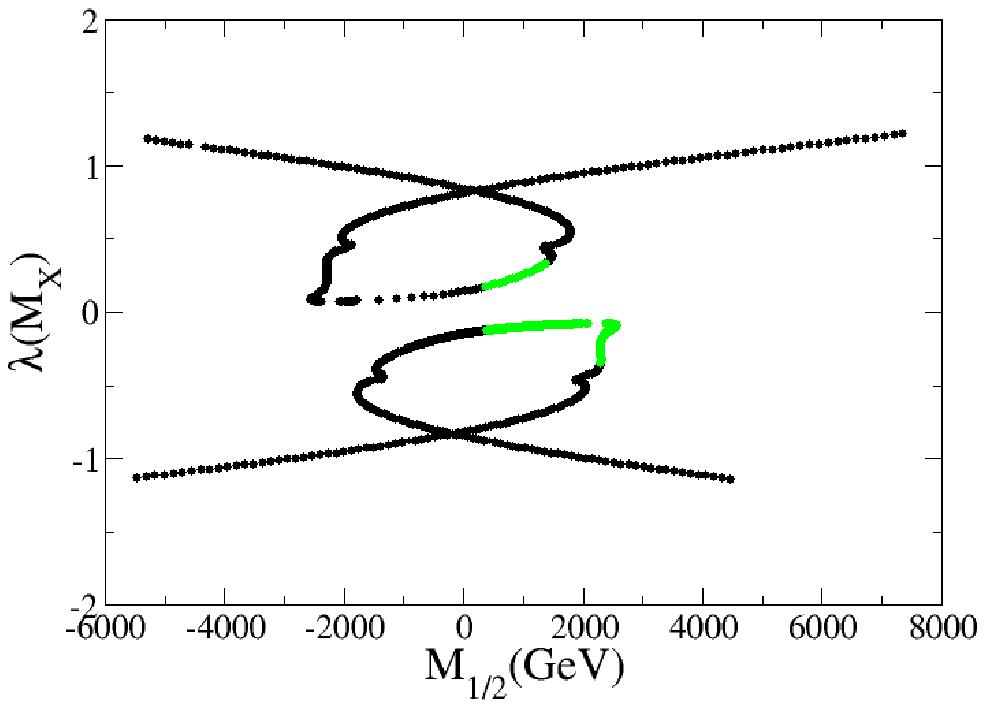}}
\resizebox{!}{6cm}{\includegraphics{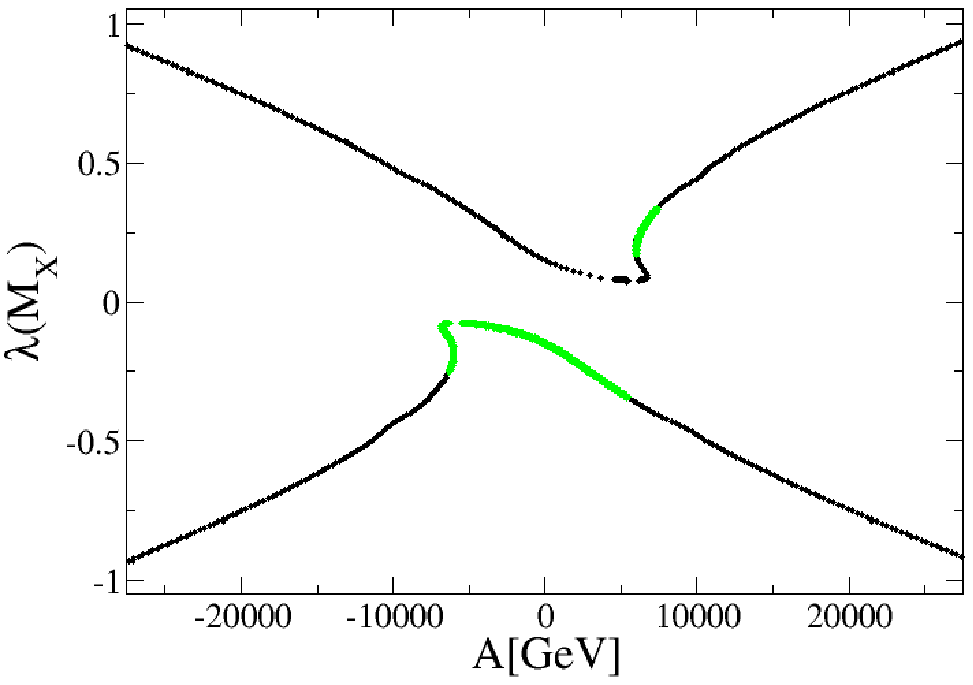}}
\caption{cE$_6$SSM solutions with $\tan \beta = 10$, $s= 20$ TeV and
$\kappa_{1,2,3} = 0.25$, $\lambda_{1,2} = 0.1$ fixed showing the
relationship between $\lambda$ and $m_0$ (top), $M_{1/2}$ (bottom left)
and $A$ (bottom right). Points in green (light gray) satisfy all
experimental constraints from LEP and Tevatron data, while points
in black are ruled out. \label{Cessm:lambda_dependence1} }
\end{center}
\end{figure}

The first results we found were for a very large singlet VEV,
$s \approx 10-20$ TeV, and this leads to a very heavy particle
spectrum where many of the new particles would be out of reach of
current collider experiments.  This can be seen in
Fig.~\ref{Cessm:lambda_dependence1} where the dependencies of the
soft mass parameters $m_0$, $M_{1/2}$ and $A_0$ on $\lambda$ for
$s = 20$ TeV and a particular value of $\kappa = 0.25$ are plotted.
One can see that for each value of $\lambda$ there are two different
values of each soft mass. This is because we find that for these
points, of the four solutions to our quartic equation, two are
complex, leaving only the two real solutions appearing in the plots.
We find the existence of two real solutions and two complex solutions
to be typical for the parameter space we have examined.

 %%

Notice also that the solutions presented above possess a certain symmetry.
This is because there is an invariance under the transformation
$A_0 \rightarrow - A_0, \, M_{1/2}\rightarrow -M_{1/2}, \lambda \rightarrow -\lambda $.
However,  we exploit this symmetry to adopt a convention whereby $M_{1/2} \geq 0$
is fixed, and therefore are only admitting physical solutions with $M_{1/2} \geq 0$,
with the result that this symmetry is not apparent for our valid solutions
shown in green (light gray).

\begin{figure}[h]
\begin{center}
\resizebox{!}{6cm}{%
\includegraphics{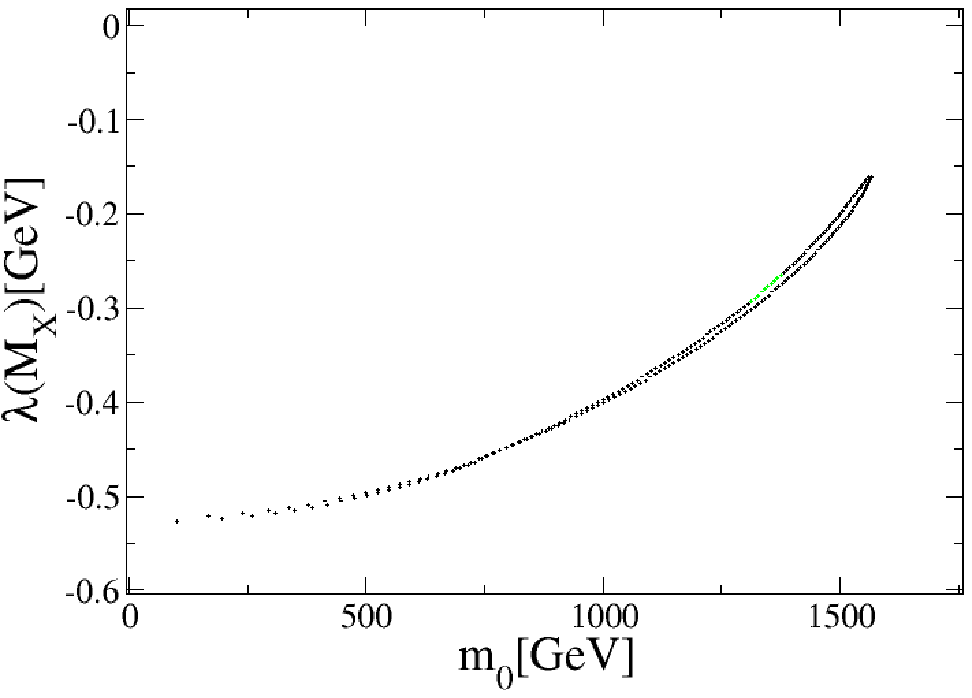}}
\resizebox{!}{6cm}{%
\includegraphics{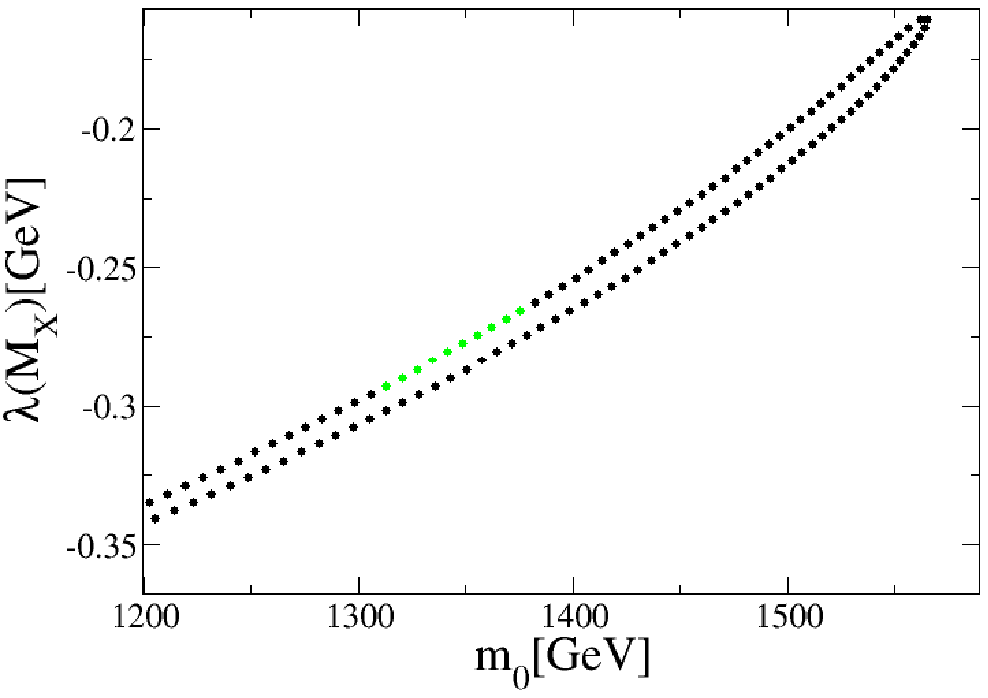}}  \\
\resizebox{!}{6cm}{%
\includegraphics{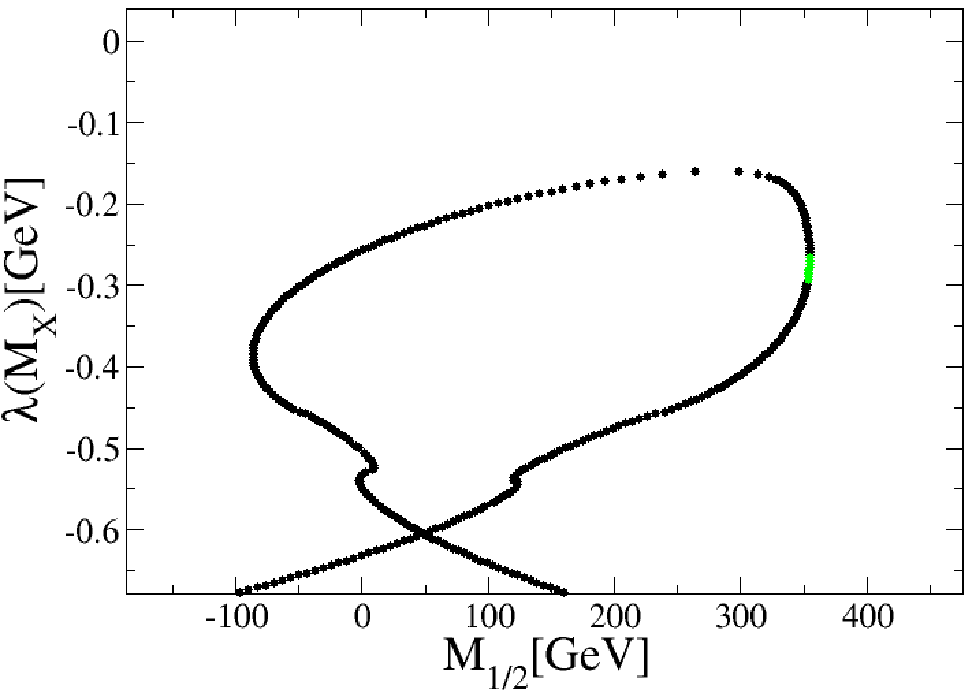}}
\resizebox{!}{6cm}{\includegraphics{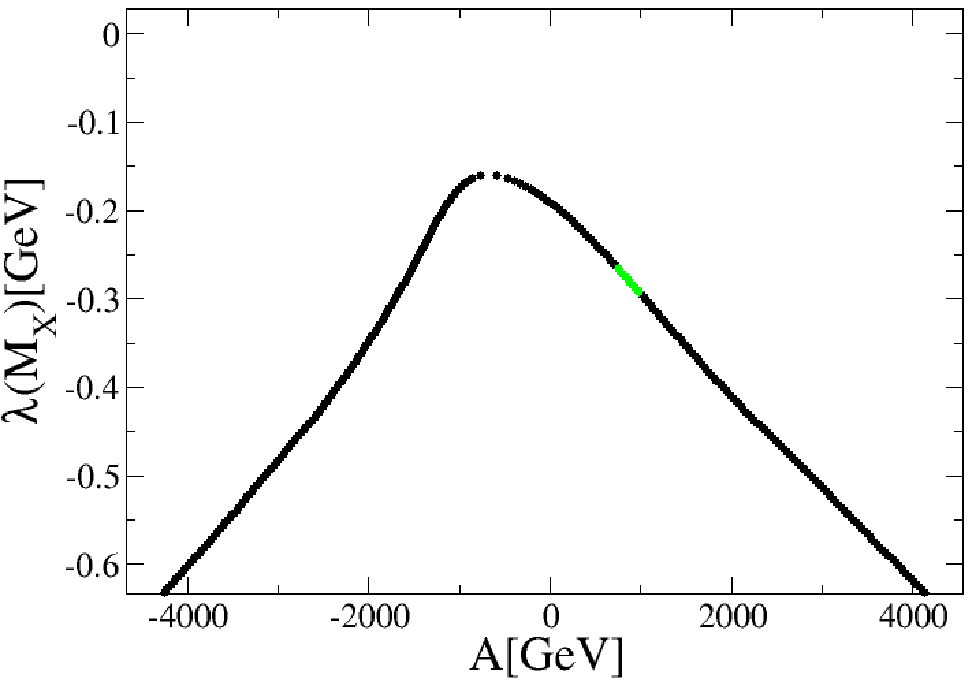}
}
\caption{cE$_6$SSM solutions with $\tan \beta = 10$, $s= 5$ TeV and $\kappa_{1,2,3} = 0.25$,
$\lambda_{1,2} = 0.1$ fixed showing the relationship between $\lambda$ and
$m_0$ (top left and magnified top right)  $M_{1/2}$ (bottom left) and
$A$ (bottom right). Points in green (light gray) satisfy all experimental constraints
from LEP and Tevatron data, while points in black are ruled out. \label{Cessm:lambda_dependence2} }
\end{center}
\end{figure}

After further study, we also discovered solutions that are allowed by all experimental
constraints and have a significantly lighter $s$ for a smaller range of $\lambda_3$
and our universal $\kappa$. This is illustrated in Fig.~\ref{Cessm:lambda_dependence2}
where the soft mass dependencies  on $\lambda$ for $s = 5$ TeV and $\kappa = 0.25$
(which is within this narrow range allowing $s$ to be relatively light). Since many
particles in the cE$_6$SSM have their masses set by the singlet VEV it is of clear
phenomenological interest to study the parameter space with low values of $s$.

 %%

To further explore this interesting region of the cE$_6$SSM parameter
space, for different fixed values of $\tan \beta = 3, 10, 30$, we scan
over $s, \kappa_i$ and $\lambda$.  From these input parameters, the sets
of soft mass parameters, $A_0$, $M_{1/2}$ and $m_0$ which are consistent
with the correct breakdown of the EW symmetry are found.

We find that for fixed values of the Yukawas the soft mass parameters
scale with $s$, while if $s$ and  $\tan \beta$ are fixed, varying the
Yukawas, $\lambda$ and $\kappa_i$ then produces a bounded region of
allowed points.

\begin{figure}[h]
\begin{center}
\resizebox{!}{10cm}{%
\includegraphics{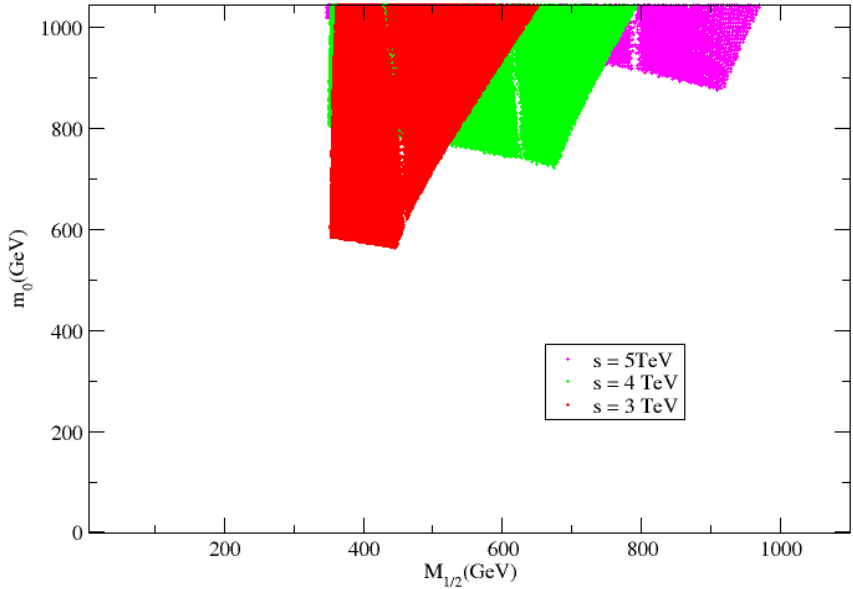}
}
\caption{Physical solutions with $\tan \beta = 10$, $\lambda_{1,2} = 0.1$,
$s= \{3, 4, 5\}$ TeV fixed and $\lambda \equiv \lambda_3$ and
$\kappa \equiv \kappa_{1,2,3}$ varying, which pass experimental
constraints from LEP and Tevatron data.  On the left hand side of each
allowed region the chargino mass is less than $100$ GeV, while underneath
the Inert Higgses are less than $100$ GeV or becoming tachyonic.
The region ruled out immediately to the right of the allowed points
is due to $m_h < 114 $ GeV.\label{tb10_s3TeV_Valid} }
\end{center}

\end{figure}

The value of $s$ determines the location and extent of the bounded regions.
As $s$ is increased the lowest values $m_0$ and $M_{1/2}$, consistent with
experimental searches and EWSB requirements, increase. This is shown in
Fig.~\ref{tb10_s3TeV_Valid} where the allowed regions for three different
values of the singlet VEV, $s = 3$ TeV, $4$ TeV and $5$ TeV, are compared,
with the allowed regions in red (dark grey), green (light grey), 
magenta (medium grey) respectively and the excluded regions in white. 
Note that these regions overlap since we are finding soft masses consistent 
with EWSB conditions that have a non-linear dependence on the VEVs and Yukawas.

%%

\begin{figure}[h!]
\begin{center}
\resizebox{!}{6cm}{%
\includegraphics{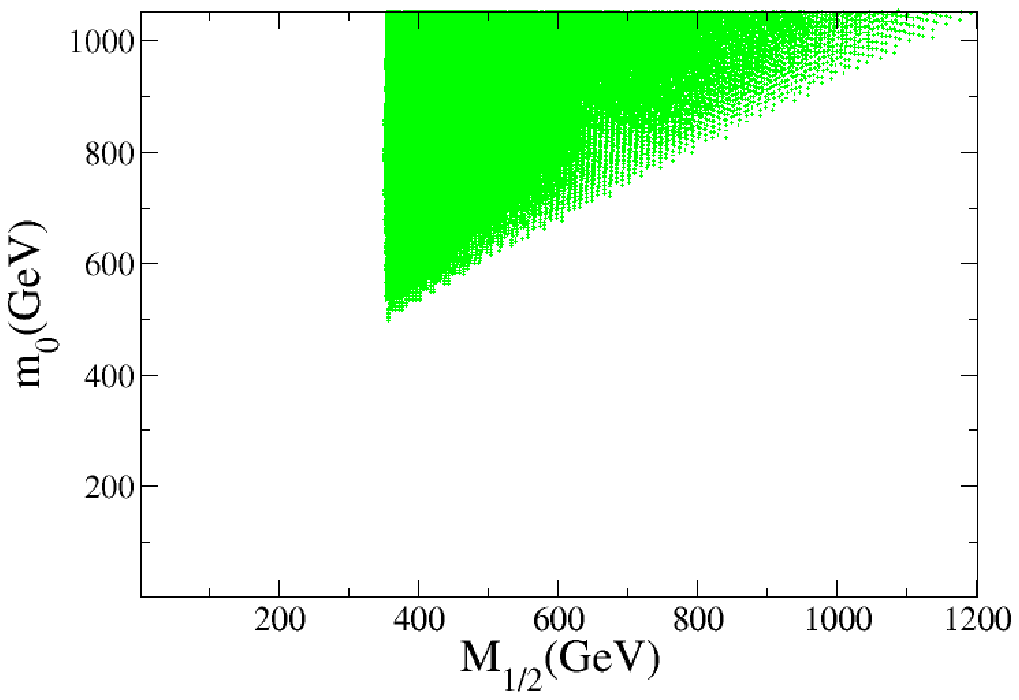}}
\resizebox{!}{6cm}{\includegraphics{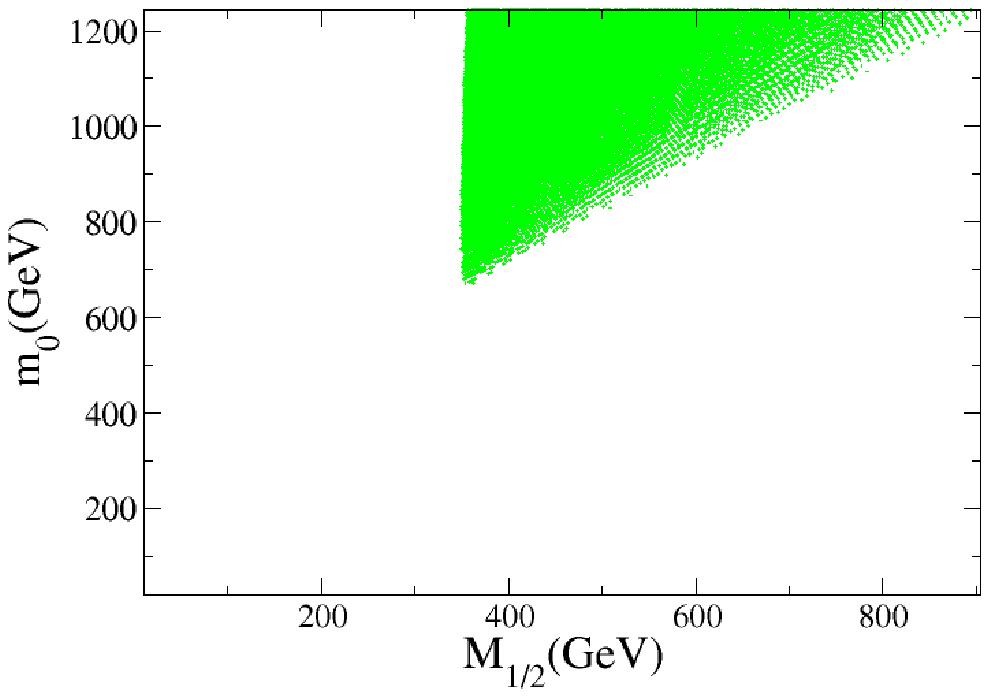}}
\resizebox{!}{6cm}{\includegraphics{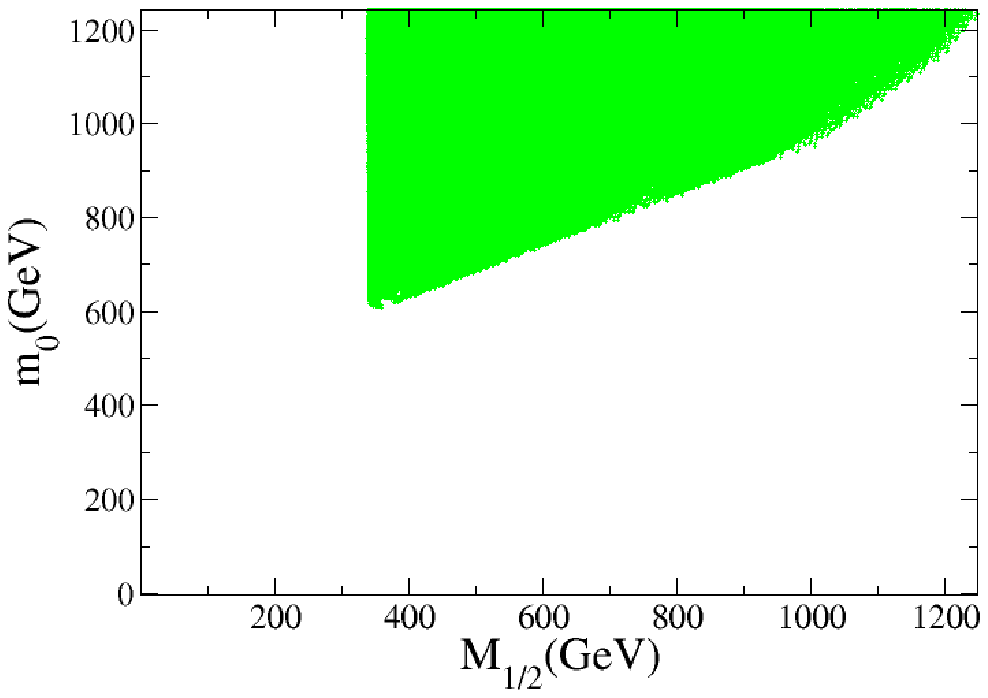}}
\caption[cE$_6$SSM exclusion plot for $\tan \beta = 10$]{Physical
solutions $\tan \beta = 10$ (top, left), $\tan \beta = 30$ (top,
right) and $\tan \beta = 3$ (bottom) with $\lambda_{1,2} = 0.1$,
fixed and $\lambda \equiv \lambda_3$ and $\kappa \equiv \kappa_{1,2,3}$
and $s$ all varying, which pass experimental constraints from LEP
and Tevatron data.\label{tb10_Valid}}
\end{center}
\end{figure}

Further scanning over $s$, leaving only $\tan \beta$ fixed, we find a
lower limit on the ratio $m_0 / M_{1/2}$ which is a weak function of
the singlet VEV $s$.  For example, consider Fig.~\ref{tb10_Valid} (top,
left). The region to the left of the allowed space is ruled out by
the lightest chargino mass, $m_{\chi_1^\pm} < 100$ GeV, while the
lower right region is ruled out by Inert Higgs bosons with masses
below experimental bounds or tachyonic. This boundary implies that
for $\tan \beta = 10$, over the allowed ranges shown, $m_0 / M_{1/2}$
varies from $\approx 1.4$ to $\approx 0.8$.

 %%

This boundary can be understood as follows.   For fixed $m_0$,
maximizing $M_{1/2}$ requires the singlet VEV $s$ to be increased,
as well as varying the Yukawas, $\lambda$ and $\kappa$. However,
the squared masses of the Inert Higgs bosons receive a positive
contribution from $m_0^2$ and a negative contribution from the
auxiliary D--term which varies with $s^2$ (see Eqs.~(\ref{cessm39})
and (\ref{cessm40})). Due to this D--term contribution the mass of the 
lightest Inert Higgs boson decreases with $s$ and at some point falls
below experimental limits, bounding $M_{1/2}$ from above.
The larger $m_0$ is, the larger the negative contribution must be
in order to drive the Inert Higgs mass below its lower limit.
Further, if one assumes that $m_0 \sim s$ and $A_\lambda \sim M_{1/2}$
then EWSB conditions imply $s \sim M_{1/2} \tan \beta$. This suggests
not only the observed limit on  $m_0 / M_{1/2}$ but also that it
will be more severe for large $\tan \beta$ and shallower for
low $\tan \beta$.

The allowed region for $\tan \beta = 30$ in Fig.~\ref{tb10_Valid}
(top, right) has a similar shape but in this case $m_0 / M_{1/2}$
varies from $\approx 1.9$ to $\approx1.4$, so for this larger $\tan
\beta = 30$ the limit on ratio $m_0 / M_{1/2}$ is enhanced. For $\tan
\beta = 3$ in Fig.~\ref{tb10_Valid} (bottom) the situation is somewhat
different.  The region to the left of the allowed parameter space is
still ruled out by experimental limits on the chargino mass. However
the lower-right region is ruled out, not by the Inert Higgs masses,
but by a light Higgs which is lower than the LEP limit.  This change
has two underlying reasons. Firstly the Inert Higgs bosons obtain
positive contributions to their masses from $m_0$ (with a coefficient
of $\approx 1$) and $M_{1/2}$, while, due to the auxiliary D--term
contribution, the Inert Higgs masses decrease with $s$.  Since
decreasing $\tan \beta$ reduces the hierarchy between $s$ and
$M_{1/2}$, this negative contribution to the mass of the Inert Higgs is
smaller and does not decrease their mass as rapidly when $m_0$ is
reduced. Secondly we observe that the lightest Higgs mass reduces 
with $\tan\beta$ as in the MSSM. At $\tan\beta=3$ the maximal value of 
the mass of the lightest Higgs boson is rather close to the LEP bound. 
As a result the variations of parameters can result in the increase of 
the mixing in the CP--even Higgs sector, which provides a negative 
contribution to the lightest Higgs boson mass, so that it becomes lower 
than the LEP limit of $114\,\mbox{GeV}$.

\subsection{Benchmark Scenarios}

A remarkable feature of the cE$_6$SSM is that the low energy gluino mass
parameter $M_3$ is driven to be smaller than $M_{1/2}$ by RG running.
The reason for this is that the E$_6$SSM has a much larger (super)field
content than the MSSM (three 27's instead of three 16's) so much so that
at one--loop order the QCD beta function (accidentally) vanishes in the
E$_6$SSM, and at two loops it loses asymptotic freedom (though the gauge
couplings remain perturbative at high energy). This implies that the low
energy gaugino masses are all less than $M_{1/2}$ in the cE$_6$SSM,
being given as roughly $M_3 \sim 0.7M_{1/2}$, $M_2 \sim 0.25M_{1/2}$,
$M_1 \sim 0.15M_{1/2}$. These should be compared to the corresponding low
energy values in the MSSM, $M_3 \sim 2.7M_{1/2}$, $M_2 \sim 0.8M_{1/2}$,
$M_1 \sim 0.4M_{1/2}$.

\begin{figure}[htbp]
\begin{center}
\resizebox{!}{10cm}{%
\includegraphics{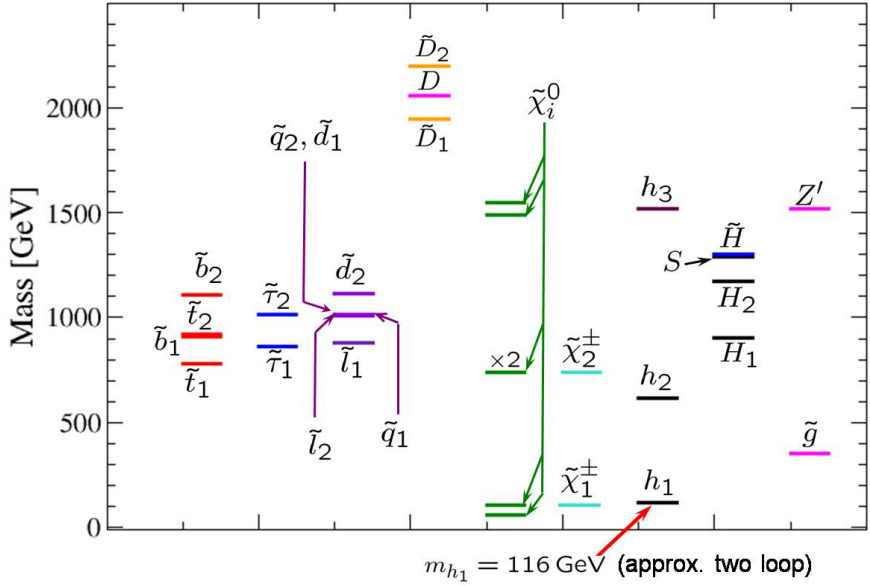}
}
\vspace{-5mm}
\caption{The particle mass spectra for cE$_6$SSM Benchmark
Point 1, with $\tan \beta = 10$, $s = 4.0 \, \rm{TeV}$,
$M_{1/2} = 389\,\rm{GeV}$,  $m_0 = 725 \, \rm{GeV}$,
$A =  -1528  \,\rm{GeV}$,  $\lambda_{1,2}(M_X) = 2.6$,
$\lambda_3(M_X) = -2.0$,   $\lambda_3(\mu_S) = -0.259$,
$\kappa_{1,2,3}= 2.5$, $\kappa_3(\mu_S) = 0.728$.
\label{fig:Spectra_bpf2}}
\vspace{-5mm}
\end{center}
\end{figure}%%

Thus, in the cE$_6$SSM, since the low energy gaugino masses $M_i$ are
driven by RG running to be small, the lightest SUSY states will
generally consist of a light gluino of mass $\sim M_3$, a light
wino-like neutralino and chargino pair of mass $\sim M_2$, and a light
bino-like neutralino of mass $\sim M_1$, which are typically all much
lighter than the Higgsino masses of order $\mu = \lambda s/\sqrt{2}$,
where $\lambda$ cannot be too small for correct EWSB. The remaining
neutralinos are mainly a superposition of the $U(1)_{N}$ gaugino and
singlet Higgsino. Their masses are governed by $M_{Z'}$. The mass of
the $Z'$ is set by the singlet VEV, i.e.\ $M_{Z'}\approx g'_1 Q_S s$
($g'_1\approx g_1$) and therefore is also much heavier than gluino,
lightest neutralino and chargino. The heaviest CP--even Higgs state is
degenerate with the $Z'$ while another CP--even Higgs, CP--odd and
charged Higgs bosons have almost the same masses which are relatively
close to the masses of charged and neutral Higgsinos. Since $m_0$
tends to be larger than $M_{1/2}$ for each value of $s$ (as may be
seen in Fig.~\ref{tb10_s3TeV_Valid}) the Superpartners of ordinary
quarks and leptons are considerably heavier than the light gauginos as
well. This is a general prediction of the cE$_6$SSM. Moreover as follows 
from benchmark 1 (Fig.~\ref{fig:Spectra_bpf2}) all extra exotic particles 
in the cE$_6$SSM can be also relatively heavy so that the light sector 
of the sparticle spectrum includes only gluino, two light neutralinos and
light chargino.  Nonetheless, even the pessimistic scenario described by
benchmark 1 leads to the striking collider signature. Indeed, because
gluino, two light neutralinos and light chargino have relatively
small masses in the considered case the pair production of
$\chi_2^0\chi_2^0$, $\chi_2^0\chi_1^\pm$, $\chi_1^\pm \chi_1^\mp$ and
$\tilde g \tilde g$ should be possible at the LHC.

With increasing VEV of the SM-singlet field the structure of the particle
spectrum becomes more hierarchical. Due to the hierarchical spectrum the
gluinos can be relatively narrow states because $\Gamma_{\tilde{g}}\propto
M_{\tilde{g}}^5/m_{\tilde{q}}^4$.  In particular their width can be
comparable to that of $W^{\pm}$ and $Z$ bosons.  They will
decay through $\tilde{g} \rightarrow q \tilde{q}^* \rightarrow q
\bar{q} + E_T^{\rm miss}$, so gluino pair production will result in an
appreciable enhancement of the cross section for $pp \rightarrow q \bar q
q \bar q + E_T^{\rm miss} + X$, where $X$ refers, hereafter, to any number
of light quark/gluon jets. The second lightest neutralino decays through
$\chi_2^0 \rightarrow \chi_1^0 + l \bar l $ and so would produce an excess
in $pp \rightarrow l \bar l l \bar l + E_T^{\rm miss} + X$, which could be
observed at the LHC.

Notice however that, while these are general predictions of the model, it is
also possible that more exciting signatures could originate in the cE$_6$SSM.
For example, when the Yukawa couplings $\kappa_i$ of the exotic fermions
$D_i$ and $\overline{D}_i$ have a hierarchical structure, some of them can
be relatively light so that their production cross section at the LHC can
be comparable with the cross section of $t\bar{t}$ production \cite{King:2005jy}.
In the E$_6$SSM the $D_i$ and $\overline{D}_i$ fermions are SUSY particles with
negative $R$--parity so they must be pair produced and decay into quark--squark
(if diquarks) or quark--slepton, squark--lepton (if leptoquarks). Assuming that
$D_i$ and $\overline{D}_i$ fermions couple most strongly to the third family
(s)quarks and (s)leptons the presence of light exotic quarks in the particle
spectrum can lead to a substantial enhancement of the cross section of either
$pp\to t\bar{t}b\bar{b}+E^{\rm miss}_{T}+X$ if exotic quarks are diquarks
or $pp\to t\bar{t}\tau \bar{\tau}+E^{\rm miss}_{T}+X$ and
$pp\to b\bar{b}+ E^{\rm miss}_{T}+X$ if new quark states are leptoquarks.
The scenarios with light exotic quarks, light stop and a TeV scale $Z'$,
which have early discovery potential at the LHC, are considered in our
companion paper \cite{Athron:2009ue}.

In this work we concentrate on the various scenarios with universal $\kappa$
couplings which have distinctive phenomenology and could provide interesting
novel signatures at the LHC. In Tab.~\ref{table:benchmarks} we specify a set
of benchmark points, which demonstrate different patterns of the particle
spectrum that can be obtained in the considered case. The first block of
Tab.~\ref{table:benchmarks} shows the input parameters which define the
benchmark points. These benchmarks cover three different values of
$\tan\beta = 3,10,30$. We deliberately restricted ourselves here to
$s=3.4-5.5\,\mbox{TeV}$ and $(m_0,M_{1/2})<(1100,950)\,\mbox{GeV}$ in order
to get a relatively light particle spectrum that can be observed at the LHC.
Since we focus on the solutions with $s=3.4-5.5\,\mbox{TeV}$, the allowed
range of the cE$_6$SSM parameter space remains rather narrow and the lightest
Higgs boson mass is always relatively close to the LEP limit of $114\,\mbox{GeV}$.
Because we have taken the $\kappa_i$ to be universal at the GUT scale these
couplings have to be large enough to trigger EWSB. Since the $\kappa_i$'s control
the exotic coloured fermion masses, this implies that all the $D_i$ and
$\overline{D}_i$ fermions are all very heavy in the considered cases.
For benchmarks presented in Tab.~\ref{table:benchmarks} the exotic coloured
fermions have masses in the range $1.2-2.2\,\mbox{TeV}$.

\begin{figure}[h!]
\begin{center}
\resizebox{!}{10cm}{%
\includegraphics{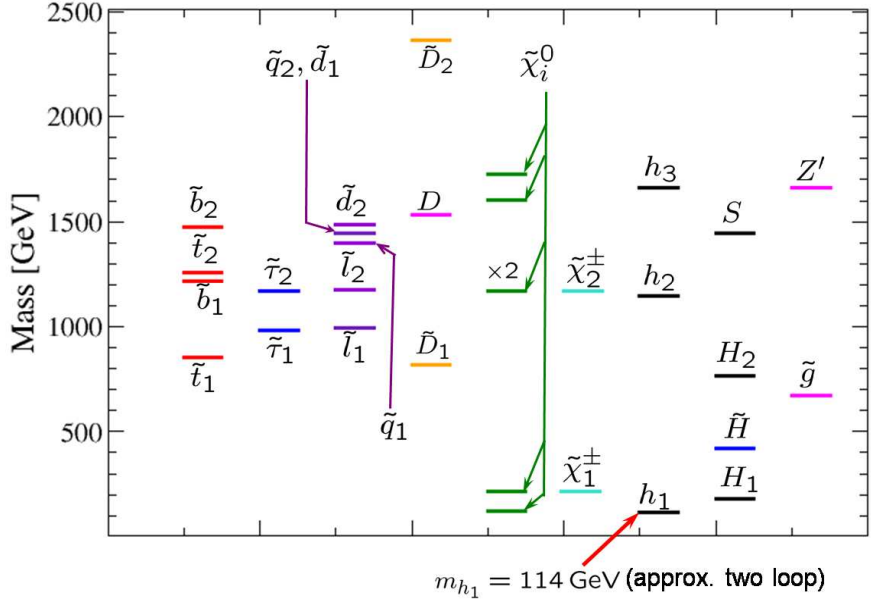}
}
\vspace{-5mm}
\caption[cE$_6$SSM Benchmark point A3]{Benchmark point 2, with
$\tan \beta = 10$,  $s = 4.4$ \rm{TeV},   $M_{1/2} = 775\,\rm{GeV}$,
$m_0 = 799 \, \rm{GeV}$,  $A = 919  \,\rm{GeV}$, $\lambda(M_X) =  -0.3698$,
$\lambda(\mu_S) =  -0.3736$, $\lambda_{1,2}(M_X) = 0.1$,
$\kappa_{1,2,3}(M_X) = 0.1780$,  $\kappa_{1,2,3}(\mu_S) = 0.4935$.
\label{fig:Spectra_bp3a} }
\vspace{-5mm}
\end{center}
\end{figure}%%

In all of the scans carried out in the previous section and for the most of
benchmark scenarios we have chosen $\lambda_{1,2}(M_X)=0.1$ so that
$|\lambda_3(M_X)|\gg \lambda_{1,2}(M_X)$. Low values of $\lambda_{1,2}(M_X)$
result in relatively light Inert Higgsinos (see benchmarks 2--6) because their
masses are proportional to the corresponding couplings.
For benchmarks 2--6 the Inert Higgsinos are much lighter than squarks,
sleptons and the exotic coloured fermions and have masses below
$400-500\,\mbox{GeV}$.  In contrast, the Inert Higgs bosons
can be light or heavy depending on the free parameters.

Benchmark 2 (shown in Fig.~\ref{fig:Spectra_bp3a}) is a scenario with
very light Inert Higgs bosons ($m_{H_{\alpha\,1}} = 182$ GeV) and fairly
light Inert Higgsinos ($\mu_{\tilde{H}} = 418$ GeV).
The presence of light Inert Higgs bosons in the particle spectrum is
caused by the large mixing effects in the Inert Higgs sector.
The negative contributions from the $U(1)_N$ D--term to the diagonal
entries of the Inert Higgs mass matrices also reduce masses of the
corresponding mass eigenstates. The light Inert Higgs bosons decay via the
$Z_2^H$ violating terms \mbox{$h_{i\alpha k}^N \hat{N}_i^c \hat{H}^u_{\alpha}\hat{L}_k$}, \linebreak
\mbox{$h_{i\alpha k}^U \hat{u}_i^c \hat{H}^u_{\alpha}\hat{Q}_k$},
\mbox{$h_{i\alpha k}^D \hat{d}_i^c \hat{H}^d_{\alpha}\hat{Q}_k$}
and  $h_{i\alpha k}^E \hat{e}_i^c \hat{H}^d_{\alpha}\hat{L}_k$, where the
Inert Higgs Superfields are $SU(2)$ doublets with
$\hat{H}^d_{\alpha} =  (\hat{H}_{\alpha}^{d\, 0}, \hat{H}_{\alpha}^{d\, -})$ and
$\hat{H}^u_{\alpha} =  (\hat{H}_{\alpha}^{u\, +}, \hat{H}_{\alpha}^{u\, 0})$.
These interactions are analogous to the Yukawa interactions of
the Higgs Superfields, $\hat{H}_u$ and $\hat H_d$.
So the neutral Inert Higgs bosons decay predominantly into third
generation fermion--anti-fermion pairs, like
$H^{0}_{\alpha\,1} \rightarrow b \bar b$.
The charged Inert Higgs bosons decays are also into fermion--anti-fermion
pairs, but in this case it is the antiparticle of the fermions' EW  partner
e.g.~$H^{-}_{\alpha\,1} \rightarrow \tau  \bar{\nu}_{\tau}$.

\begin{figure}[h!]
\begin{center}
\resizebox{!}{10cm}{%
\includegraphics{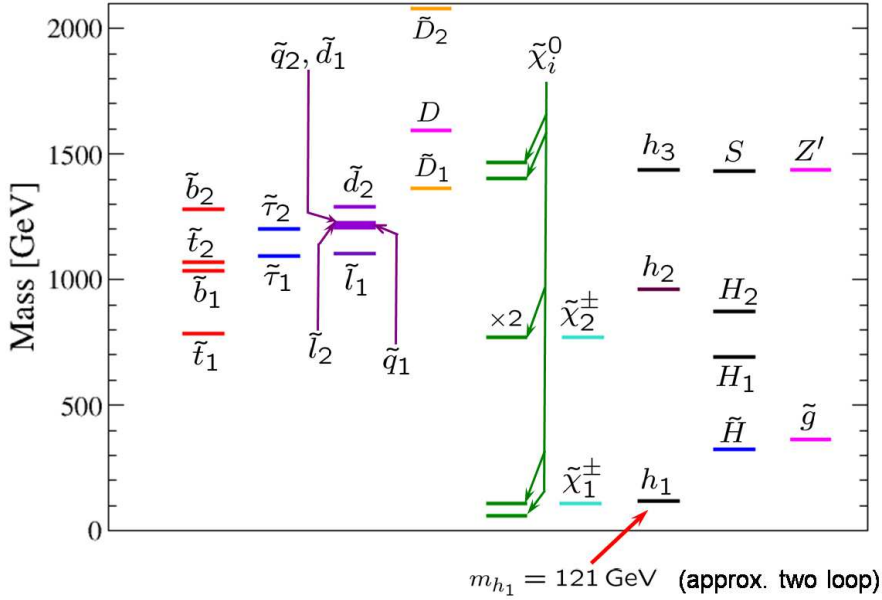}
}
\vspace{-5mm}
\caption[cE$_6$SSM Benchmark point A2]{Benchmark point 3, with $\tan \beta = 10$,
$s = 3.8 \, \rm{TeV}$,   $M_{1/2} = 390\,\rm{GeV}$,  $m_0 = 998 \, \rm{GeV}$,
$A =   768  \,\rm{GeV}$,  $\lambda(M_X) = - 0.3066$,   $\lambda(\mu_S) =  -0.2845$,
$\lambda_{1,2}(M_X) = 0.1$, $\kappa_{1,2,3}(M_X) = 0.2463$,
$\kappa_{1,2,3}(\mu_S) = 0.5935$.
\label{fig:Spectra_bp2a} }
\vspace{-5mm}
\end{center}
\end{figure}%%

The Inert Higgs bosons may also be quite heavy, so that the only light
exotic particles are the Inert Higgsinos. Benchmark 3 (Fig.~\ref{fig:Spectra_bp2a})
is an example of this, emphasising the need to search for both the Inert
Higgsinos as well as the Inert Higgs bosons at future colliders.

The $Z_2^H$ symmetry violating couplings mentioned above also govern the
decays of the Inert Higgsinos. The electromagnetically neutral Higgsinos
predominantly decay into fermion anti-sfermion pairs (e.g.
$\tilde{H}_{\alpha}^0 \rightarrow t  \tilde{\bar{t}}^*$,
$\tilde{H}_{\alpha}^0 \rightarrow \tau  \tilde{\bar{\tau}}^*$).
The charged Higgsino decays are similar, but in this case the sfermion is
the Supersymmetric partner of the EW partner of the fermion,
(e.g. $\tilde{H}_{\alpha}^+ \rightarrow t  \tilde{\bar{b}}^*$,
$\tilde{H}_{\alpha}^- \rightarrow \tau  \tilde{\bar{\nu}}_{\tau}^*$).

\begin{figure}[htbp]
\begin{center}
\resizebox{!}{10cm}{%
\includegraphics{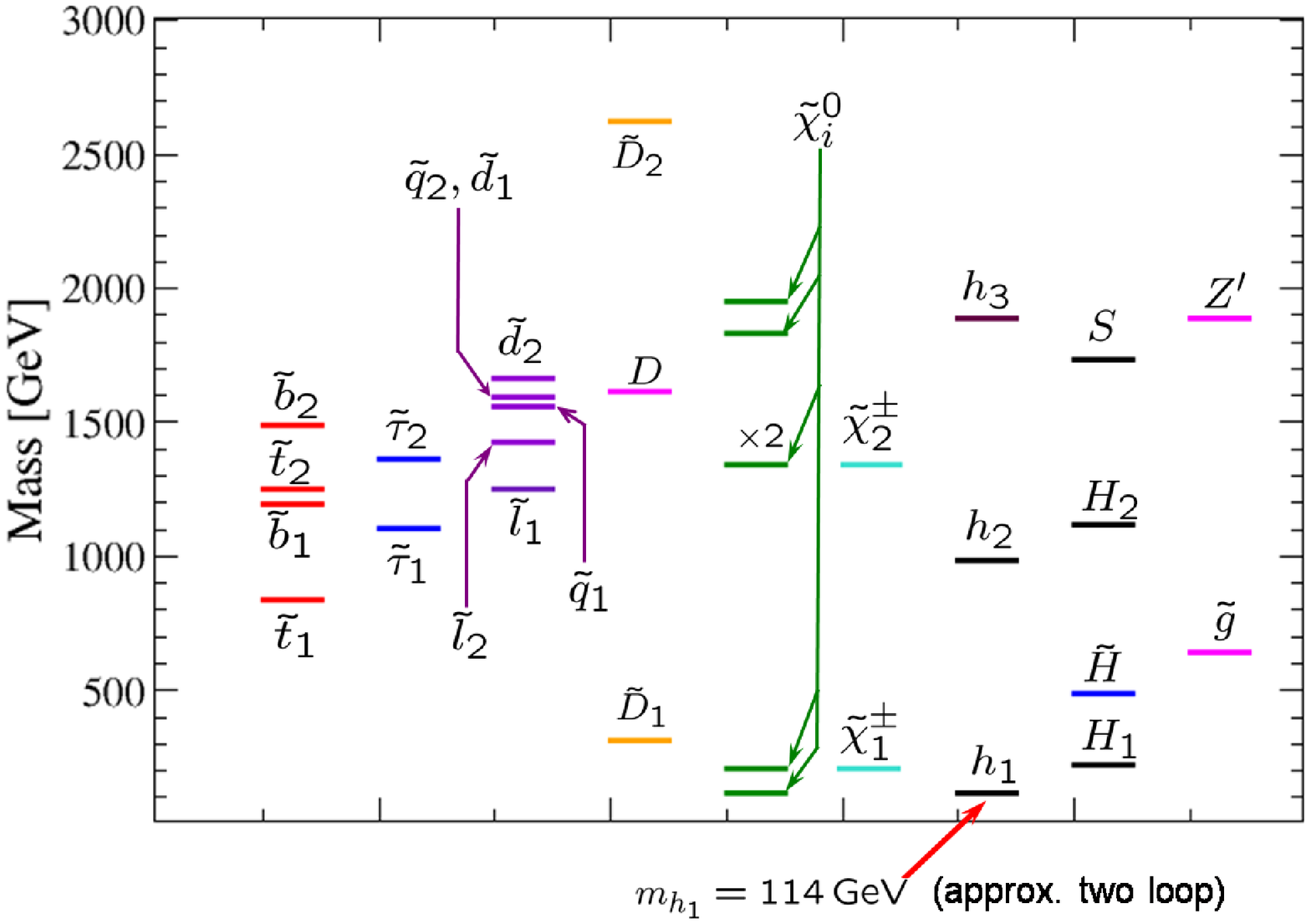}
}
\vspace{-5mm}
\caption[cE$_6$SSM Benchmark point B3]{Benchmark point 4, with $\tan \beta = 30$,
$s = 5.0 \, \rm{TeV}$, $M_{1/2} = 725\,\rm{GeV}$, $m_0 = 1074 \, \rm{GeV}$,
$A =   1726  \,\rm{GeV}$, $\lambda(M_X) = -0.3847$,  $\lambda(\mu_S) = -0.3788$,
$\lambda_{1,2}(M_X) = 0.1$, $\kappa_{1,2,3}(M_X) = 0.1579$,
$\kappa_{1,2,3}(\mu_S) = 0.4559$.
\label{fig:Spectra_bp3b} }
\vspace{-5mm}
\end{center}
\end{figure}%%

Unfortunately the production cross sections of the Inert Higgs bosons and Inert
Higgsinos at the LHC will not be large because they do not participate in strong
interactions. In this context it is more interesting to study scenarios with
light coloured particles. Benchmark 4 represents such a scenario (spectra shown
in Fig.~\ref{fig:Spectra_bp3b}). In this case the lightest exotic squarks have
masses $312\,\mbox{GeV}$ and can be efficiently produced at the LHC. Once again
the presence of light exotic squarks in the particle spectrum is caused by the
mixing effects in the exotic squark sector. The RGEs for the soft
SUSY--breaking masses, $m_{D_i}^2$ and $m_{\bar{D}_i}^2$, are very
similar with $\frac{d}{dt} (m_{D_i}^2 -m_{D_i}^2)=g_1'^2 M_1'^2 $,
resulting in comparatively small splitting between these soft masses.
Therefore, although the diagonal entries of the exotic squark mass matrices
acquire large contributions proportional to $s^2$ that
come from the F--term quartic interactions in the scalar 
potential\footnote{Note that in this case positive contributions to the diagonal
entries of the exotic squark mass matrices from the F--terms dominate over
negative contributions that originate from $U(1)_N$ D--term quartic interactions
in the scalar potential.}, mixing can be large even for moderate values of
$A_0$, leading to a large mass splitting between the two scalar partners
of the exotic coloured fermion. Recent, as yet unpublished, results from
Tevatron searches for di-jet resonances \cite{CDFtevDijet} rule out scalar
diquarks with a mass less than $630$ GeV, however, scalar leptoquarks may
be as light as $300$ GeV since at hadron colliders they are pair produced
through gluon fusion.

\begin{figure}[h!]
\begin{center}
\resizebox{!}{10cm}{%
\includegraphics{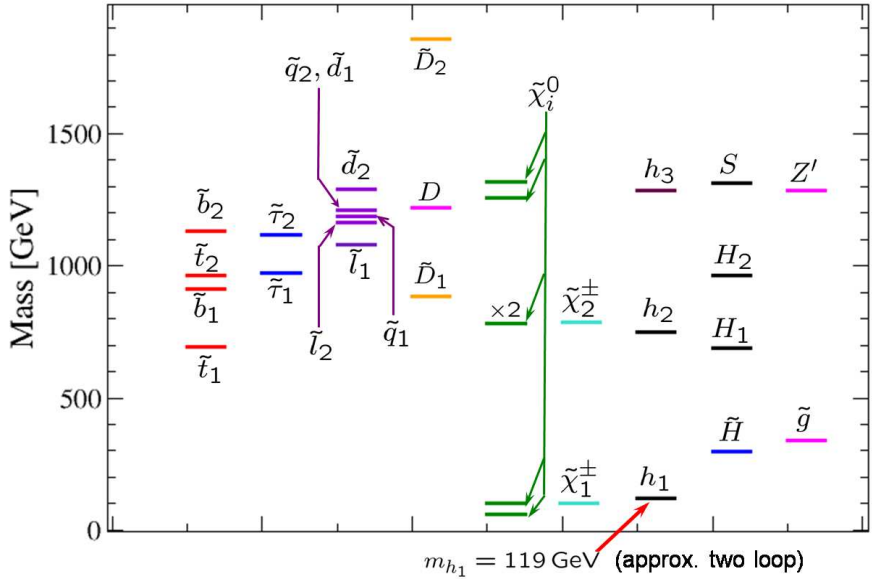}
}
\vspace{-5mm}
\caption[cE$_6$SSM Benchmark point B2]{Benchmark point 5, with $\tan \beta = 30$,
$s = 3.4 \, \rm{TeV}$,   $M_{1/2} = 361\,\rm{GeV}$,  $m_0 = 993 \, \rm{GeV}$,
$A = 1121 \,\rm{GeV}$,  $\lambda(M_X) = -0.33$, $\lambda(\mu_S) =  -0.32$,
$\lambda_{1,2}(M_X) = 0.1$,   $\kappa_{1,2,3}(M_X) = 0.18$,
$\kappa_{1,2,3}(\mu_S) = 0.51$. \label{fig:Spectra_bp2b} }
\vspace{-5mm}
\end{center}
\end{figure}%%

Scalar leptoquarks decay through $Z_2^H$ violating terms, $g^N_{ijk}\hat{N}_i^c \hat{D}_j \hat{d}^c_k$,
$g^E_{ijk} \hat{e}^c_i \hat{D}_j \hat{u}^c_k$ and $g^D_{ijk} (\hat{Q}_i \hat{L}_j) \hat{\overline{D}}_k$.
Thus in the cE$_6$SSM light scalar leptoquarks decay into quark--lepton final states.
If the $Z_2^{H}$ symmetry is mostly broken by the operators involving quarks and
leptons of the third generation each scalar leptoquark gives one top quark
and one $\tau$--lepton in the final state. Since scalar leptoquarks can be pair
produced through gluon fusion, light scalar leptoquarks should lead to an enhancement
of $pp \rightarrow t \bar t l \bar{l}+X$ at the LHC \cite{Accomando:2006ga}.
Notice that SM production of $ t \bar t \tau^+ \tau ^-$ is $(\alpha_W / \pi)^2$
suppressed in comparison to the light scalar leptoquark production cross section.
Therefore light scalar leptoquark should produce a strong signal with low SM
background at the LHC. 

\begin{figure}[htbp]
\begin{center}
\resizebox{!}{10cm}{%
\includegraphics{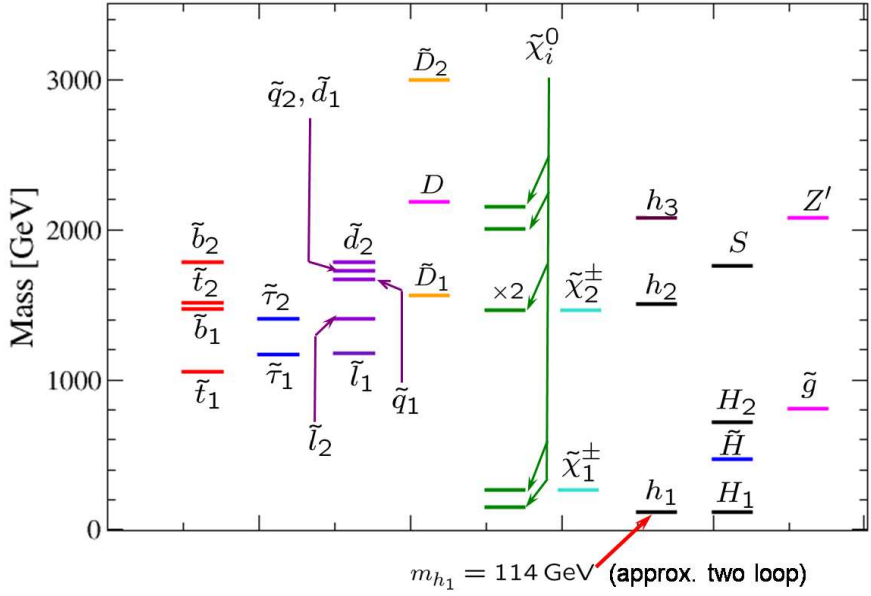}
}
\vspace{-5mm}
\caption[cE$_6$SSM Benchmark point C3]{Benchmark point 6, with
$ \tan \beta = 3$, $s = 5.5 \, \rm{TeV}$, $M_{1/2} = 931\,\rm{GeV}$,
$m_0 = 918 \, \rm{GeV}$,  $A =  751  \,\rm{GeV}$, $\lambda(M_X) = -0.434$,
$\lambda(\mu_S) = -0.375$, $\lambda_{1,2}(M_X) = 0.1$,
$\kappa_{1,2,3}(M_X) =  0.23$,  $\kappa_{1,2,3}(\mu_S) = 0.56 $.
\label{fig:Spectra_bp3c} }
\vspace{-5mm}
\end{center}
\end{figure}%%

The decays of the lightest scalar diquarks are induced by the $Z_2^H$ symmetry
violating symmetry operators $g^Q_{ijk}\hat{D}_{i} (\hat{Q}_j \hat{Q}_k)$ and
$g^{q}_{ijk}\hat{\overline{D}}_i \hat{d}^c_j \hat{u}^c_k$ in the Superpotential. This results in
the decays of $\tilde{D}_{i\,1}$ into quark--quark final states. Assuming that
exotic squarks couple most strongly to the third family quarks each $\tilde{D}_{i\,1}$
gives $t$ and $b$ quarks in the final state. It is worth to emphasise here
that exotic squarks are particles with positive $R$--parity. Therefore
they can decay without missing energy from the LSP.
The production and decay of isosinglet charge -1/3 quark and
its scalar partner were explored in \cite{Kang:2007ib}.

Another very intriguing feature of the cE$_6$SSM is the presence of a $U(1)_N$
$Z'$ gauge boson.  In our benchmark point 5 (Fig.~\ref{fig:Spectra_bp2b}) this $Z'$
is fairly light ($M_{Z'} = 1.285$ TeV) and within the reach of the LHC.  Signatures for
the $Z'$ have already been discussed in \cite{King:2005jy} and are to be explored
further in a follow up study \cite{Ce6SSM_Pheno}.
However, the $Z'$ mass can be significantly heavier, of order 2 TeV,
as is shown in our final benchmark point 6
(Fig.~\ref{fig:Spectra_bp3c}). The spectrum for this point is
rather heavy with even the lightest chargino being as heavy
as $m_{\chi_1^\pm}=262\,{\rm GeV}$ and the lightest neutralino having
$m_{\chi_1^0} = 148\,{\rm GeV}$.

The full spectrum for each of the benchmark points is given in
Tab.~\ref{table:benchmarks}. The Higgs spectrum for all the benchmark points
contains a very light SM--like CP--even Higgs boson $h_1$ with a mass
close to the LEP limit of 114 GeV. Other Higgs states have masses
in the range 600--2100 GeV making them difficult to discover. The benchmark
points all exhibit the characteristic SUSY spectrum described above
containing a relatively light gluino, a light wino-like neutralino and
chargino pair, and a light bino-like neutralino, with other sparticle masses
being much heavier.

\newpage
\begin{table}[h!]
\begin{center}
\begin{tabular}{|c|c|c|c|c|c|c|}
\hline  & \textbf{\footnotesize BM 1} & \textbf{\footnotesize BM 2}   & \textbf{\footnotesize BM 3} & \textbf{\footnotesize BM 4}  & \textbf{\footnotesize BM 5} & \textbf{\footnotesize BM 6} \\
\hline
\footnotesize $\tan \beta$             &  \footnotesize  10    &\footnotesize 10     &\footnotesize 10    &\footnotesize 30     & \footnotesize    30           & \footnotesize  3    \\[-1.5mm]
\footnotesize $\lambda_3(M_X)$         &\footnotesize   -2.0   &\footnotesize -0.37  &\footnotesize -0.31  &\footnotesize -0.38   &\footnotesize   -0.33             & \footnotesize -0.43       \\[-1.5mm]
\footnotesize $\lambda_{1,2}(M_X)$     &\footnotesize    2.6   &\footnotesize 0.1    &\footnotesize 0.1   &\footnotesize 0.1    &\footnotesize    0.1     & \footnotesize 0.1        \\[-1.5mm]
\footnotesize $\kappa_{1,2,3}(M_X)$          &\footnotesize    2.5   &\footnotesize 0.18    &\footnotesize 0.25  &\footnotesize 0.16   &\footnotesize    0.18         & \footnotesize 0.23    \\[-1.5mm]

\footnotesize $s$[TeV]                 &\footnotesize    4  &\footnotesize 4.4   &\footnotesize 3.8  &\footnotesize 5.0   &\footnotesize    3.4                  & \footnotesize  5.5     \\[-1.5mm]
\footnotesize $M_{1/2}$[GeV]                &\footnotesize    389   &\footnotesize 775   &\footnotesize 390   &\footnotesize 725    &\footnotesize    361               & \footnotesize  931 \\[-1.5mm]
\footnotesize $m_0$ [GeV]                       &\footnotesize    725   &\footnotesize 799   &\footnotesize 998   &\footnotesize 1074    &\footnotesize    993     &\footnotesize  918  \\ [-1.5mm]
\footnotesize $A$[GeV]                          &\footnotesize    -1528 &\footnotesize 919   &\footnotesize 768   &\footnotesize 1726   &\footnotesize    1121 &\footnotesize  751 \\
\hline
\footnotesize $m_{\tilde{D}_{1}}(1,2,3)$[GeV]      &\footnotesize    1948  &\footnotesize 821   &\footnotesize 1363 &\footnotesize  312     &\footnotesize    884 &\footnotesize  1567   \\[-1.5mm]

\footnotesize $m_{\tilde{D}_{2}}(1,2,3)$[GeV]   &\footnotesize    2200  &\footnotesize 2363  &\footnotesize 2077 &\footnotesize 2623       &\footnotesize   1860  &\footnotesize 2997 \\[-1.5mm]
\footnotesize $\mu_D(3)$[GeV]                &\footnotesize    2060  &\footnotesize 1535  &\footnotesize 1595 &\footnotesize 1612      &\footnotesize    1221 &\footnotesize  2187  \\
\hline
\footnotesize $|m_{\chi^0_6}|$[GeV]          &\footnotesize    1548  &\footnotesize  1727 &\footnotesize 1496 &\footnotesize 1950       &\footnotesize    1316  &\footnotesize 2155 \\[-1.5mm]
\footnotesize $m_{h_3}\simeq M_{Z'}$[GeV]    &\footnotesize    1518  &\footnotesize  1664 &\footnotesize 1437 &\footnotesize 1890        &\footnotesize    1285   &\footnotesize 2079  \\[-1.5mm]
$|m_{\chi^0_5}|$[GeV]                         &\footnotesize    1490  &\footnotesize  1603  &\footnotesize 1405 &\footnotesize 1832    &\footnotesize    1256  &\footnotesize  2006  \\
\hline
\footnotesize $m_S(1,2)$[GeV]                &\footnotesize    1290  &\footnotesize 1446   &\footnotesize 1430  &\footnotesize 1732      &\footnotesize    1351  &\footnotesize  1763 \\[-1.5mm]
\footnotesize $m_{H^u}(1,2)$[GeV]            &\footnotesize    1172  &\footnotesize 765    &\footnotesize 875   &\footnotesize  1117      &\footnotesize    966  &\footnotesize  714  \\[-1.5mm]
\footnotesize $m_{H^d}(1,2)$[GeV]            &\footnotesize    903   &\footnotesize 182    &\footnotesize 694   &\footnotesize 220    &\footnotesize    689  &\footnotesize  121 \\[-2mm]
\footnotesize $\mu_{\tilde{H}}(1,2)$[GeV]    &\footnotesize    1302  &\footnotesize 418    &\footnotesize 324   &\footnotesize   491        &\footnotesize    323  &\footnotesize  471\\
\hline
\footnotesize $m_{\tilde{u}_1}(1,2)$[GeV]    &\footnotesize    1007  &\footnotesize 1398  &\footnotesize 1211  &\footnotesize 1557   &\footnotesize    1173       &\footnotesize  1666 \\[-1.5mm]
\footnotesize $m_{\tilde{d}_1}(1,2)$[GeV]
                                             &\footnotesize    1023  &\footnotesize 1446 &\footnotesize 1225   &\footnotesize 1595   &\footnotesize 1186   &\footnotesize  1724 \\[-1.5mm]
\footnotesize $m_{\tilde{u}_2}(1,2)$[GeV]
                                           &\footnotesize    1023  &\footnotesize 1446 &\footnotesize 1225   &\footnotesize 1595   &\footnotesize 1186  &\footnotesize  1724 \\[-1.5mm]

\footnotesize $m_{\tilde{d}_2}(1,2)$[GeV]  &\footnotesize    1113  &\footnotesize 1488 &\footnotesize 1292   &\footnotesize 1664 &\footnotesize     1241  &\footnotesize  1785 \\[-1.5mm]

\footnotesize $m_{\tilde{e}_2}(1,2)$[GeV]    &\footnotesize    1015  &\footnotesize 1176 &\footnotesize 1207   &\footnotesize 1427  &\footnotesize     1165  &\footnotesize 1409  \\[-1.5mm]
\footnotesize $m_{\tilde{e}_1}(1,2)$[GeV]    &\footnotesize     873  &\footnotesize 992 &\footnotesize 1105   &\footnotesize 1254  &\footnotesize     1080  &\footnotesize  1173 \\[-1.5mm]

\footnotesize $m_{\tilde{\tau}_2}$[GeV]    &\footnotesize    1012  &\footnotesize 1172 &\footnotesize 1203   &\footnotesize 1363  &\footnotesize     1117  &\footnotesize 1409  \\[-1.5mm]
\footnotesize $m_{\tilde{\tau}_1}$[GeV]    &\footnotesize     867  &\footnotesize 982 &\footnotesize 1095  &\footnotesize 1102  &\footnotesize     973  &\footnotesize  1172 \\[-1.5mm]

\footnotesize $m_{\tilde{b}_2}$[GeV]         &\footnotesize    1108  &\footnotesize 1473  &\footnotesize 1282  &\footnotesize 1491 &\footnotesize      1133   &\footnotesize  1784 \\[-1.5mm]
\footnotesize $m_{\tilde{b}_1}$[GeV]         &\footnotesize     907  &\footnotesize 1216  &\footnotesize 1036  & \footnotesize 1193  &\footnotesize    914  &\footnotesize  1472\\[-1.5mm]
\footnotesize $m_{\tilde{t}_2}$[GeV]         &\footnotesize     921  &\footnotesize 1259  &\footnotesize 1070  & \footnotesize 1248   &\footnotesize     964  &\footnotesize  1511 \\[-1.5mm]
\footnotesize $m_{\tilde{t}_1}$[GeV]         &\footnotesize     777  &\footnotesize 853  &\footnotesize 787  & \footnotesize  837    &\footnotesize     694  &\footnotesize  1056 \\
\hline
\footnotesize $|m_{\chi^0_3}|\simeq |m_{\chi^0_4}|\simeq |m_{\chi^{\pm}_2}|$[GeV]
                                           &\footnotesize     739  &\footnotesize 1168   &\footnotesize 771    &\footnotesize  1343       &\footnotesize    784  &\footnotesize 1463 \\ [-1.5mm]
\footnotesize $m_{h_2}\simeq m_A \simeq m_{H^{\pm}}$[GeV]
                                           &\footnotesize     615  &\footnotesize 1145   &\footnotesize 963    &\footnotesize 998        &\footnotesize    748  &\footnotesize   1508 \\[-1.5mm]
\footnotesize $m_{h_1}$[GeV]               &\footnotesize     116  &\footnotesize 114   &\footnotesize 121  &\footnotesize 114   &\footnotesize    119 &\footnotesize  114 \\ \hline
\footnotesize $m_{\tilde{g}}$[GeV]         &\footnotesize     350  &\footnotesize 673   &\footnotesize 362 &\footnotesize 642  &\footnotesize  338  &\footnotesize    805 \\[-1.5mm]
\footnotesize $|m_{\chi^{\pm}_1}|\simeq |m_{\chi^0_2}|$[GeV]
                                           &\footnotesize     106  &\footnotesize 217   &\footnotesize 110      &\footnotesize 206   &\footnotesize    102  &\footnotesize  262 \\[-1.5mm]
\footnotesize $|m_{\chi^0_1}|$[GeV]          &\footnotesize  59   &\footnotesize 122   &\footnotesize 62  &\footnotesize 116         &\footnotesize    58   &\footnotesize  148 \\
\hline
\end{tabular}
\caption{Particle spectra for our constrained E$_6$SSM benchmark points.}
\label{table:benchmarks}
\end{center}
\end{table}
\vspace{5cm}

\section{Conclusions}
In this paper we have considered the constrained version of the
Exceptional Supersymmetric Standard Model (E$_6$SSM). The E$_6$SSM is
based on the $SU(3)_C\times SU(2)_W\times U(1)_Y\times$ $U(1)_N$ gauge
group, which can originate from the breakdown of the $E_6$ symmetry at
high energies.  In this $E_6$ inspired SUSY model the right-handed
neutrino does not participate in gauge interactions, allowing it to be
used for both the see--saw mechanism and leptogenesis.  To ensure
anomaly cancellation and gauge coupling unification, the particle
content of the E$_6$SSM includes three complete fundamental $27$
representations of $E_6$ as well as the doublet $H'$ and anti-doublet
$\overline{H}'$ from extra $27'$ and $\overline{27}'$
representations. Thus, in addition to a $Z'$ corresponding to the
$U(1)_N$ symmetry, the E$_6$SSM involves extra matter beyond the MSSM
that form three families of new exotic charge 1/3 quarks and squarks,
three generations of $SU(2)$ doublets of Inert Higgs bosons and Inert
Higgsinos, as well as three SM-singlet bosons and their fermionic
Superpartners, which carry $U(1)_{N}$ charges. The baryon number
conservation requires exotic quarks and squarks to be either diquarks
(E$_6$SSM Model I) or leptoquarks (E$_6$SSM Model II).

The extra $U(1)_{N}$ gauge symmetry forbids the term $\mu \hat{H}_d \hat{H}_u$ 
in the Superpotential.  Nevertheless one of the SM-singlet bosons $S$
develops a VEV $\langle S \rangle= s/\sqrt{2}$, breaking the extra
$U(1)_N$ symmetry and providing the effective $\mu$ term for the Higgs
doublets, as well as masses for exotic quarks, Inert Higgsinos and
$Z'$. 

In general, the $E_6$ inspired SUSY models involves lots of new
Yukawa couplings in comparison to the SM and MSSM. Some of these new
couplings give rise to unacceptably large non--diagonal flavour
transitions, which have not been observed. To suppress flavour
changing processes, we have imposed an approximate $Z^{H}_2$ symmetry under
which only the Higgs Superfields $H_u$, $H_d$ and $S$ are even while all
other Supermultiplets are odd. This discrete symmetry can only be
an approximate one because it forbids all terms that allow the lightest
exotic quarks to decay. 

The number of new couplings is further reduced within the {\it
constrained} E$_6$SSM \linebreak[4] (cE$_6$SSM).  The cE$_6$SSM
demands that all soft scalar masses, gaugino masses and trilinear
scalar couplings are universal at the GUT scale. We analysed the RG
flow of the gauge and Yukawa couplings, as well as soft SUSY breaking
terms, using two--loop RGEs for the gauge and Yukawa couplings
together with two--loop RGEs for the gaugino masses and trilinear
scalar couplings and one--loop RGEs for the soft scalar masses. Since
the E$_6$SSM has a much larger Superfield content than the MSSM, the
RG flow of the gauge and Yukawa couplings and soft SUSY breaking terms
is entirely different from the minimal SUSY model. For example, due to
the presence of three families of exotic quarks and squarks, the QCD
beta function vanishes at one loop and at two loops it loses
asymptotic freedom (though the gauge couplings remain perturbative at
high energy). Thus, the E$_6$SSM gauge couplings are considerably
larger at high energies than in the MSSM, and the RG flows of gaugino
and soft scalar masses are entirely different. For the same values of
$M_{1/2}$, the gaugino masses in the cE$_6$SSM are much smaller than in
the cMSSM at low energies. One remarkable feature is
that the low energy gluino mass parameter $M_3$ is driven to be
smaller than $M_{1/2}$ by RG running.

For each set of $\tan\beta$ and SUSY preserving couplings we
established semi--analytic relations between the soft SUSY breaking
terms at the SUSY breaking scale and their values at the GUT
scale. Then we imposed EWSB constraints, which can be considered as a
system of non--linear algebraic equations with respect to $A_0$, $m_0$
and $M_{1/2}$ and found the solutions of these equations for fixed
$\tan\beta$, $s$ and Yukawa couplings. At the last stage of our analysis,
we varied the Yukawa couplings, $\tan\beta$ and $s$ to establish the
qualitative pattern of the particle spectrum.  To avoid any conflict
with present and former collider experiments, as well as recent
cosmological observations, we imposed a set of experimental and
theoretical constraints which restrict the allowed region of parameter
space.

The results of our analysis indicate that $m_0$ tends to be
considerably larger than $M_{1/2}$ in the allowed region. As a
consequence, the Superpartners of ordinary quarks and leptons are
significantly heavier than the gluino and lightest neutralino and
chargino, which are predominantly gaugino. Some of the exotic
squarks can also be relatively light due to large mixing effects 
induced by the corresponding Yukawa couplings and $A_0$.  
The substantial mixing and negative $U(1)_N$ D--term contributions
can lead to the presence of light Inert Higgs bosons as well.  
The masses of exotic quarks and Inert Higgsinos which originate 
from complete 27 plets are controlled by the corresponding
Yukawa couplings and can be relatively light if some of these
couplings are small.

The mass terms of the right--handed neutrinos and survival components
of $27'$ and $\overline{27}'$ are not forbidden by the gauge symmetry
and therefore the scalar and fermion components of these
Supermultiplets are expected to be rather heavy, so they decouple from
the rest of the particle spectrum. The mass of the $Z'$ is set by the
singlet VEV, i.e.\ $M_{Z'}\approx g'_1 Q_S s$ where $Q_S\simeq
5/\sqrt{40}$ and $g'_1\approx g_1$. As a result, the $Z'$ is
considerably heavier than the gluino, lightest neutralino and
chargino. The lightest neutralino, $\chi_1^0$, is essentially pure
bino, while the second lightest neutralino $\chi_2^0$ and the lightest
chargino $\chi_1^\pm$ are the degenerate components of the wino. The
Higgsino states are degenerate and much heavier with the masses given
by the effective $\mu$ term. The remaining neutralinos are mainly a
superposition of the $U(1)_{N}$ gaugino and singlet Higgsino. Their
masses are governed by $M_{Z'}$.  The heaviest CP--even Higgs state
is degenerate with the $Z'$ while another CP--even Higgs, CP--odd and
charged Higgs bosons have almost the same masses and are considerably
heavier than the lightest SUSY particles. For $s=3-5\,\mbox{TeV}$, the
lightest Higgs boson mass is rather close to the LEP limit of
$114\,\mbox{GeV}$. In this work we specified a set of benchmark points
that illustrate all the features of the particle spectrum discussed
above. 
%%and presented the allowed range of parameter space for different
%%values of $\tan\beta$ and $s$. With increasing VEV of the SM-singlet field, the
%%allowed region of parameter space grows, whereas the structure of
%%the particle spectrum becomes more hierarchical.

Thus, throughout all cE$_6$SSM regions of parameter space, there is a
general prediction that the lightest sparticles always include the
gluino $\tilde g$, two lightest neutralinos $\chi_1^0,\chi_2^0$, and
the lightest chargino $\chi_1^\pm$, which are considerably lighter than
all the sfermions of ordinary matter.  The corresponding hierarchical
structure of the particle spectrum is caused by the RG flow.  As a
consequence, at the LHC one should observe pair production of
$\chi_2^0\chi_2^0$, $\chi_2^0\chi_1^\pm$, $\chi_1^\pm \chi_1^\mp$ and
$\tilde g \tilde g$. Due to the hierarchical spectrum, the gluinos can
be relatively narrow states so their width can be comparable to
that of $W^{\pm}$ and $Z$ bosons.  Gluino pair production would
result in an appreciable enhancement of the cross section for $pp
\rightarrow q \bar q q \bar q + E_T^{\rm miss} + X$. Since the second
lightest neutralino decays through $\chi_2^0 \rightarrow \chi_1^0 + l
\bar l $, its pair production would produce an excess in $pp
\rightarrow l \bar l l \bar l + E_T^{\rm miss} + X$, which can be also
observed at the LHC.

Other possible manifestations of the E$_6$SSM at the LHC are related
to the presence of a $Z'$ and exotic multiplets of matter. A TeV
scale $Z'$ will provide an unmistakable signal that can be observed
soon after the LHC starts. If exotic quarks are relatively light,
their production cross sections can be comparable with the cross
section of $t\bar{t}$ production. The lifetime and decays of light
exotic quarks are determined by the $Z_2^H$ violating Yukawa
couplings. If $D_i$ and $\overline{D}_i$ couple most strongly to the
third family of (s)quarks and (s)leptons, then light exotic quarks
lead to a substantial enhancement of the cross section of either
$pp\to t\bar{t}b\bar{b}+E^{\rm miss}_{T}+X$ (if they are diquarks) or
$pp\to t\bar{t}\tau \bar{\tau}+E^{\rm miss}_{T}+X$ and $pp\to
b\bar{b}+ E^{\rm miss}_{T}+X$ (if they are leptoquarks). When scalar
exotic quarks are light, they can decay into quark--quark (if
diquarks) or quark--lepton (if leptoquarks) without missing energy
from the LSP. As a result, their pair production leads to the
enhancement of the cross section of either $pp\to t\bar{t}b\bar{b}+X$
or $pp\to t\bar{t}\tau \bar{\tau}+X$.  Since the SM
production cross sections of $pp\to t\bar{t}b\bar{b}+X$ or $pp\to
t\bar{t}\tau \bar{\tau}+X$ are suppressed by many orders of magnitude
compared to the cross section for $t\bar{t}$ production, the
light exotic quarks and squarks should produce a strong signal with
low SM background at the LHC.

The production cross sections of the Inert Higgs bosons and Inert
Higgsinos will be much smaller at the LHC than the exotic (s)quark
one. Nevertheless, their detection might also be possible if the
corresponding states are light. Assuming that Inert Higgs bosons and
Inert Higgsinos couple most strongly to the third family (s)quarks and
(s)leptons, the lightest Inert Higgs bosons decay predominantly into
third generation fermion--anti-fermion pairs like $H^{0}_{\alpha\,1}
\rightarrow b \bar b$ and $H^{-}_{\alpha\,1} \rightarrow \tau
\bar{\nu}_{\tau}$, while Inert Higgsinos predominantly decay into third
generation fermion-anti-sfermion pairs. At an ILC the production rates
of the light exotic (s)quarks and Inert Higgs bosons (Higgsinos) can
be comparable, allowing their simultaneous observation.

We have not considered the question of cosmological cold dark
matter (CDM) relic abundance due to the neutralino LSP and so one
may be concerned that a bino-like lightest neutralino mass of
around 100 GeV might give too large a contribution to
$\Omega_{CDM}$. Indeed a recent calculation of $\Omega_{CDM}$ in
the USSM \cite{Kalinowski:2008iq}, which includes the effect of
the MSSM states plus the extra $Z'$ and the active singlet $S$,
together with their superpartners, indicates that for the
benchmarks considered here that $\Omega_{CDM}$ would be too large.
However the USSM does not include the effect of the extra inert
Higgs and Higgsinos that are present in the E$_6$SSM. While we
have considered the inert Higgsino masses given by
$\mu_{\tilde{H}_{\alpha}} = \lambda_{\alpha} s/\sqrt{2}$, we have
not considered the mass of the inert singlinos which are generated
by mixing with the Higgs and inert Higgsinos, and are thus of
order $fv^2/s$ where their masses are controlled by additional
Yukawa couplings $f$ which we have not specified in our analysis.
Since $s\gg v$ it is quite likely that the LSP neutralino in the
cE$_6$SSM will be an inert singlino with a mass lighter than 100
GeV. This would imply that the state $\chi_1^0$ considered here is
not cosmologically stable but would decay into lighter singlinos.
The question of the calculation of the relic abundance of such an
LSP singlino within the framework of the cE$_6$SSM is beyond the
scope of this article and will be considered elsewhere. In summary,
it is clear that one should not regard the benchmark points with
$|m_{\chi^0_1}|\approx 100$ GeV as being excluded by
$\Omega_{CDM}$.

The discovery of $Z'$ and new exotic particles predicted by the
E$_6$SSM at future colliders will open a new era in elementary
particle physics. It will represent a possible indirect signature of
an underlying $E_6$ gauge structure at high energies and may provide a
window into string theory.

\section*{Acknowledgements}
\vspace{0mm} We would like to thank A.~Belyaev, C.~D.~Froggatt and
D.~Sutherland for fruitful discussions. RN is also grateful to
E.~Boos, D.~I.~Kazakov, M.~Sher and P.~M.~Zerwas for valuable comments
and remarks. PA would like to thank D.~Stockinger for helpful
discussions during the preparation of this manuscript. DJM
acknowledges support from the STFC Advanced Fellowship Grant
PP/C502722/1. RN acknowledges support from the SHEFC grant HR03020
SUPA 36878. SFK acknowledges partial support from the following
grants: STFC Rolling Grant ST/G000557/1 (also SM); EU Network
MRTN-CT-2004-503369; NATO grant PST.CLG.980066 (also SM); EU ILIAS
RII3-CT-2004-506222. SM is also partially supported by the FP7 RTN
MRTN-CT-2006-035505 and by the scheme `Visiting Professor - Azione D -
Atto Integrativo tra la Regione Piemonte e gli Atenei Piemontesi'.

\clearpage
\begin{appendix}
\section{One--loop corrections to the Higgs masses}

\setcounter{equation}{0}
\def\theequation{A.\arabic{equation}}

Higgs masses are obtained by taking double derivatives of the
effective potential with respect to the Higgs fields.

The tree--level Higgs masses for the CP--even Higgs sector were
presented in section \ref{sec:Higgs}, Eq.~(\ref{cessm47}).  The
expression for the one--loop contribution, $\Delta V ^{(1)}$, to the
effective potential also appears in Eq.~(\ref{cessm12}) and the
physical masses of the stops, appearing in this equation, are
calculated in the tree--level approximation,

\begin{equation}
\hspace{-10mm} m^2_{\tilde{t}_1,\tilde{t}_2}
=\dfrac{1}{2}\left\{ \lefteqn{\phantom{\sqrt{\left[\frac{1}{2}\right]^2}}}
m^2_{Q_3}+m^2_{u^c_3}+\dfrac{1}{2} M_Z^2 \cos2\beta+
\Delta_{Q}+\Delta_{u^c}+2m_t^2 \mp \sqrt{M_{QQ}^4 + 4 m_t^2 X_t^2}\, \right\}\,, \label{cessm27}
\end{equation} where \bea\Delta_Q &=&  \frac{g_1'}{80}(-3 v_1^2 -2 v_2^2 + 5s^2),  \quad\quad
\Delta_{u^c} =  \frac{g_1'}{80}(-3 v_1^2 -2 v_2^2 + 5s^2),
\eea  \bea M_{QQ}^2 &=&m_{Q_3}^2 - m_{u^c_3}^2 + \left[\dfrac{1}{2}-\dfrac{4}{3}\sin^2 \theta_W\right]M_Z^2 \cos2\beta  + \Delta_Q - \Delta_U, \\ X_t & =&  A_t - \frac{\lambda s }{\sqrt{2} \tan \beta},   \eea  and for further convenience defining, \beq r_t   \equiv  M_{QQ}^4 + 4m_t^2 X_t^2 \quad \quad {\rm and }\quad \quad R_{QQ} \equiv  M_{QQ}^2 (g_2^2 -g_1^2 ), \eeq \beq \mu_{eff} \equiv \frac{\lambda s}{\sqrt{2}} \quad \quad {\rm and } \quad\quad \bar{g} \equiv \sqrt{g_2^2 + \frac{3 g_1^2}{5}}\eeq

\noindent  Including only stop/top contributions we find,
\begin{equation} \frac{\partial \Delta V}{\partial x} =  \frac{3}{32 \pi^2} \left[2 a_0 (m_{\tilde{t}_1}) \frac{\partial }{\partial x } m_{\tilde{t}_1}^2  + 2 a_0 (m_{\tilde{t}_2}) \frac{\partial }{\partial x } m_{\tilde{t}_2}^2  -  4 a_0 (m_t) \frac{\partial }{\partial x } m_t^2 \right], \end{equation}
where
\bea a_0(m) \equiv m^2\left[\ln{\frac{m^2}{Q^2}} -1\right]. \eea  Now, defining \bea\Delta_x m_i \equiv a_0 (m_i) \frac{\partial }{\partial x } m_i^2, \eea
it follows that \bea \frac{\partial^2 \Delta V}{\partial y \partial x }& = & \frac{3}{32 \pi^2} \Big[2 \frac{\partial }{\partial y }\Delta_x m_{\tilde{t_1}} + 2 \frac{\partial}{\partial y }\Delta_x m_{\tilde{t_2}} - 4  \frac{\partial }{\partial y }\Delta_x m_t \Big], \\ \frac{\partial }{ \partial x }\Delta_x m & = & (\frac{\partial }{\partial x } m^2)^2\ ln{\frac{m^2}{Q^2}}  + a_0(m) \frac{\partial^2 }{\partial x^2 } m^2,  \\
\frac{\partial }{ \partial y }\Delta_x m & = &(\frac{\partial }{\partial y } m^2) (\frac{\partial }{\partial x } m^2)\ ln{\frac{m^2}{Q^2}}  + a_0(m) \frac{\partial^2 }{\partial y \partial x } m^2.\eea

Here we present the corrections in the basis ($v_1$, $v_2$, $v_3 \equiv s$), with $\Delta_{ij}^{\prime} = \frac{\partial^2}{\partial v_i \partial v_j} \Delta V$  such that $\Delta_{11}^{\prime} = \frac{\partial^2}{\partial v_1^2} \Delta V$ etc.  The corrections, $\Delta_{ij}$ appearing in Eq.~(\ref{cessm47}) can be obtained from these using the relations,

%%\newpage
%%Now we obtain single and double derivatives of the masses.

\bea \Delta_{11} &=&  \cos^2\beta \Delta_{11}^\prime - 2 \sin\beta \cos \beta \Delta_{12}^\prime  + \sin^2\beta \Delta_{22}^\prime \\
\Delta_{22} &=&  \sin^2\beta \Delta_{11}^\prime - 2 \sin\beta \cos \beta \Delta_{12}^\prime  + \cos^2\beta \Delta_{22}^\prime\\
\Delta_{33} &=& \Delta_{33}^\prime \\
\Delta_{12} &=&  (\cos^2\beta -  \sin^2\beta) \Delta_{12}^\prime  + \sin\beta \cos \beta( \Delta_{22}^\prime -  \Delta_{11}^\prime) \\
\Delta_{31} &=&  \cos\beta \Delta_{13}^\prime +  \sin \beta \Delta_{23}^\prime \\
\Delta_{32} &=&   \cos \beta \Delta_{23}^\prime - \sin \beta\Delta_{13}^\prime
\eea

\begin{eqnarray} \hspace{-20mm}\Delta_{11}^\prime\hspace{-3mm} &=&\hspace{-3mm} \frac{3}{16 \pi^2}\left\{ \left[ \left(\frac{\bar{g}^2 }{8} -  \frac{3g_1^{\prime\,2}}{40}\right)^2 v_1^2   + \frac{1}{r_t}  \left(\frac{v_1}{8}  R_{QQ} -2m_t^2 X_t \frac{s \lambda}{\sqrt{2} v_2}\right)^2  \right] \ln \frac{ m^2_{\tilde{t}_1}m^2_{\tilde{t}_2}}  {Q^4}  \right.\nonumber \\  &&\hspace{-6mm} \left. +\frac{v_1}{32}\left(\bar{g}^2  -  \frac{3}{5}g_1^{\prime\,2} \right) r_t^{-\frac{1}{2}}\left( v_1 R_{QQ}-16m_t^2 X_t \frac{s \lambda}{\sqrt{2} v_2} \right)\ln \frac{ m^2_{\tilde{t}_2}}{ m^2_{\tilde{t}_1}}  \right. \nonumber \\ && \hspace{-6mm}\left.+ \left(\frac{\bar{g}^2}{8}  - \frac{3g_1'^2}{40}\right)\left(a_0(m_{\tilde{t}_1}) +a_0(m_{\tilde{t}_2}) \right) + \frac{1}{32}\left[ r_t^{-\frac{1}{2}}\left( 4 R_{QQ}   + (g_2^2 -g_1^2 )^2 v_1^2 \right.\right. \right.\nonumber \\ &&\hspace{-6mm} \left.\left.\left. + 16 y_t^2 s^2 \lambda^2 \right)  - \left( v_1 R_{QQ} -16m_t^2 X_t \frac{s \lambda}{\sqrt{2} v_2}\right)^2  r_t^{-\frac{3}{2}}\right]\left(a_0(m_{\tilde{t}_2}) - a_0(m_{\tilde{t}_1})\right) \right\} \phantom{(A.12)}\end{eqnarray}

\begin{eqnarray}\hspace{-20mm} \Delta_{22}^\prime\hspace{-3mm} &=& \hspace{-3mm}\frac{3}{16 \pi^2}\left\{ \left[  \left(y_t^2 - \frac{\bar{g}^2}{8}  -  \frac{g_1^{\prime \, 2}}{20}\right)^2   + \frac{\left( 8 X_t A_t  y_t^2 - R_{QQ} \right)^2 }{64 r_t } \right]  v_2^2\ln \frac{ m^2_{\tilde{t}_1}m^2_{\tilde{t}_2}}{Q^4} \right.\nonumber \\  &&\hspace{-6mm} \left. +\frac{v_2^2}{4\sqrt{r_t} }\left( y_t^2 - \frac{\bar{g}^2 }{8} - \frac{g_1^{\prime \, 2}}{20} \right)\left( 8y_t^2 X_t A_t - R_{QQ} \right) \ln \frac{ m^2_{\tilde{t}_2}}{ m^2_{\tilde{t}_1}}  \right.\nonumber \\  && \left. \hspace{-6mm} + \left(y_t^2 - \frac{\bar{g}^2 }{8} - \frac{g_1^{\prime \, 2}}{20}\right)\left(a_0(m_{\tilde{t}_1}) +a_0(m_{\tilde{t}_2}) \right)+ \frac{1}{\sqrt{r_t}}\left[ \frac{(g_2^2 -g_1^2 )^2 v_2^2}{32}  - \frac{R_{QQ}}{8}   + y_t^2 A_t^2 \right. \right. \nonumber \\ && \hspace{-6mm}\left.\left.  -  \frac{\left( 8 X_t A_t  y_t^2 - R_{QQ} \right)^2v_2^2 }{32 r_t} \right]\left(a_0(m_{\tilde{t}_1}) - a_0(m_{\tilde{t}_1})\right)   -2 y_t^4 v_2^2\ln\frac{m^2_{t}}{Q^2}-2 y_t^2 a_0(m_t)\right\} \phantom{(A.22)} \end{eqnarray}

\begin{eqnarray}\hspace{-20mm}\Delta_{33}^\prime\hspace{-3mm} &=& \hspace{-3mm}\frac{3}{16 \pi^2}\left\{  \left[\frac{g_1^{\prime \, 4} s^2}{64} + \frac{2 m_t^4X_t^2 \lambda^2}{r_t \tan^2 \beta} \right] \ln  \frac{ m^2_{\tilde{t}_1}m^2_{\tilde{t}_2}}{Q^4}
 - \frac{g_1^{\prime \, 2} m_t^2 X_t \mu_{eff}}{2 \sqrt{r_t}  \tan \beta} \ln \frac{ m^2_{\tilde{t}_2}}{ m^2_{\tilde{t}_1}}  \right.\nonumber \\ &&\hspace{-6mm} \left.  +  \frac{g_1^{\prime \, 2}}{8}\left(a_0(m_{\tilde{t}_1}) +a_0(m_{\tilde{t}_2})\right) +  \frac{\lambda^2 m_t^2}{ \sqrt{r_t}\tan^2 \beta}\left[1   -  \frac{4 X_t^2 m_t^2} {r_t} \right]\left(a_0(m_{\tilde{t}_2}) - a_0(m_{\tilde{t}_1})\right)     \right\} \phantom{A.232} \end{eqnarray}

\begin{eqnarray}\hspace{-20mm} \Delta_{12}^\prime\hspace{-3mm} &=& \hspace{-3mm}\frac{3}{16 \pi^2}\left\{ \left[\left(\frac{\bar{g}^2 }{8} -  \frac{3}{40}g_1^{\prime \, 2}\right) \left(y_t^2 - \frac{\bar{g}^2}{8}  -  \frac{g_1^{\prime \, 2}}{20}\right)  +  \left( \frac{R_{QQ}}{8}  - y_t^2 X_t \mu_{eff}\tan \beta \right)\frac{1}{r_t}  \right.\right.\nonumber \\  &&\hspace{-6mm} \left.\left.\times \left(y_t^2  X_t A_t    -\frac{ R_{QQ}}{8} \right)  \right] v_1v_2 \ln\frac{ m^2_{\tilde{t}_1}m^2_{\tilde{t}_2}}{Q^4}
+ \left[ \left(\frac{\bar{g}^2 }{8}  -  \frac{3 g_1^{\prime\,2}}{40}\right)  \left(y_t^2 X_tA_t    -\frac{  R_{QQ}}{8} \right)   \right.\right.\nonumber \\  && \hspace{-6mm}\left.\left.
+\left(\frac{R_{QQ}}{8} -y_t^2 X_t \mu_{eff}\tan \beta  \right) \left(y_t^2 - \frac{1}{8}\bar{g}^2  -  \frac{1}{20}g_1^{\prime \, 2}\right)\right]\frac{v_1 v_2}{\sqrt{ r_t}}\ln \frac{ m^2_{\tilde{t}_2}}{ m^2_{\tilde{t}_1}} \right.\nonumber \\  && \left.\hspace{-6mm}
-\left[ \left(\frac{(g_2^2 - g_1^2)^2}{32} v_1 v_2 + A_t y_t^2\mu_{eff} \right)  +   \left(\frac{ R_{QQ}}{8}   -y_t^2 X_t \mu_{eff}\tan\beta \right)\left(2 y_t^2  X_t A_t    \right.  \right.\right.\nonumber \\  &&\hspace{-6mm} \left.\left. \left.  -  \frac{R_{QQ}}{4}\right)\frac{v_1 v_2}{r_t} \right]  \left(\frac{a_0(m_{\tilde{t}_2}) - a_0(m_{\tilde{t}_1})}{\sqrt{r_t}}\right)
\right\}  \end{eqnarray}

\begin{eqnarray}\hspace{-20mm} \Delta_{13}^\prime\hspace{-3mm} &=& \hspace{-3mm}\frac{3}{16 \pi^2}\left\{  \left[\left(\frac{\bar{g}^2 }{8}  -  \frac{3 g_1^{\prime\,2}}{40}\right) \frac{g_1^{\prime \, 2}}{8} s v_1 - \left(\frac{ R_{QQ}}{8} v_1 -  y_t^2 X_t \mu_{eff}v_2 \right) \frac{2 m_t^2X_t\lambda }{r_t\sqrt{2}\tan\beta} \right] \ln\frac{ m^2_{\tilde{t}_1}m^2_{\tilde{t}_2}}{Q^4}  \right.\nonumber \\  && \hspace{-6mm}\left. -\left[\left(\frac{\bar{g}^2 }{8} -  \frac{3 g_1^{\prime\,2}}{40}\right)\frac{2 v_1 m_t^2 X_t\lambda}{ \sqrt{2}\tan\beta \sqrt{r_t}} -   \frac{g_1^{\prime \, 2 }s}{8 \sqrt{r_t} }\left(\frac{R_{QQ} }{8} v_1 - 2m_t^2 X_t \frac{s \lambda}{\sqrt{2} v_2} \right)    \right]\ln \frac{ m^2_{\tilde{t}_2}}{ m^2_{\tilde{t}_1}} \right.\nonumber \\  && \hspace{-6mm}\left.
 +\frac{y_t^2 v_1 \lambda \mu_{eff}}{ \sqrt{2 r_t}}\left[1 - \frac{X_t\tan \beta}{\mu_{eff}   }
 - \frac{ 4 X^2_tm_t^2}{r_t } + \frac{ v_1 v_2 R_{QQ} X_t}{4 \mu_{eff} r_t }
\right]\left(a_0(m_{\tilde{t}_2}) - a_0(m_{\tilde{t}_1})\right)
 \right\}  \end{eqnarray}

\begin{eqnarray}\hspace{-20mm} \Delta_{23}^\prime\hspace{-3mm} &=&\hspace{-3mm} \frac{3}{16 \pi^2}\left\{ \left[ \frac{g_1^{\prime \, 2}}{8} \left(y_t^2 - \frac{\bar{g}^2 }{8} -  \frac{g_1^{\prime \, 2}}{20}\right) s -  \frac{2 m_t^2X_t\lambda  }{r_t\sqrt{2}\tan \beta} \left(y_t^2 X_t A_t -\frac{R_{QQ}}{8} \right)   \right]v_2 \ln\frac{ m^2_{\tilde{t}_1}m^2_{\tilde{t}_2}}{Q^4}   \right.\nonumber \\  &&\hspace{-6mm} \left.
+r_t^{-\frac{1}{2}} \left[  \frac{g_1^{\prime \, 2}}{8}  s  \left(y_t^2 X_t A_t   -\frac{ R_{QQ}}{8} \right)
-  \frac{2 m_t^2X_t\lambda }{\sqrt{2}\tan\beta}\left(y_t^2 - \frac{\bar{g}^2 }{8} -  \frac{g_1^{\prime \, 2}}{20}\right) \right]v_2\ln \frac{ m^2_{\tilde{t}_2}}{ m^2_{\tilde{t}_1}}  \right.\nonumber \\  && \hspace{-6mm}\left.
  - \frac {y_t^2 \lambda v_1 A_t}{\sqrt{2 r_t}}\left[ 1   - \frac{4m_t^2 X_t^2 }{r_t}  + \frac{v_2^2 X_t R_{QQ}}{4 A_t r_t}\right]\left(a_0(m_{\tilde{t}_2}) - a_0(m_{\tilde{t}_1})\right)
\right\}
\end{eqnarray}

These complicated expressions can be simplified by keeping only the dominant contributions.  Neglecting those auxiliary D-term contributions to the stop masses which are proportional to $v_1^2$ and $v_2^2$ we obtain the following simpler expressions, \begin{eqnarray} \hspace{-8mm}\Delta_{11}^\prime &\approx& \frac{3y_t^2}{16 \pi^2}\left\{ \frac{m_t^2 X_t^2s^2 \lambda^2}{r_t}  \ln \frac{ m^2_{\tilde{t}_1}m^2_{\tilde{t}_2}}  {Q^4}  + \frac{\mu_{eff}^2}{ \sqrt{r_t}}\left[1   -  \frac{4 m_t^2 X_t^2 } {r_t} \right]\left(a_0(m_{\tilde{t}_2}) - a_0(m_{\tilde{t}_1})\right) \right\} \phantom{A.23}\end{eqnarray} \begin{eqnarray}\hspace{-12mm} \Delta_{22}^\prime &\approx& \frac{3y_t^2}{16 \pi^2}\left\{2m_t^2\left(1+\frac{X_t^2 A_t^2}{ r_t } \right) \ln \frac{ m^2_{\tilde{t}_1}m^2_{\tilde{t}_2}}{Q^4} +   \frac{ A_t^2}{\sqrt{r_t}} \left[1 - \frac{4m_t^2X_t^2}{r_t}\right]\left(a_0(m_{\tilde{t}_1}) - a_0(m_{\tilde{t}_1})\right) \right.\nonumber \\  && \left. \hspace{-6mm}   +  \left(a_0(m_{\tilde{t}_1}) +a_0(m_{\tilde{t}_2}) -2 a_0(m_t)\right) +\frac{4 m_t^2}{\sqrt{r_t} } X_t A_t \ln \frac{ m^2_{\tilde{t}_2}}{ m^2_{\tilde{t}_1}}   -4 m_t^2 \ln\frac{m^2_{t}}{Q^2}\right\} \phantom{(A.22)} \end{eqnarray}\begin{eqnarray}\hspace{-20mm}\Delta_{33}^\prime &\approx& \frac{3}{16 \pi^2}\left\{  \left[\frac{g_1^{\prime \, 4} s^2}{64} + \frac{2 m_t^4X_t^2 \lambda^2}{r_t \tan^2 \beta} \right] \ln  \frac{ m^2_{\tilde{t}_1}m^2_{\tilde{t}_2}}{Q^4}
 - \frac{g_1^{\prime \, 2} m_t^2 X_t \mu_{eff}}{2 \sqrt{r_t}  \tan \beta} \ln \frac{ m^2_{\tilde{t}_2}}{ m^2_{\tilde{t}_1}}  \right.\nonumber \\ &&\hspace{-10mm} \left.  +  \frac{g_1^{\prime \, 2}}{8}\left(a_0(m_{\tilde{t}_1}) +a_0(m_{\tilde{t}_2})\right) +  \frac{\lambda^2 m_t^2}{ \sqrt{r_t}\tan^2 \beta}\left[1   -  \frac{4 X_t^2 m_t^2} {r_t} \right]\left(a_0(m_{\tilde{t}_2}) - a_0(m_{\tilde{t}_1})\right)     \right\} \phantom{A.232} \end{eqnarray}\begin{eqnarray} \hspace{-20mm}\Delta_{12}^\prime &\approx& \frac{3 y_t^2}{16 \pi^2} \left\{  - \frac{ 2 m_t^2 \mu_{eff}A_t X_t^2 }{r_t} \ln\frac{ m^2_{\tilde{t}_1}m^2_{\tilde{t}_2}}{Q^4} -\frac{ 2m_t^2 X_t \mu_{eff} }{\sqrt{ r_t}}\ln \frac{ m^2_{\tilde{t}_2}}{ m^2_{\tilde{t}_1}}  \right.\nonumber \\  && \hspace{-6mm}\left.  - \frac{\mu_{eff} A_t}{\sqrt{r_t}}\left[1 -
\frac{ 4  m_t^2 X_t^2 }{r_t}  \right]\left(a_0(m_{\tilde{t}_2}) - a_0(m_{\tilde{t}_1})\right) \right\}\phantom{A.232} \\\nonumber \end{eqnarray}
\vspace{-10mm}\begin{eqnarray}\hspace{-20mm} \Delta_{13}^\prime &\approx& \frac{3}{16 \pi^2}\left\{  \frac{2 m_t^4 X_t^2\lambda^2 s}{r_t\tan\beta v_2} \ln\frac{ m^2_{\tilde{t}_1}m^2_{\tilde{t}_2}}{Q^4} - \frac{g_1^{\prime \, 2}s^2 m_t^2 X_t\lambda}{4 v_2 \sqrt{2 r_t} } \ln \frac{ m^2_{\tilde{t}_2}}{ m^2_{\tilde{t}_1}}  \right.\nonumber \\  && \left.
+\frac{y_t^2 v_1 \lambda \mu_{eff}}{ \sqrt{2 r_t}}\left[1 - \frac{X_t\tan \beta}{\mu_{eff}   }
 - \frac{ 4 X^2_tm_t^2}{r_t }
\right] \left(a_0(m_{\tilde{t}_2}) - a_0(m_{\tilde{t}_1})\right)
 \right\}  \end{eqnarray}  \begin{eqnarray}\hspace{-20mm} \Delta_{23}^\prime &\approx& \frac{3y_t^2}{16 \pi^2}\left\{\left(\frac{g_1^{\prime \, 2 } s v_2}{8}   -  \frac{2 m_t^2X_t^2\lambda v_2 }{r_t\sqrt{2}\tan\beta} A_t\right) \ln\frac{ m^2_{\tilde{t}_1}m^2_{\tilde{t}_2}}{Q^4}+\frac{ (g_1^{\prime \, 2}  A_t v_2 s - 8\sqrt{2} m_t^2  \lambda v_1)X_t }{8\sqrt{r_t}} \ln \frac{ m^2_{\tilde{t}_2}}{ m^2_{\tilde{t}_1}}  \right.\nonumber \\  && \left.
- \frac {\lambda v_1 A_t}{\sqrt{2 r_t}}\left[ 1   - \frac{4m_t^2 X_t^2 }{r_t} \right]\left(a_0(m_{\tilde{t}_2}) - a_0(m_{\tilde{t}_1})\right)
\right\}
\end{eqnarray}

\newpage

\section{RGEs}
\label{Appendix:RGEs}
\setcounter{equation}{0}
\def\theequation{B.\arabic{equation}}

The running of the gauge couplings from the GUT scale to the EW scale is determined
by a set of RGEs. In our analysis, we use two--loop RGEs for the gauge and Yukawa couplings
together with two--loop RGEs for the gaugino masses $M_a(\mu)$ and trilinear scalar couplings $A_i(\mu)$,
as well as one--loop RGEs for the soft scalar masses $m_i^2(\mu)$. A simplified set of one--loop RG
equations may be found in \cite{Keith:1997zb}. The two--loop RGEs can be derived using general
results presented in \cite{two--loop-rge}.

In the E$_6$SSM the RGEs for the gauge couplings can written,
\be
\ds\frac{d G}{d t}=G\times B\,,\qquad\qquad \frac{d g_2}{dt}=\ds\frac{\beta_2 g_2^3}{(4\pi)^2}\,,\qquad\qquad
\frac{d g_3}{dt}=\frac{\beta_3 g_3^3}{(4\pi)^2}\,,
\label{eq1}
\ee
where $t=\ln\left[Q/M_X\right]$, while $B$ and $G$ are $2\times 2$ matrices describing the RG flow
of the Abelian gauge couplings, which is affected by the kinetic term mixing,
\bea
\hspace{-9mm} G=\left(
\ba{cc}
g_1 & g_{11}\\[2mm]
0   & g'_1
\ea
\right), \:
B=\left(
\ba{cc}
B_1 & B_{11} \\[2mm]
0   & B'_1
\ea
\right)= \lefteqn{ \ds\frac{1}{(4\pi)^2}
\left(
\ba{cc}
\beta_1 g_1^2 & 2g_1g'_1\beta_{11}+2g_1g_{11}\beta_1\\[2mm]
0 & g^{'2}_1\beta'_1+2g'_1 g_{11}\beta_{11}+g_{11}^2\beta_1
\ea
\right).} \nonumber \\[2mm]
\label{eq2} \\[-10mm] \nonumber
\eea
In the one--loop approximation $\beta_{11}=\ds\frac{\sqrt{6}}{5}$.
The two--loop diagonal $\beta$--functions of the gauge couplings are given by
\be
\ba{rcl}
\beta_3&=&-9+3N_g+\ds\frac{1}{16\pi^2}\Biggl[g_3^2(-54+34 N_g)+3 N_g\,g_2^2+ N_g\, g_1^2\\[3mm]
&&+N_g\,g_1^{'2}-4h_t^2-4h_b^2-2\Sigma_{\kappa}\Biggr]\,,\\[3mm]
\beta_2&=&-5+3N_g+\ds\frac{1}{16\pi^2}\Biggl[8N_g g_3^2+(-17+21 N_g)g_2^2+ \left(\ds\frac{3}{5}+N_g\right) g_1^2\\[3mm]
&&+\left(\ds\frac{2}{5}+N_g\right) g_1^{'2}
-6 h_t^2-6 h_b^2-2h_{\tau}^2-2\Sigma_{\lambda}\Biggr]\,,\\[3mm]
\beta_1&=&\ds\frac{3}{5}+3N_g+\ds\frac{1}{16\pi^2}\Biggl[8N_g g_3^2+\left(\ds\frac{9}{5}+3N_g\right)g_2^2+
\left(\ds\frac{9}{25}+3 N_g\right) g_1^2\\[3mm]
&&+\left(\ds\frac{6}{25}+N_g\right) g_1^{'2}-\ds\frac{26}{5} h_t^2-\ds\frac{14}{5}h_b^2-
\ds\frac{18}{5}h_{\tau}^2-\ds\frac{6}{5}\Sigma_{\lambda}-\ds\frac{4}{5}\Sigma_{\kappa}\Biggr]\,,\\[3mm]
\beta'_1&=&\ds\frac{2}{5}+3N_g+\ds\frac{1}{16\pi^2}\Biggl[8N_g g_3^2+\left(\ds\frac{6}{5}+3N_g\right)g_2^2+
\left(\ds\frac{6}{25}+ N_g\right) g_1^2\\[3mm]
&&+\left(\ds\frac{4}{25}+3N_g\right) g_1^{'2}-\ds\frac{9}{5} h_t^2-\ds\frac{21}{5}h_b^2-\ds\frac{7}{5}h_{\tau}^2-
\ds\frac{19}{5}\Sigma_{\lambda}-\ds\frac{57}{10}\Sigma_{\kappa}\Biggr]\,,\\[3mm]
\Sigma_{\lambda}&=&\lambda_1^2+\lambda_2^2+\lambda_3^2\,,\qquad\qquad\qquad\Sigma_{\kappa}=\kappa_1^2+\kappa_2^2+\kappa_3^2\,.
\ea
\label{eq3}
\ee

The Yukawa couplings appearing in the Superpotential of the cE$_6$SSM obey the following system
of two--loop RGEs:
\be
\ba{rcl}
\ds\frac{d\lambda_i}{dt}&=&\ds\frac{\lambda_i}{(4\pi)^2}\biggl[2\lambda_i^2+2\Sigma_{\lambda}+3\Sigma_{\kappa}+
\biggl(3h_t^2+3h_b^2+h_{\tau}^2\biggr)\delta_{i3}\\[4mm]
&&-3g_2^2-\dfrac{3}{5}g_1^2-\dfrac{19}{10} g^{'2}_1+\dfrac{\beta^{(2)}_{\lambda_i}}{(4\pi)^2}\biggr]\,,\\[4mm]
\ds\frac{d\kappa_i}{dt}&=&\ds\frac{\kappa_i}{(4\pi)^2}\biggl[2\kappa^2_i+2\Sigma_{\lambda}+3\Sigma_{\kappa}-\ds\frac{16}{3}g_3^2-
\frac{4}{15}g_1^2-\frac{19}{10} g^{'2}_1+\frac{\beta^{(2)}_{\kappa_i}}{(4\pi)^2}\biggr]\,,\\[4mm]
\ds\frac{dh_t}{dt}&=&\ds\frac{h_t}{(4\pi)^2}\biggl[\lambda^2+6h_t^2+h_b^2-\ds\frac{16}{3}g_3^2-3g_2^2-\ds\frac{13}{15}g_1^2-
\frac{3}{10} g^{'2}_1+\frac{\beta^{(2)}_{h_t}}{(4\pi)^2}\biggr]\,,\\[4mm]
\ds\frac{dh_b}{dt}&=&\ds\frac{h_b}{(4\pi)^2}\biggl[\lambda^2+h_t^2+6h_b^2+h_{\tau}^2-\ds\frac{16}{3}g_3^2-3g_2^2-\ds\frac{7}{15}g_1^2-
\frac{7}{10} g^{'2}_1+\frac{\beta^{(2)}_{h_b}}{(4\pi)^2}\biggr]\,,\\[4mm]
\ds\frac{dh_{\tau}}{dt}&=&\ds\frac{h_{\tau}}{(4\pi)^2}\biggl[\lambda^2+3h_b^2+4h_{\tau}^2-3g_2^2-\frac{9}{5}g_1^2-
\frac{7}{10} g^{'2}_1+\frac{\beta^{(2)}_{h_{\tau}}}{(4\pi)^2}
\biggr]\,,
\ea
\label{eq4}
\ee
where the two--loop contributions to the corresponding $\beta$--functions are given by
\be
\ba{rcl}
\beta^{(2)}_{\lambda_i}&=&-2\lambda_i^2\biggl(\lambda_i^2+2\Sigma_{\lambda}+3\Sigma_{\kappa}\biggr)-4\Pi_{\lambda}-6\Pi_{\kappa}\\[4mm]
&&-\lambda^2\biggl(3h_t^2+3h_b^2+h_{\tau}^2\biggr)(2+\delta_{i3})-\biggl[9h_t^4+9h_b^4+6h_t^2h_b^2+3h_{\tau}^4\biggr]\delta_{i3}\\[4mm]
&&+16g_3^2\Sigma_{\kappa}+6g_2^2\Sigma_{\lambda}+g_1^2\left(\ds\frac{4}{5}\Sigma_{\kappa}+\ds\frac{6}{5}\Sigma_{\lambda}\right)
+g_1^{'2}\left(\ds\frac{5}{2}\lambda_i^2-\ds\frac{9}{5}\Sigma_{\kappa}-\ds\frac{6}{5}\Sigma_{\lambda}\right)\\[4mm]
&&+\biggl[16g_3^2\biggl(h_t^2+h_b^2\biggr)+g_1^2\left(\ds\frac{4}{5}h_t^2-\ds\frac{2}{5}h_b^2+\ds\frac{6}{5}h_{\tau}^2\right)\\[4mm]
&&+g_1^{'2}\left(-\ds\frac{3}{10}h_t^2-\ds\frac{1}{5}h_b^2-\ds\frac{1}{5}h_{\tau}^2\right)\biggr]\delta_{i3}+
3g_2^4\biggl(3N_g-\ds\frac{7}{2}\biggr)\\[4mm]
&&+\ds\frac{3}{5}g_1^4\biggl(3N_g+\ds\frac{9}{10}\biggr)+\ds\frac{19}{10}g_1^{'4}\biggl(3N_g+
\ds\frac{27}{20}\biggr)+\ds\frac{9}{5}g_2^2g_1^2\\[4mm]
&&+\ds\frac{39}{20}g_2^2g_1^{'2}+\ds\frac{39}{100}g_1^2g_1^{'2}\,,\\[7mm]
\beta^{(2)}_{\kappa_i}&=&-2\kappa_i^2\biggl(\kappa_i^2+2\Sigma_{\lambda}+3\Sigma_{\kappa}\biggr)-4\Pi_{\lambda}-6\Pi_{\kappa}-
2 \lambda^2\biggl(3h_t^2+3h_b^2+h_{\tau}^2\biggr)\\[4mm]
&&+16g_3^2\Sigma_{\kappa}+6g_2^2\Sigma_{\lambda}+g_1^2\left(\ds\frac{4}{5}\Sigma_{\kappa}+\ds\frac{6}{5}\Sigma_{\lambda}\right)+
g_1^{'2}\left(\ds\frac{5}{2}\kappa_i^2-\ds\frac{9}{5}\Sigma_{\kappa}-\ds\frac{6}{5}\Sigma_{\lambda}\right)\\[4mm]
&&+\ds\frac{16}{3}g_3^4\left(3N_g-\ds\frac{19}{3}\right)+\ds\frac{4}{15}g_1^4\left(3N_g+\ds\frac{11}{15}\right)+
\ds\frac{19}{10}g_1^{'4}\left(3N_g+\ds\frac{27}{20}\right)\\[4mm]
&&+\ds\frac{64}{45}g_3^2g_1^2+\ds\frac{52}{15}g_3^2g_1^{'2}+\ds\frac{13}{75}g_1^2g_1^{'2}\,,
\ea
\label{eq5}
\ee
$$
\ba{rcl}
\beta^{(2)}_{h_t}&=&-22h_t^4-5h_b^4-5h_t^2h_b^2-h_b^2h_{\tau}^2
-\lambda^2\biggl(\lambda^2+3h_t^2\qquad\qquad\\[2mm]
&&+4h_b^2+h_{\tau}^2+2\Sigma_{\lambda}+3\Sigma_{\kappa}\biggr)+16g_3^2h_t^2+6g_2^2h_t^2+
g_1^2\left(\ds\frac{6}{5}h_t^2+\ds\frac{2}{5}h_b^2\right)\\[2mm]
&&+g_1^{'2}\left(\ds\frac{3}{2}\lambda^2+\ds\frac{3}{10}h_t^2+\ds\frac{3}{5}h_b^2\right)+
\ds\frac{16}{3}g_3^4\left(3N_g-\ds\frac{19}{3}\right)+3g_2^4\biggl(3N_g-\ds\frac{7}{2}\biggr)\\[2mm]
&&+\ds\frac{13}{15}g_1^4\left(3N_g+\ds\frac{31}{30}\right)+\ds\frac{3}{10}g_1^{'4}\left(3N_g+\ds\frac{11}{20}\right)+
8g_3^2g_2^2+\ds\frac{136}{45}g_3^2g_1^2\\[2mm]
&&+\ds\frac{8}{15}g_3^2g_1^{'2}+g_2^2g_1^2+\ds\frac{3}{4}g_2^2g_1^{'2}+\ds\frac{53}{300}g_1^2g_1^{'2}\,,\\[2mm]
\beta^{(2)}_{h_b}&=&-5h_t^4-22h_b^4-5h_t^2h_b^2-3h_b^2h_{\tau}^2-3h_{\tau}^4
-\lambda^2\biggl(\lambda^2+4h_t^2\\[2mm]
&&+3h_b^2+2\Sigma_{\lambda}+3\Sigma_{\kappa}\biggr)+16g_3^2h_b^2+6g_2^2h_b^2+
g_1^2\left(\ds\frac{4}{5}h_t^2+\ds\frac{2}{5}h_b^2+\ds\frac{6}{5}h_{\tau}^2\right)\\[2mm]
&&+g_1^{'2}\left(\lambda^2+\ds\frac{1}{5}h_t^2+h_b^2-\ds\frac{1}{5}h_{\tau}^2\right)+
\ds\frac{16}{3}g_3^4\left(3N_g-\ds\frac{19}{3}\right)+3g_2^4\biggl(3N_g-\ds\frac{7}{2}\biggr)\\[2mm]
&&+\ds\frac{7}{15}g_1^4\left(3N_g+\ds\frac{5}{6}\right)+\ds\frac{7}{10}g_1^{'4}\left(3N_g+\ds\frac{3}{4}\right)+
8g_3^2g_2^2+\ds\frac{8}{9}g_3^2g_1^2+\ds\frac{4}{3}g_3^2g_1^{'2}\\[2mm]
&&+g_2^2g_1^2+\ds\frac{3}{2}g_2^2g_1^{'2}+\ds\frac{49}{150}g_1^2g_1^{'2}\,,\\[2mm]
\beta^{(2)}_{h_{\tau}}&=&-9h_b^4-3h_t^2h_b^2-9h_b^2h_{\tau}^2-10h_{\tau}^4-\lambda^2\biggl(\lambda^2+3h_t^2\\[2mm]
&&+3h_{\tau}^2+2\Sigma_{\lambda}+3\Sigma_{\kappa}\biggr)+16g_3^2h_b^2+6g_2^2h_{\tau}^2+
g_1^2\left(-\ds\frac{2}{5}h_b^2+\ds\frac{6}{5}h_{\tau}^2\right)\\[2mm]
&&+g_1^{'2}\left(\lambda^2-\ds\frac{1}{5}h_b^2+\ds\frac{13}{10}h_{\tau}^2\right)+
3g_2^4\biggl(3N_g-\ds\frac{7}{2}\biggr)+\ds\frac{9}{5}g_1^4\left(3N_g+\ds\frac{3}{2}\right)\\[2mm]
&&+\ds\frac{7}{10}g_1^{'4}\left(3N_g+\ds\frac{3}{4}\right)+\ds\frac{9}{5}g_2^2g_1^2+\ds\frac{39}{20}g_2^2g_1^{'2}+
\ds\frac{51}{100}g_1^2g_1^{'2}\,,
\ea
$$
and
$$
\Pi_{\lambda}=\lambda_1^4+\lambda_2^4+\lambda_3^4\,,\qquad\qquad\qquad\Pi_{\kappa}=\kappa_1^4+\kappa_2^4+\kappa_3^4\,.
$$

Using the two--loop $\beta$--functions for the gauge and Yukawa couplings and a method proposed in \cite{Kazakov:1998uj},
one can obtain the two--loop RGEs for the gaugino masses and trilinear scalar couplings:
$$
\ba{rcl}
\ds\frac{d M_3}{dt}&=&\ds\frac{g_3^2}{16\pi^2}\Biggl[(-18+6N_g)M_3+\ds\frac{1}{16\pi^2}\Biggl(
(-216+136 N_g) g_3^2 M_3+6 N_g\,g_2^2 (M_2+M_3)\\[2mm]
&+& 2 N_g\, g_1^2 (M_1+M_3)+2 N_g\,g_1^{'2} (M'_1+M_3)
-8 h_t^2 (A_t+M_3)-8 h_b^2 (A_b+M_3)\\[2mm]
&-&4 \Sigma_{A_{\kappa}}-4\Sigma_{\kappa} M_3\Biggr)\Biggr]\,,\\[2mm]
\ds\frac{d M_2}{dt}&=&\ds\frac{g_2^2}{16\pi^2}\Biggl[(-10+6N_g)M_2+\ds\frac{1}{16\pi^2}\Biggl(
16 N_g g_3^2 (M_3+M_2)+(-68+84 N_g)g_2^2 M_2\\[2mm]
&+& \left(\ds\frac{6}{5}+2 N_g\right) g_1^2 (M_1+M_2)+
\left(\ds\frac{4}{5}+2 N_g\right) g_1^{'2} (M'_1+M_2)-12 h_t^2 (A_t+M_2)\\[2mm]
&-& 12 h_b^2 (A_b+M_2)-4 h_{\tau}^2 (A_{\tau}+M_2)-4 \Sigma_{A_{\lambda}}-
4 \Sigma_{\lambda}M_2\Biggr)\Biggr]\,,
\ea
$$
\be
\ba{rcl}
\ds\frac{d M_1}{dt}&=&\ds\frac{g_1^2}{16\pi^2}\Biggl[\left(\frac{6}{5}+6N_g\right)M_1+
\ds\frac{1}{16\pi^2}\Biggl(16 N_g g_3^2 (M_3+M_1)\\[3mm]
&+& \left(\ds\frac{18}{5}+6N_g\right)g_2^2(M_2+M_1) + \left(\ds\frac{36}{25}+12 N_g\right) g_1^2 M_1\\[3mm]
&+& \left(\ds\frac{12}{25}+2 N_g\right) g_1^{'2} (M'_1+M_1) - \ds\frac{52}{5} h_t^2 (A_t+M_1)-
\ds\frac{28}{5}h_b^2 (A_b+M_1)\\[3mm]
&-&\ds\frac{36}{5}h_{\tau}^2 (A_{\tau}+M_1) - \ds\frac{12}{5}\Sigma_{A_{\lambda}}-\ds\frac{12}{5} \Sigma_{\lambda} M_1-
\ds\frac{8}{5} \Sigma_{A_{\kappa}}- \ds\frac{8}{5} \Sigma_{\kappa} M_1\Biggr)\Biggr]\,,\\[4mm]
\ds\frac{d M'_1}{dt}&=&\ds\frac{g_1^{'2}}{16\pi^2}\Biggl[\left(\ds\frac{4}{5}+6N_g\right)M'_1
+\ds\frac{1}{16\pi^2}\Biggl(16 N_g g_3^2 (M_3+M'_1)\\[3mm]
&+& \left(\ds\frac{12}{5}+6N_g\right)g_2^2 (M_2+M'_1)+\left(\ds\frac{12}{25}+ 2 N_g\right) g_1^2(M_1+M'_1)\\[3mm]
&+&\left(\ds\frac{16}{25}+
12N_g\right) g_1^{'2}M'_1 - \ds\frac{18}{5} h_t^2 (A_t+M'_1)-\ds\frac{42}{5}h_b^2 (A_b+M'_1)\\[3mm]
&-&\ds\frac{14}{5}h_{\tau}^2 (A_{\tau}+M'_1) -\ds\frac{38}{5}\Sigma_{A_{\lambda}}-\ds\frac{38}{5} \Sigma_{\lambda} M'_1-
\ds\frac{57}{5} \Sigma_{A_{\kappa}}- \ds\frac{57}{5} \Sigma_{\kappa} M'_1\Biggr)\Biggr]\,,
\ea
\label{eq6}
\ee
\vspace{7mm}
\be
\ba{rcl}
\ds\frac{dA_{\lambda_i}}{dt}&=&\ds\frac{1}{(4\pi)^2}\biggl[4\lambda^2_i A_{\lambda_i}+4\Sigma_{A_\lambda}+6\Sigma_{A_{\kappa}}+
(6h_t^2 A_t+6h_b^2 A_b+2h_{\tau}^2 A_{\tau})\,\delta_{i3}\qquad\\[3mm]
&&\ds-6g_2^2 M_2-\frac{6}{5}g_1^2 M_1-\frac{19}{5}g_1^{'2} M'_1+\frac{\beta^{(2)}_{A_{\lambda_i}}}{(4\pi)^2}
\biggr]\,,\\[3mm]
\ds\frac{dA_{\kappa_i}}{dt}&=&\ds\frac{1}{(4\pi)^2}\biggl[4\kappa_i^2 A_{\kappa_i}+4\Sigma_{A_\lambda}
+6\Sigma_{A_{\kappa}}-\ds\frac{32}{3}g_3^2M_3-\frac{8}{15}g_1^2M_1\qquad\\[3mm]
&&\ds-\frac{19}{5}g_1^{'2} M'_1+\frac{\beta^{(2)}_{A_{\kappa_i}}}{(4\pi)^2}\biggr]\,,\\[3mm]
\ds\frac{dA_t}{dt}&=&\ds\frac{1}{(4\pi)^2}\biggl[2\lambda^2 A_{\lambda}+12h_t^2A_t+2h_b^2A_b
-\ds\frac{32}{3}g_3^2M_3-6g_2^2M_2\\[3mm]
&&-\ds\frac{26}{15}g_1^2M_1-\frac{3}{5}g_1^{'2} M'_1+\frac{\beta^{(2)}_{A_t}}{(4\pi)^2}\biggr]\,,\\[3mm]
\ds\frac{dA_b}{dt}&=&\ds\frac{1}{(4\pi)^2}\biggl[2\lambda^2 A_{\lambda}+2 h_t^2 A_t+12 h_b^2 A_b+2 h_{\tau}^2 A_{\tau}-
\ds\frac{32}{3}g_3^2M_3-6g_2^2M_2\\[3mm]
&&-\ds\frac{14}{15}g_1^2 M_1-\frac{7}{5}g_1^{'2} M'_1+\frac{\beta^{(2)}_{A_b}}{(4\pi)^2}\biggr]\,,\\[3mm]
\ds\frac{dA_{\tau}}{dt}&=&\ds\frac{1}{(4\pi)^2}\biggl[2\lambda^2 A_{\lambda}+6h_b^2 A_{b}+8h_{\tau}^2 A_{\tau}-
6g_2^2M_2-\frac{18}{5}g_1^2M_1-\frac{7}{5}g_1^{'2} M'_1+\frac{\beta^{(2)}_{A_{\tau}}}{(4\pi)^2}
\biggr]\,,
\ea
\ee
where the two--loop contributions to the $\beta$--functions of trilinear scalar couplings are given by
$$
\ba{rcl}
\beta^{(2)}_{A_{\lambda_i}}&=&-4\lambda_i^2\biggl(\lambda_i^2+2\Sigma_{\lambda}+3\Sigma_{\kappa}\biggr)A_{\lambda_i}-
4\lambda_i^2\biggl(\lambda_i^2 A_{\lambda_i}+2\Sigma_{A_{\lambda}}+3\Sigma_{A_{\kappa}}\biggr)-16\Pi_{A_{\lambda}}-24\Pi_{A_{\kappa}}\\[2mm]
&&-2\lambda^2\biggl(3h_t^2+3h_b^2+h_{\tau}^2\biggr)(2+\delta_{i3})A_{\lambda}
-2\lambda^2\biggl(3h_t^2 A_t + 3h_b^2 A_b + h_{\tau}^2 A_{\tau}\biggr)(2+\delta_{i3})\\[2mm]
&&-12\biggl[3 h_t^4 A_t + 3 h_b^4 A_b+ h_t^2 h_b^2(A_t+A_b)+ h_{\tau}^4 A_{\tau}\biggr]\delta_{i3}
+ 32 g_3^2 \biggl(\Sigma_{\kappa} M_3+\Sigma_{A_{\kappa}}\biggr)\\[2mm]
&&+ 12 g_2^2 \biggl(\Sigma_{\lambda} M_2 + \Sigma_{A_{\lambda}}\biggr)+
2 g_1^2\biggl[\left(\ds\frac{4}{5}\Sigma_{\kappa}+\ds\frac{6}{5}\Sigma_{\lambda}\right)M_1+\ds\frac{4}{5}\Sigma_{A_{\kappa}}+
\ds\frac{6}{5}\Sigma_{A_{\lambda}}\biggr]\\[2mm]
&&+ 2 g_1^{'2}\biggl[\left(\ds\frac{5}{2}\lambda_i^2-
\ds\frac{9}{5}\Sigma_{\kappa}-\ds\frac{6}{5}\Sigma_{\lambda}\right)M'_1+\dfrac{5}{2}\lambda_i^2 A_{\lambda_i}-
\dfrac{9}{5}\Sigma_{A_{\kappa}}-\dfrac{6}{5}\Sigma_{A_{\lambda}}\biggr]\\[2mm]
&&+32 g_3^2\biggl[\biggl(h_t^2+h_b^2\biggr)M_3 + h_t^2 A_t+ h_b^2 A_b\biggr]\delta_{i3}
+2 g_1^2\biggl[\left(\ds\frac{4}{5}h_t^2-\ds\frac{2}{5}h_b^2+\ds\frac{6}{5}h_{\tau}^2\right)M_1 +\dfrac{4}{5}h_t^2 A_t\\[2mm]
&&-\dfrac{2}{5} h_b^2 A_b + \dfrac{6}{5}h_{\tau}^2 A_{\tau}\biggr]\delta_{i3}
+ g_1^{'2}\biggl[\left(-\ds\frac{3}{5}h_t^2-\ds\frac{2}{5}h_b^2-\ds\frac{2}{5}h_{\tau}^2\right)M'_1-\dfrac{3}{5}h_t^2 A_t -\dfrac{2}{5}h_b^2
A_b\\[2mm]
&&- \dfrac{2}{5}h_{\tau}^2 A_{\tau}\biggr]\delta_{i3}+
12 g_2^4\biggl(3N_g-\ds\frac{7}{2}\biggr)M_2+\ds\frac{12}{5}g_1^4\biggl(3N_g+\ds\frac{9}{10}\biggr)M_1
+\ds\frac{38}{5}g_1^{'4}\biggl(3N_g+\ds\frac{27}{20}\biggr)M'_1\\[2mm]
&&+\dfrac{18}{5}g_2^2g_1^2\biggl(M_2+M_1\biggr)+\ds\frac{39}{10}g_2^2g_1^{'2}\biggl(M_2+M'_1\biggr)+
\ds\frac{39}{50}g_1^2g_1^{'2}\biggl(M_1+M'_1\biggr)\,,
\ea
$$
$$
\ba{rcl}
\beta^{(2)}_{A_{\kappa_i}}&=&-4\kappa_i^2\biggl(\kappa_i^2+2\Sigma_{\lambda}+3\Sigma_{\kappa}\biggr)A_{\kappa_i}
-4\kappa_i^2\biggl(\kappa_i^2 A_{\kappa_i} + 2 \Sigma_{A_{\lambda}} + 3 \Sigma_{A_{\kappa}}\biggr)-16\Pi_{A_{\lambda}}-24\Pi_{A_{\kappa}}\\[2mm]
&&-4 \lambda^2\biggl(3h_t^2+3h_b^2+h_{\tau}^2\biggr) A_{\lambda}
-4 \lambda^2\biggl(3h_t^2 A_t + 3 h_b^2 A_b + h_{\tau}^2 A_{\tau}\biggr)+32 g_3^2 \biggl(\Sigma_{\kappa} M_3+\Sigma_{A_{\kappa}}\biggr)\\[2mm]
&&+ 12 g_2^2 \biggl(\Sigma_{\lambda} M_2+ \Sigma_{A_{\lambda}}\biggr)+ 2 g_1^2\biggl[\left(\ds\frac{4}{5}\Sigma_{\kappa}
+\ds\frac{6}{5}\Sigma_{\lambda}\right)M_1 + \dfrac{4}{5}\Sigma_{A_{\kappa}} + \dfrac{6}{5}\Sigma_{A_{\lambda}}\biggr]\\[2mm]
&&+2g_1^{'2}\biggl[\left(\ds\frac{5}{2}\kappa_i^2-\ds\frac{9}{5}\Sigma_{\kappa}-\ds\frac{6}{5}\Sigma_{\lambda}\right)M'_1+
\dfrac{5}{2}\kappa_i^2 A_{\kappa_i}-\dfrac{9}{5}\Sigma_{A_{\kappa}}-\dfrac{6}{5}\Sigma_{A_{\lambda}}\biggr]\\[2mm]
&&+\ds\frac{64}{3}g_3^4\left(3N_g-\ds\frac{19}{3}\right)M_3+\ds\frac{16}{15}g_1^4\left(3N_g+\ds\frac{11}{15}\right)M_1+
\ds\frac{38}{5}g_1^{'4}\left(3N_g+\ds\frac{27}{20}\right)M'_1\\[2mm]
&&+\ds\frac{128}{45}g_3^2g_1^2\biggl(M_3+M_1\biggr)+
\ds\frac{104}{15}g_3^2g_1^{'2}\biggl(M_3+M'_1\biggr)+
\ds\frac{26}{75}g_1^2g_1^{'2}\biggl(M_1+M'_1\biggr)\,,\\[2mm]
\beta^{(2)}_{A_t}&=&-88 h_t^4 A_t - 20 h_b^4 A_b - 10 h_t^2h_b^2 \biggl(A_t+A_b\biggr) - 2 h_b^2 h_{\tau}^2 \biggl(A_b+A_{\tau}\biggr)
-2\lambda^2\biggl[\biggl(2\lambda^2+3h_t^2\\[2mm]
&&+4h_b^2+h_{\tau}^2+2\Sigma_{\lambda}+3\Sigma_{\kappa}\biggr)A_{\lambda}
+ 3 h_t^2 A_t + 4 h_b^2 A_b + h_{\tau}^2 A_{\tau} + 2\Sigma_{A_{\lambda}} + 3\Sigma_{A_{\kappa}}\biggr]\\[2mm]
&&+32 g_3^2 h_t^2 \biggl( M_3+A_t \biggr) + 12 g_2^2h_t^2 \biggl( M_2 + A_t \biggr)+ 2 g_1^2 \biggl[
\left(\ds\frac{6}{5}h_t^2+\ds\frac{2}{5}h_b^2\right) M_1 + \dfrac{6}{5} h_t^2 A_t \\[2mm]
&&+\dfrac{2}{5} h_b^2 A_b\biggr] + 2 g_1^{'2}\biggl[\left(\ds\frac{3}{2}\lambda^2+\ds\frac{3}{10}h_t^2+\ds\frac{3}{5}h_b^2\right)M'_1+
\dfrac{3}{2}\lambda^2 A_{\lambda} +\dfrac{3}{10}h_t^2 A_t + \dfrac{3}{5}h_b^2 A_b\biggr]\\[2mm]
&&+\ds\frac{64}{3}g_3^4\left(3N_g-\ds\frac{19}{3}\right) M_3
+ 12 g_2^4\biggl(3N_g-\ds\frac{7}{2}\biggr) M_2 + \ds\frac{52}{15}g_1^4\left(3N_g+\ds\frac{31}{30}\right) M_1\\[2mm]
&&+\ds\frac{6}{5}g_1^{'4}\left(3N_g+\ds\frac{11}{20}\right)M'_1+ 16 g_3^2 g_2^2 \biggl( M_3+M_2 \biggr)
+\ds\frac{272}{45}g_3^2g_1^2\biggl( M_3+M_1 \biggr)\\[2mm]
&&+\ds\frac{16}{15}g_3^2g_1^{'2}\biggl( M_3+M'_1 \biggr)
+2 g_2^2g_1^2 \biggl( M_2 + M_1 \biggr)+\ds\frac{3}{2}g_2^2g_1^{'2}\biggl( M_2+M'_1 \biggr)
+\ds\frac{53}{150}g_1^2g_1^{'2}\biggl( M_1+M'_1 \biggr)\,,
\ea
$$
$$
\ba{rcl}
\beta^{(2)}_{A_b}&=&-20 h_t^4 A_t - 88 h_b^4 A_b - 10 h_t^2h_b^2 \biggl(A_t + A_b \biggr) - 6 h_b^2 h_{\tau}^2 \biggl( A_b + A_{\tau} \biggr)
-12 h_{\tau}^4 A_{\tau}\\[2mm]
&&-2\lambda^2\biggl[\biggl(2\lambda^2+4h_t^2+3h_b^2+2\Sigma_{\lambda}+3\Sigma_{\kappa}\biggr)A_{\lambda}+
4 h_t^2 A_t + 3 h_b^2 A_b + 2 \Sigma_{A_{\lambda}} + 3 \Sigma_{A_{\kappa}}\biggr]\\[2mm]
&&+ 32 g_3^2 h_b^2 \biggl( M_3 + A_b \biggr) + 12 g_2^2 h_b^2 \biggl( M_2 + A_b \biggr)
+ 4 g_1^2\biggl[ \left(\ds\frac{2}{5}h_t^2+\ds\frac{1}{5}h_b^2+\ds\frac{3}{5}h_{\tau}^2\right)M_1\\[2mm]
&&+\dfrac{2}{5}h_t^2 A_t + \dfrac{1}{5}h_b^2 A_b +\dfrac{3}{5}h_{\tau}^2 A_{\tau}\biggr]
+ 2 g_1^{'2} \biggl[\left(\lambda^2+\ds\frac{1}{5}h_t^2+h_b^2-\ds\frac{1}{5}h_{\tau}^2\right)M'_1+
\lambda^2 A_{\lambda} \\[2mm]
&& +\dfrac{1}{5} h_t^2 A_t + h_b^2 A_b - \dfrac{1}{5}h_{\tau}^2 A_{\tau}\biggr] +
\dfrac{64}{3}g_3^4\left(3N_g-\ds\frac{19}{3}\right) M_3
+12 g_2^4 \biggl(3N_g-\ds\frac{7}{2}\biggr) M_2\\[2mm]
&&+\dfrac{28}{15}g_1^4\left(3N_g+\ds\frac{5}{6}\right) M_1 +\dfrac{14}{5}g_1^{'4}\left(3N_g+\ds\frac{3}{4}\right)M'_1+
16 g_3^2g_2^2\biggl( M_3+M_2\biggr)\\[2mm]
&&+\dfrac{16}{9}g_3^2g_1^2 \biggl( M_3+M_1\biggr) + \dfrac{8}{3}g_3^2g_1^{'2} \biggl( M_3+M'_1\biggr)+
2 g_2^2g_1^2 \biggl( M_2 + M_1\biggr)\\[2mm]
&&+3 g_2^2 g_1^{'2} \biggl( M_2 + M'_1\biggr)+\dfrac{49}{75}g_1^2g_1^{'2} \biggl( M_1 + M'_1\biggr)\,,\\[2mm]
\beta^{(2)}_{A_{\tau}}&=&-36 h_b^4 A_b - 6 h_t^2h_b^2 \biggl(A_t+A_b\biggr) - 18 h_b^2h_{\tau}^2 \biggl(A_b+A_{\tau}\biggr)
- 40 h_{\tau}^4 A_{\tau} -2\lambda^2\biggl[\biggl(2\lambda^2+3h_t^2\\[2mm]
&&+3h_{\tau}^2+2\Sigma_{\lambda}+3\Sigma_{\kappa}\biggr) A_{\lambda} + 3h_t^2 A_t + 3h_{\tau}^2 A_{\tau} +
2\Sigma_{A_{\lambda}}+3\Sigma_{A_{\kappa}} \biggr] + 32 g_3^2 h_b^2 \biggl(M_3+A_b \biggr)\\[2mm]
&&+12 g_2^2 h_{\tau}^2 \biggl( M_2 + A_{\tau}\biggr) + 4 g_1^2 \biggl[\left(-\ds\frac{1}{5}h_b^2+\ds\frac{3}{5}h_{\tau}^2\right) M_1
-\dfrac{1}{5}h_b^2 A_b +\dfrac{3}{5}h_{\tau}^2 A_{\tau} \biggr]\\[2mm]
&&+ 2 g_1^{'2}\biggl[\left(\lambda^2-\ds\frac{1}{5}h_b^2+\ds\frac{13}{10}h_{\tau}^2\right)M'_1+
\lambda^2 A_{\lambda}-\dfrac{1}{5}h_b^2 A_b + \dfrac{13}{10}h_{\tau}^2 A_{\tau}\biggr]\\[2mm]
&&+ 12 g_2^4\biggl(3N_g-\ds\frac{7}{2}\biggr) M_2 + \ds\frac{36}{5}g_1^4\left(3N_g+\ds\frac{3}{2}\right) M_1
+\dfrac{14}{5}g_1^{'4}\left(3N_g+\ds\frac{3}{4}\right) M'_1\\[2mm]
&&+\ds\frac{18}{5}g_2^2g_1^2 \biggl(M_2+M_1\biggr)+\ds\frac{39}{10}g_2^2g_1^{'2} \biggl(M_2+M'_1\biggr)+
\ds\frac{51}{50}g_1^2g_1^{'2}\biggl(M_1+M'_1\biggr)\,,
\ea
$$
whereas
$$
\ba{rclcrcl}
\Sigma_{A_{\lambda}}&=&\lambda_1^2 A_{\lambda_1}+\lambda_2^2 A_{\lambda_2}+\lambda_3^2 A_{\lambda_3}\,,&\qquad\qquad &
\Sigma_{A_{\kappa}}&=&\kappa_1^2 A_{\kappa_1}+\kappa_2^2 A_{\kappa_2}+\kappa_3^2 A_{\kappa_3}\,,\\
\Pi_{\lambda} &=& \lambda_1^4 A_{\lambda_1}+\lambda_2^4 A_{\lambda_2} +\lambda_3^4 A_{\lambda_3}\,,&\qquad\qquad &
\Pi_{\kappa} &=& \kappa_1^4 A_{\kappa_1} + \kappa_2^4 A_{\kappa_2} +\kappa_3^4 A_{\kappa_3}\,.
\ea
$$

The one--loop RGEs for the soft scalar masses can be written as
$$
\ba{rcl}
\ds\frac{d m_{S_i}^2}{dt}&=&\ds\frac{1}{(4\pi)^2}\biggl[\sum_{j=1..3}
4\lambda_j^2\biggl(m_{H^u_{j}}^2+m_{H^d_{j}}^2+m_{S}^2+A_{\lambda_j}^2\biggr)\delta_{i3}\\[4mm]
&&\ds+\sum_{j=1..3} 6\kappa_j^2\biggl(m_{S}^2+m^2_{D_j}+m^2_{\overline{D}_j}+A_{\kappa_j}^2\biggr)\delta_{i3}
-5g_1^{'2}M^{'2}_1+\ds\frac{g^{'2}_1}{4}\Sigma'_1\biggr]\,,\\[4mm]
\ds\frac{d m_{H^u_{i}}^2}{dt}&=&\ds\frac{1}{(4\pi)^2}\biggl[2\lambda_i^2\biggl(m_{H^u_{i}}^2+m_{H^d_{i}}^2+m_S^2+A_{\lambda_i}^2\biggr)+
6h_t^2\biggl(m^2_{H_u}+m^2_{Q}+m^2_{t^c}+A_t^2\biggr)\delta_{i3}\qquad\qquad\qquad\qquad\\[4mm]
&&\ds-6g_2^2M_2^2-\ds\frac{6}{5}g_1^2 M_1^2-\frac{4}{5} g^{'2}_1 M^{'2}_1+\frac{3}{5}g_1^2\Sigma_1-\ds\frac{g^{'2}_1}{10}\Sigma'_1\biggr]\,,
\ea
$$
\be
\ba{rcl}
\ds\frac{d m_{H^d_{i}}^2}{dt}&=&\ds\frac{1}{(4\pi)^2}\biggl[2\lambda_i^2\biggl(m_{H^u_{i}}^2+m_{H^d_{i}}^2+m_S^2+A_{\lambda_i}^2\biggr)
+6h_b^2\biggl(m^2_{H_d}+m^2_{Q}+m^2_{b^c}+A_b^2\biggr)\delta_{i3}\\[2mm]
&&+2h_{\tau}^2\biggl(m_{H_d}^2+m_{L}^2+m^2_{\tau^c}+A_{\tau}^2\biggr)\delta_{i3}-6g_2^2M_2^2-
\ds\frac{6}{5}g_1^2 M_1^2-\ds\frac{9}{5} g^{'2}_1 M^{'2}_1\\[2mm]
&&\ds-\frac{3}{5}g_1^2\Sigma_1-\ds\frac{3}{20}g^{'2}_1\Sigma'_1\biggr]\,,\\[2mm]
\ds\frac{d m_{Q_i}^2}{dt}&=&\ds\frac{1}{(4\pi)^2}\biggl[2h_t^2\biggl(m^2_{H_u}+m^2_{Q}+m^2_{t^c}+A_t^2\biggr)\delta_{i3}+
2h_b^2\biggl(m^2_{H_d}+m^2_{Q}+m^2_{b^c}+A_b^2\biggr)\delta_{i3}\\[2mm]
&&\ds-\frac{32}{3}g_3^2M_3^2-6g_2^2M_2^2-\ds\frac{2}{15}g_1^2M_1^2-\ds\frac{1}{5} g^{'2}_1 M^{'2}_1
+\frac{1}{5}g_1^2\Sigma_1+\ds\frac{g^{'2}_1}{20}\Sigma'_1\biggr]\,,\\[2mm]
\ds\frac{d m_{u^c_i}^2}{dt}&=&\ds\frac{1}{(4\pi)^2}\biggl[4h_t^2\biggl(m^2_{H_u}+m^2_{Q}+m^2_{t^c}+A_t^2\biggr)\delta_{i3}-\frac{32}{3}g_3^2M_3^2-
\ds\frac{32}{15}g_1^2 M_1^2-\ds\frac{1}{5} g^{'2}_1 M^{'2}_1\\[2mm]
&&\ds-\frac{4}{5}g_1^2\Sigma_1+\ds\frac{g^{'2}_1}{20}\Sigma'_1\biggr]\,,\\[2mm]
\ds\frac{d m_{d^c_i}^2}{dt}&=&\ds\frac{1}{(4\pi)^2}\biggl[4h_b^2\biggl(m^2_{H_d}+m^2_{Q}+m^2_{b^c}+A_b^2\biggr)\delta_{i3}-\frac{32}{3}g_3^2M_3^2-
\ds\frac{8}{15}g_1^2 M_1^2-\ds\frac{4}{5} g^{'2}_1 M^{'2}_1\\[2mm]
&&\ds+\frac{2}{5}g_1^2\Sigma_1+\ds\frac{g^{'2}_1}{10}\Sigma'_1\biggr]\,,\\[2mm]
\ds\frac{dm_{L_i}^2}{dt}&=&\ds\frac{1}{(4\pi)^2}\biggl[2h_{\tau}^2\biggl(m^2_{H_d}+m^2_{L}+m^2_{\tau^c}+A_{\tau}^2\biggr)\delta_{i3}+
-6g_2^2M_2^2-\ds\frac{6}{5}g_1^2 M_1^2-\ds\frac{4}{5} g^{'2}_1 M^{'2}_1\\[2mm]
&&\ds-\frac{3}{5}g_1^2\Sigma_1+\ds\frac{g^{'2}_1}{10}\Sigma'_1\biggr]\,,\\[2mm]
\ds\frac{d m_{e^c_i}^2}{dt}&=&\ds\frac{1}{(4\pi)^2}\biggl[4h_{\tau}^2\biggl(m^2_{H_d}+m^2_{L}+m^2_{\tau^c}+A_{\tau}^2\biggr)\delta_{i3}-
\ds\frac{24}{5}g_1^2 M_1^2-\ds\frac{1}{5} g^{'2}_1 M^{'2}_1\\[2mm]
&&\ds+\frac{6}{5}g_1^2\Sigma_1+\ds\frac{g^{'2}_1}{20}\Sigma'_1\biggr]\,,\\[2mm]
\ds\frac{d m_{D_i}^2}{dt}&=&\ds\frac{1}{(4\pi)^2}\biggl[2\kappa_i^2\biggl(m_S^2+m^2_{D_i}+m^2_{\overline{D}_i}+A_{\kappa_i}^2\biggr)-
\ds\frac{32}{3}g_3^2M_3^2-\ds\frac{8}{15}g_1^2 M_1^2-\ds\frac{4}{5} g^{'2}_1 M^{'2}_1\\[2mm]
&&\ds-\frac{2}{5}g_1^2\Sigma_1-\ds\frac{g^{'2}_1}{10}\Sigma'_1\biggr]\,,\\[2mm]
\ds\frac{d m_{\overline{D}_i}^2}{dt}&=&\ds\frac{1}{(4\pi)^2}\biggl[2\kappa_i^2\biggl(m_S^2+m^2_{D_i}+m^2_{\overline{D}_i}+A_{\kappa_i}^2\biggr)-
\frac{32}{3}g_3^2M_3^2-\ds\frac{8}{15}g_1^2M_1^2-\ds\frac{9}{5} g^{'2}_1 M^{'2}_1\\[2mm]
&&\ds+\frac{2}{5}g_1^2\Sigma_1-\ds\frac{3}{20}g^{'2}_1\Sigma'_1\biggr]\,,\\[2mm]
\ds\frac{d m^2_{H'}}{dt}&=&\ds\frac{1}{(4\pi)^2}\biggl[-6g_2^2M_2^2-\ds\frac{6}{5}g_1^2 M_1^2-\ds\frac{4}{5} g^{'2}_1 M^{'2}_1
-\frac{3}{5}g_1^2\Sigma_1+\ds\frac{g^{'2}_1}{10}\Sigma'_1\biggr]\,,\\[2mm]
\ds\frac{d m^2_{\overline{H^{'}}} }{dt}&=&\ds\frac{1}{(4\pi)^2}\biggl[-6g_2^2M_2^2-\ds\frac{6}{5}g_1^2 M_1^2-\ds\frac{4}{5} g^{'2}_1 M^{'2}_1
+\frac{3}{5}g_1^2\Sigma_1-\ds\frac{g^{'2}_1}{10}\Sigma'_1\biggr]\,,
\ea
\label{eq8}
\ee
where
$$
\Sigma_1=\sum_{i=1}^3\biggl(m^2_{Q_i}-2m^2_{u^c_i}+m^2_{d^c_i}+m^2_{e^c_i}-m^2_{L_i}+m^2_{H^u_{i}}-m^2_{H^d_{i}}+
m^2_{\overline{D}_i}-m^2_{D_i}\biggr)-m^2_{H'}+m^2_{\overline{H}'}\,,
$$
$$
\Sigma'_1=\sum_{i=1}^3\biggl(6 m^2_{Q_i}+3 m^2_{u^c_i}+6 m^2_{d^c_i}+ m^2_{e^c_i}+4 m^2_{L_i}-4 m^2_{H^u_{i}}-6 m^2_{H^d_{i}}
+5 m^2_{S_i}-9 m^2_{\overline{D}_i}-6 m^2_{D_i}\biggr)+4 m^2_{H'}-4 m^2_{\overline{H}'}\,.
$$

\end{appendix}

%section the requirement of validity of perturbation theory up to
%the GUT scale leads to an upper bound $\lambda \leq \lambda_{\max}$.

%where $\bar{g}=\sqrt{g_2^2+g'^2}$. Instead of $v_1$ and $v_2$ it is more convenient to use $\tan\beta$ and
%$v$ defined above. To simplify the analysis of the Higgs spectrum it is worth to express the soft masses $m_1^2,\,m_2^2,\,m_{S}^2$
%in terms of $s,\, v$, $\tan\beta$ and other parameters. Because from precision measurements we know that $v=246\,\mbox{GeV}$
%the tree--level Higgs masses and couplings depend on four variables only:
%\be
%\lambda\,,\qquad s\,,\qquad \tan\beta\,,\qquad A_{\lambda}\,.
%\label{43}
%\ee

\end{document}